\documentclass[aoas,preprint]{imsart}

\usepackage{tikz}
\usetikzlibrary{fit,positioning}
\usepackage{graphicx,psfrag,epsf}
\usepackage{enumerate}
\usepackage{url} 
\usepackage{epstopdf}
\usepackage{subcaption}
\usepackage{algorithm}
\usepackage{algpseudocode}
\usepackage{subcaption}

\usepackage{afterpage}

\RequirePackage{amsthm,amsmath,amsfonts,amssymb}
\RequirePackage[authoryear]{natbib}
\RequirePackage[colorlinks,citecolor=blue,urlcolor=blue]{hyperref}
\RequirePackage{graphicx}

\theoremstyle{remark}
\newtheorem{remark}{Remark}

\startlocaldefs

\newtheorem{theorem}{Theorem}[section]

\theoremstyle{remark}

\endlocaldefs

\usepackage{amsmath,amsfonts,amssymb}
\usepackage{graphicx,psfrag,epsf}
\usepackage{enumerate}

\newcommand{\bbR}{\mathbb{R}}
\newcommand{\sC}{\mathcal{C}}
\newcommand{\sD}{\mathcal{D}}

\newcommand{\sI}{\mathcal{I}}
\newcommand{\sK}{\mathcal{K}}

\newcommand{\sL}{\mathcal{L}}
\newcommand{\sN}{\mathcal{N}}

\newcommand{\vb}{\mathbf{b}}

\newcommand{\vd}{\mathbf{d}}

\newcommand{\vf}{\mathbf{f}}

\newcommand{\vh}{\mathbf{h}}

\newcommand{\vm}{\mathbf{m}}

\newcommand{\vx}{\mathbf{x}}
\newcommand{\vy}{\mathbf{y}}

\newcommand{\valpha}{\boldsymbol{\alpha}}
\newcommand{\vbeta}{\boldsymbol{\beta}}
\newcommand{\vtheta}{\boldsymbol{\theta}}

\newcommand{\vmu}{\boldsymbol{\mu}}

\newcommand{\vtau}{\boldsymbol{\tau}}
\newcommand{\vgamma}{\boldsymbol{\gamma}}

\newcommand{\vzero}{\mathbf{0}}
\newcommand{\vone}{\mathbf{1}}

\newcommand{\mB}{\mathbf{B}}
\newcommand{\mC}{\mathbf{C}}

\newcommand{\mH}{\mathbf{H}}
\newcommand{\mI}{\mathbf{I}}

\newcommand{\mV}{\mathbf{V}}

\newcommand{\mX}{\mathbf{X}}

\newcommand{\mTheta}{\boldsymbol{\Theta}}
\newcommand{\mSigma}{\boldsymbol{\Sigma}}

\newcommand{\mOmega}{\boldsymbol{\Omega}}

\newcommand{\Expect}{\mathbb{E}}
\newcommand{\intd}{\,\mathrm{d}}
\newcommand{\diag}{\mathrm{diag}}
\newcommand{\Var}{\mathrm{Var}}

\newcommand{\tr}{\mathrm{tr}}

\begin{document}

\begin{frontmatter}
\title{Simultaneous inference of periods and period-luminosity relations for Mira variable stars}
\runtitle{Simultaneous inference of periods and PLRs}

\begin{aug}
\author[A]{\fnms{Shiyuan} \snm{He}\ead[label=e1]{heshiyuan@ruc.edu.cn}},
\author[B]{\fnms{Zhenfeng} \snm{Lin}\ead[label=e2]{zhenfeng.lin@microsoft.com}},
\author[C]{\fnms{Wenlong} \snm{Yuan}\ead[label=e3]{wyuan10@jhu.edu}}, \\
\author[D]{\fnms{Lucas M.} \snm{Macri}\ead[label=e4]{lmacri@tamu.edu}}
\and
\author[E]{\fnms{Jianhua Z.} \snm{Huang}\ead[label=e5]{jianhua@stat.tamu.edu}}

\address[A]{ Institute of Statistics and Big Data, Renmin University of China,
\href{mailto:heshiyuan@ruc.edu.cn}{heshiyuan@ruc.edu.cn} }
\address[B]{Microsoft, \href{mailto:zhenfeng.lin@microsoft.com}{zhenfeng.lin@microsoft.com}}
\address[C]{Department of Physics and Astronomy, Johns Hopkins University,
	\href{mailto:wyuan10@jhu.edu}{wyuan10@jhu.edu}}
\address[D]{Department of Physics and Astronomy,
	 Texas A\&M University
	\href{mailto:lmacri@tamu.edu}{lmacri@tamu.edu}}
\address[E]{Department of Statistics, Texas A\&M University
	\href{mailto:jianhua@stat.tamu.edu}{jianhua@stat.tamu.edu}}
\end{aug}

\begin{abstract}
The Period--Luminosity relation (PLR) of Mira variable stars is an important tool to determine astronomical distances. The common approach of estimating the PLR is a two-step procedure that first estimates the Mira periods and then runs a linear regression of magnitude on log period. When the light curves are sparse and noisy, the accuracy of period estimation decreases and can suffer from aliasing effects. Some methods improve accuracy by incorporating complex model structures at the expense of significant computational costs. Another drawback of existing methods is that they only provide point estimation without proper estimation of uncertainty. To overcome these challenges, we develop a hierarchical Bayesian model that simultaneously models the quasi-periodic variations for a collection of Mira light curves while estimating their common PLR. By borrowing strengths through the PLR, our method automatically reduces the aliasing effect, improves the accuracy of period estimation, and is capable of characterizing the estimation uncertainty. We develop a scalable stochastic variational inference algorithm for computation that can effectively deal with the multimodal posterior of period. The effectiveness of the proposed method is demonstrated through simulations, and an application to observations of Miras in the Local Group galaxy M33. Without using ad-hoc period correction tricks, our method achieves a distance estimate of M33 that is consistent with published work. Our method also shows superior robustness to downsampling of the light curves.
\end{abstract}

\begin{keyword}
\kwd{variational inference}
\kwd{hierarchical Bayesian modeling}
\kwd{astrostatistics}
\kwd{Mira variables}
\end{keyword}

\end{frontmatter}

\section{Introduction}
\label{sec:introduction}

One essential step to determine the age and composition of our Universe is to calibrate its current expansion rate (commonly referred to as the ``Hubble constant'' or $H_0$ in astronomy) using the cosmological distance ladder. Observational astronomers have extensively relied on the Period--Luminosity relations (PLRs) of classical Cepheid variable stars to measure distances to galaxies within 150 million light years of the Milky Way, enabling the calibration of farther-reaching techniques (such as type Ia supernovae) to obtain increasingly more robust estimates of $H_0$ \citep{freedman2001final,riess20113, Riess2016,riess2019large}. Recent studies \citep{whitelock2014vizier, whitelock2014gaia, yuan2017large, huang2018near} have shown that Mira variable stars (hereafter, Miras) also exhibit tight PLRs at near-infrared (NIR) wavelengths and thus can serve as independent distance indicators in lieu of Cepheids. Miras are highly-evolved stars with two subclasses based on their atmospheric chemistry: oxygen- and carbon-rich (O- and C-rich, respectively). They constitute a class of long-period variable stars whose luminosity changes dramatically (in a quasi-periodic manner) as a function of time. One pulsation cycle of Miras usually lasts more than one hundred days, with extreme cases beyond several years.

In a given system, such as the Triangulum Galaxy (also known as Messier 33; hereafter, M33), all its member stars essentially are at the same distance ($D$) from us. Compared to their intrinsic luminosity, their apparent brightness is scaled by the same factor of $1/D^2$. As a result, the observed PLR of Miras in different galaxies located at various distances are expected to follow the same functional form, up to a zero-point offset. Astronomical time-series imaging surveys, using modern CCD cameras on meter-class or larger telescopes, routinely perform accurate brightness measurements (called {\it photometry}) of millions of stars in other galaxies. These measurements are usually reported using units of \textit{magnitude}, which are inversely proportional to the logarithm of the light flux. Thus,  brighter objects have smaller magnitudes.  Repeated imaging of a given field containing a large number of stars yields time series measurements of stellar brightness called \textit{light curves}. To determine the colors of stars, images are usually taken using filters (also called {\it bands}) which only allow certain wavelengths to pass. As a result, several light curves in various colors are collected for each star; these constitute the so-called multi-band light curve data.

\begin{figure}[t]
	\centering
	\begin{subfigure}[b]{0.49\textwidth}
		\includegraphics[width = \textwidth]{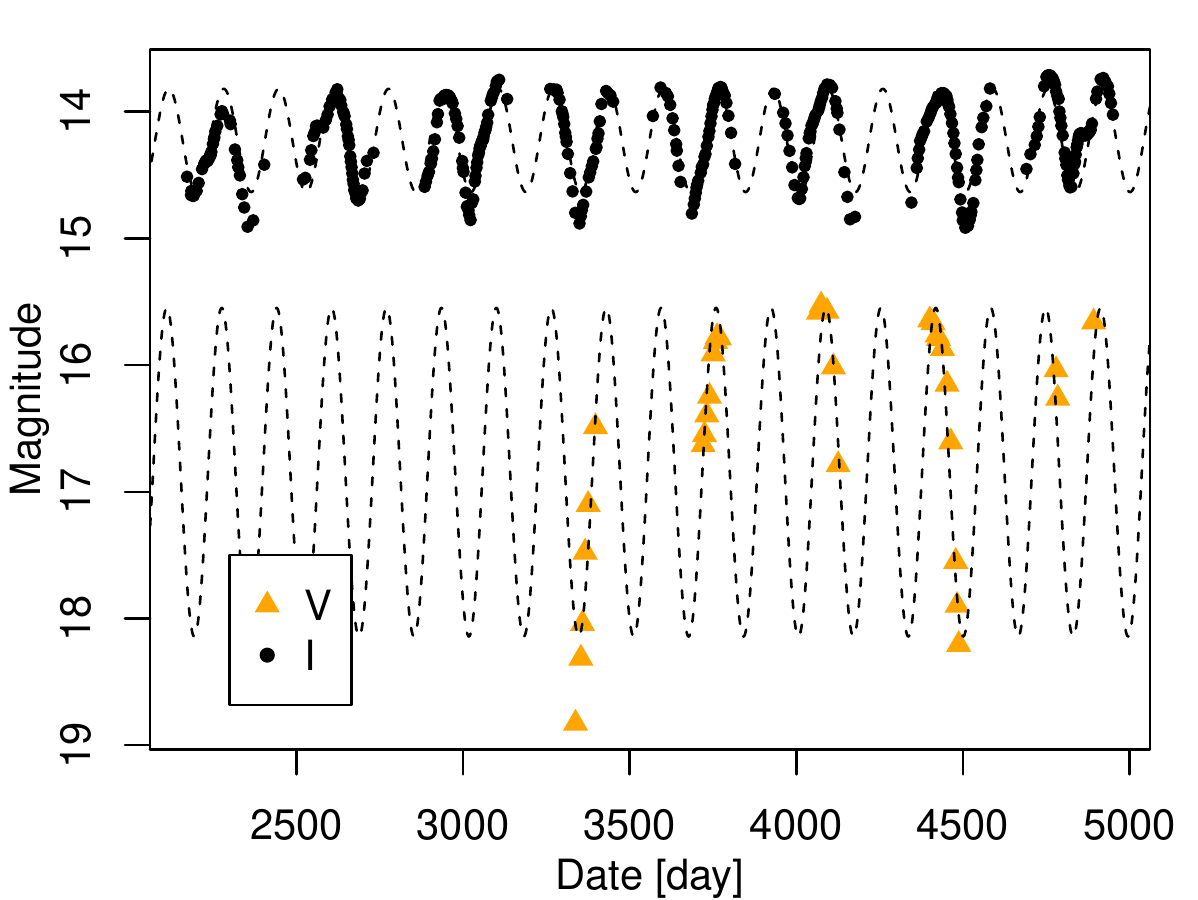}
		\caption{One LMC O-rich Mira.}
		\label{fig:firstMira}
	\end{subfigure}
	\begin{subfigure}[b]{0.49\textwidth}
		\includegraphics[width=\textwidth]{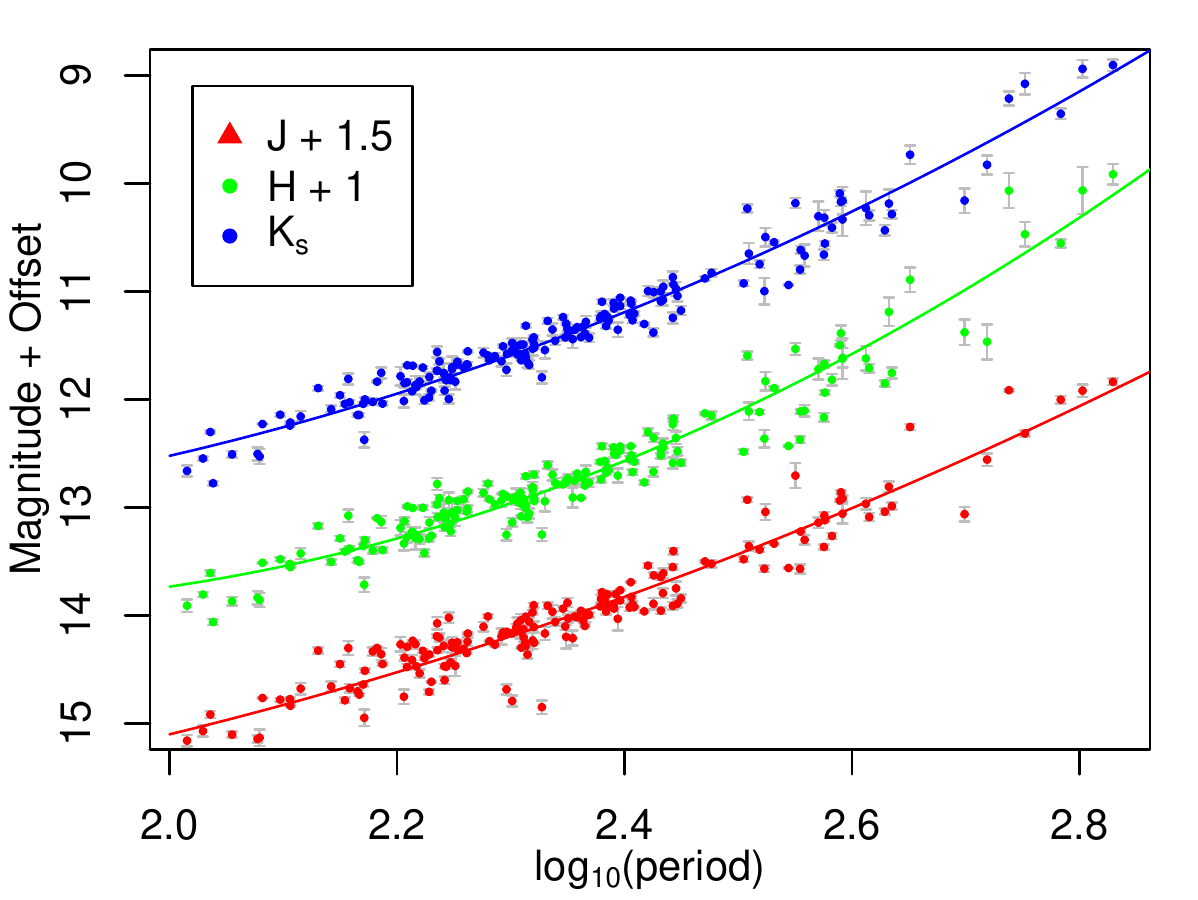}
		\caption{PLRs for LMC.} \label{fig:lmcPLR}
	\end{subfigure}
	\caption{The left panel illustrates observations of the O-rich Mira by the OGLE survey (ID:  \texttt{OGLE-LMC-LPV-00082}). The right panel shows the $J$-, $H$- and $K_s$-band PLRs for O-rich Miras in the LMC, along with the best-fit quadratic polynomials.}
\end{figure}

The third phase of the OGLE survey \citep{Udalski2008} obtained densely-sampled light curves of 1,663 Miras \citep{Soszynski2009} in the Large Magellanic Cloud (LMC), a satellite galaxy of the Milky Way. The survey was conducted using two filters ($V$ and $I$). The OGLE-III $I$- and $V$-band light curves have a median of about 470 and 45 photometric measurements per object, respectively. Figure~\ref{fig:firstMira} shows the light curve obtained by this survey for one O-rich Mira with a period of $164.84$ days. The horizontal axis represents time, while the vertical axis shows the magnitude. Note that as brighter objects have smaller magnitude values, the vertical axis is flipped. The $I$- and $V$-band light curves are plotted using black circles and orange triangles, respectively. The dashed curves are the best sinusoidal fit to each band separately. Using additional measurements in the NIR $J$, $H$ and $K_s$ bands, \cite{yuan2017large} obtained tight PLRs shown in Figure~\ref{fig:lmcPLR}. In this figure, each point represents one O-rich Mira. The vertical axis is the average magnitude of each star, while the horizontal axis is the period of each object, using a logarithmic scale (with base $10$).  The solid lines in Figure~\ref{fig:lmcPLR} are least-squares fits to each band, using a quadratic polynomial in the logarithm of the period. The zero-order terms of the polynomials are used for astronomical distance determination. 

\begin{figure}
	\centering
	\includegraphics[width=0.49\textwidth]{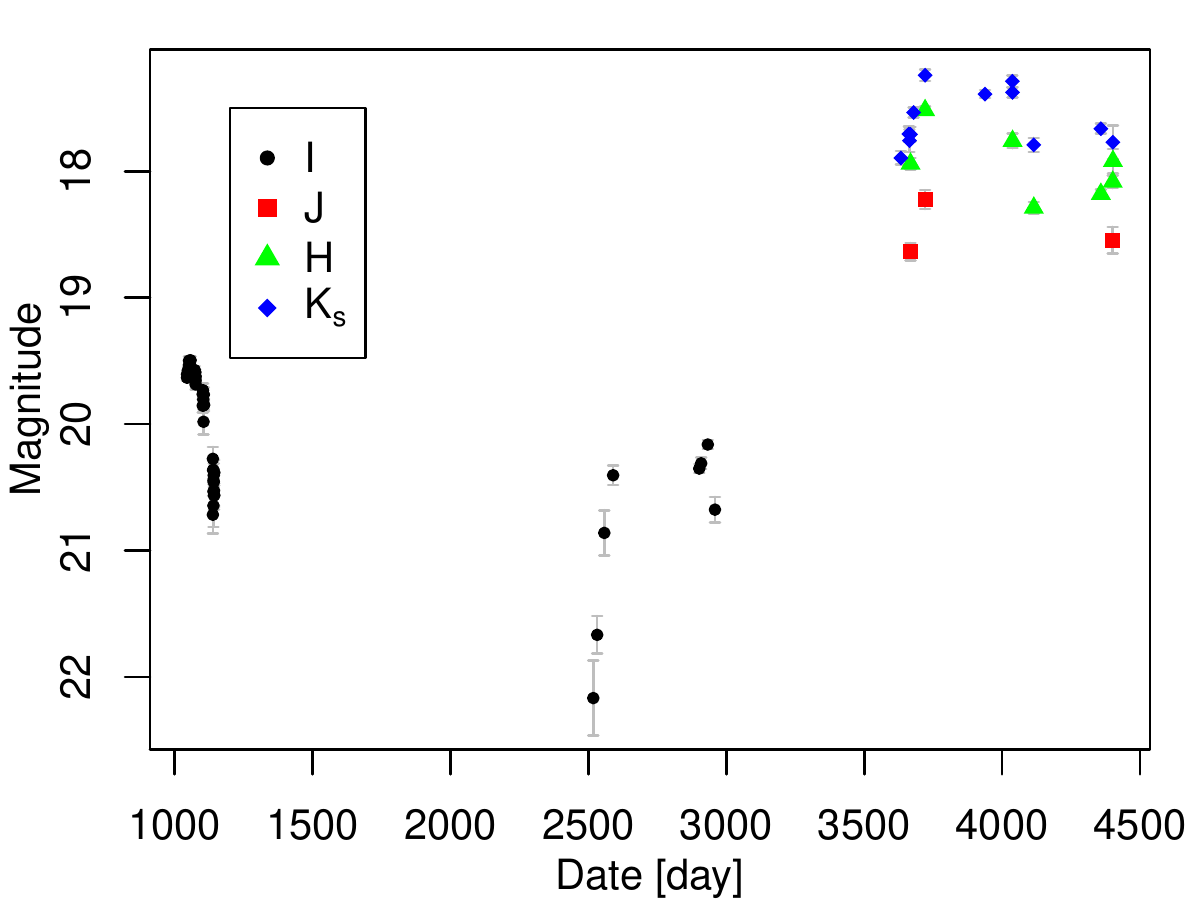}
	\includegraphics[width=0.49\textwidth]{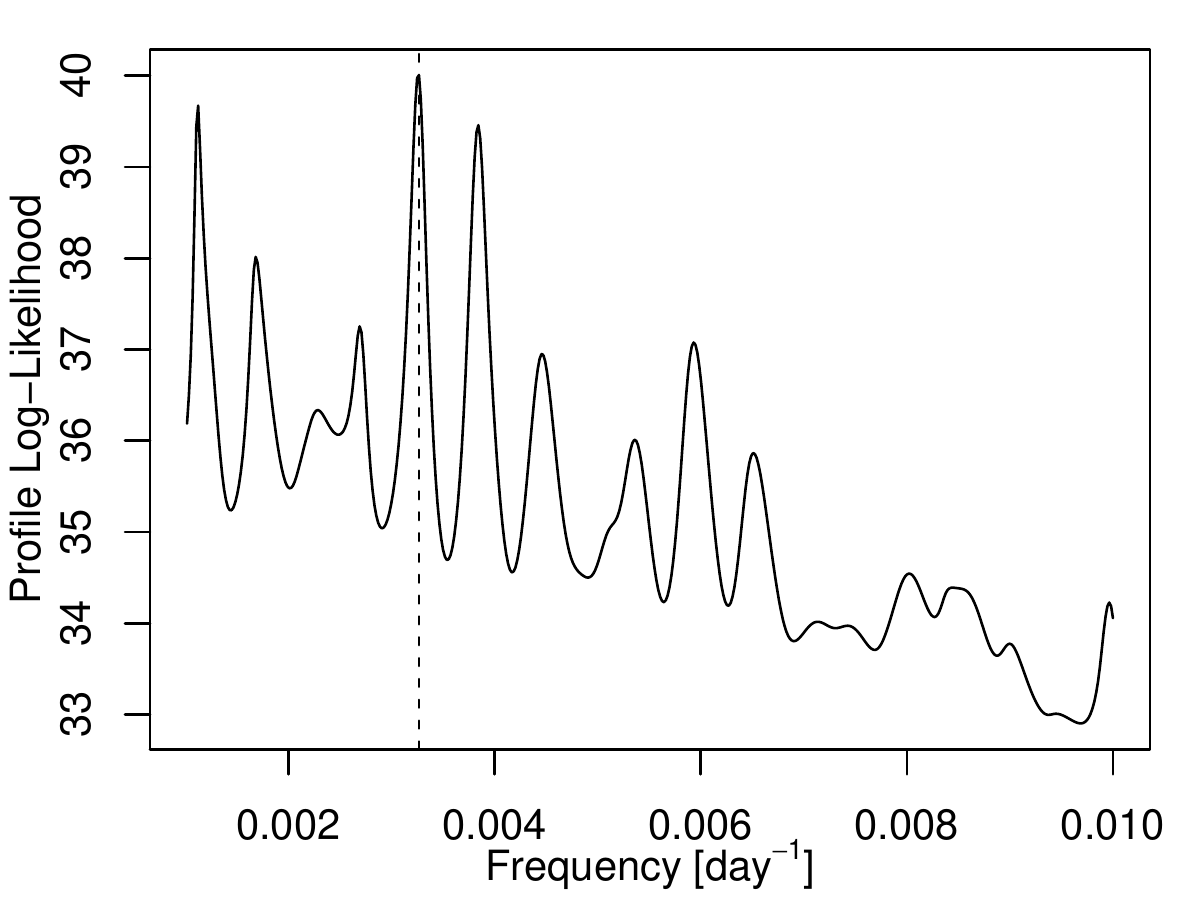}
	\caption{In the left panel is a typical light curve for a Mira in M33. 
		In the right panel is resulting periodogram applying the SP method \citep{he2016period} to its $I$-band light curve. The vertical dashed line indicates the estimated frequency.}\label{fig:firstM33}
\end{figure}

The OGLE survey has obtained light curves of unusual high quality thanks to decades-long observations with a dedicated telescope. More typical astronomical time-series surveys have lower data quality which hinders accurate period estimation and PLR construction.  \cite{yuan2018near} analyzed Miras in M33, a galaxy in our Local Group located $\sim 20\times$ farther than the LMC.  Each object was measured in four bands ($I$, $J$, $H$, and $K_s$) with  a median number of observations of 68, 5, 6, and 11, respectively. The left panel of Figure~\ref{fig:firstM33} shows the light curves of one of these Miras. It is evident that both the number of data points and their signal-to-noise ratio are limited. This dataset is not alone in terms of the difficulty to accurately recover periods. For example, the second data release of the \textit{Gaia} survey \citep{mowlavi2018gaia} consists of 550,737 variable stars observed in three filters each with an median of about 25 data points, while the second data release of the Pan-STARRS survey \citep{Heather2018}, contained an average of less than 10 measurements in each of the five bands it used. 

When light curves are sparsely sampled and subject to a high level of noise, the task of period estimation is no doubt challenging. The Lomb-Scargle method \citep{lomb1976least,Scargle1982} is the most commonly used for this purpose. For a general periodic star, the method fits a simple sinusoidal curve to a single-band light curve. The fitting is performed over a sequence of trial frequencies (the inverse of periods), and the frequency with the minimal residual sum of squares is selected as the estimate. However, a light curve of one Mira variable is not exactly periodic. From Figure~\ref{fig:firstMira},  it is evident that either the $I$- or the $V$-band light curves have additional fluctuations that are not accounted for by the fitted sinusoids (dashed curve).  The semi-parametric (SP) model of \cite{he2016period} addresses the issue by fitting a Gaussian Process to the residuals.  The Lomb-Scargle  method has also been extended by some authors \citep{vanderplas2015periodograms,long2016estimating} via exploiting multi-band light curves of a periodic star. As the number of parameters usually grows with the number of bands, \cite{long2016estimating} employed explicit regularization to constrain the model degree of freedom. For multi-band Mira observations, \cite{yuan2018near} enhanced the SP model and have achieved the best performance for Mira variables to date.

After all these efforts and explorations, the next bottleneck in raising estimation accuracy even further has been encountered: existing methods are susceptible to the so-called ``aliasing effect'' due to the discrete sampling of a continuous function. For illustration,  we apply the SP model \citep{he2016period}  to the $I$-band light curve shown in the left panel of Figure~\ref{fig:firstM33}.  The model output is in the right panel of Figure~\ref{fig:firstM33}, with profile log-likelihood on the vertical axis and frequency on the horizontal axis. Notice the evaluated log-likelihood is highly multimodal, and the frequency with maximum log-likelihood (the primary peak marked by the vertical dashed line) is identified as the estimate.  For a light curve with a weak signal, a false frequency can show stronger evidence (larger likelihood) than the true one.  When this happens, an incorrect period estimation is acquired. The work of \cite{yuan2018near}  devised an ad-hoc trick for remedy: instead of the primary peak (with the largest log-likelihood value), they selected the secondary peak (with the second largest log-likelihood value) as the final estimate if it better fits the PLR.

In addition to the estimation accuracy, computational cost is another curb on the application of some existing methods. Due to the multimodal nature of the likelihood, a grid search over the frequency domain seems inevitable to find the optimal frequency. This is especially troublesome for the methods of \cite{he2016period} and \cite{yuan2018near}. When these methods estimate the period of a Mira, the related Gaussian process likelihood has to be computed repeatedly over the grid of trial frequencies. Moreover, this computationally intensive process has to be carried out for each Mira in the dataset. 

In this work, we propose a novel period estimation framework to address these challenges.  Based on the state-of-the-art method of \cite{he2016period} and \cite{yuan2018near}, a Bayesian hierarchical model is constructed for Mira variables. The model embeds the PLR as one of its model components, and simultaneously models all Mira light curves in a dataset.  The resulting model is capable of automatically reducing the aliasing effect, thereby increasing estimation accuracy.  Under this framework, each Mira supplies information to update their common PLR. In return, the PLR component guides parameter estimation of individual Miras. In our numerical studies, this mechanism is demonstrated to be key for improving estimation accuracy. 

To reduce the computational cost, we develop an efficient algorithm for model inference. Given a Bayesian model, Monte Carlo Markov Chain (MCMC) is the standard tool for inference. However, computing MCMC will be prohibitively expensive for our model. As we are fitting a collection of Mira light curves, tens of thousands of parameters will be involved even for a dataset of moderate size. Moreover, the multimodal posteriors  will also obstruct the convergence of the Markov chain. Fortunately, advances in variational inference \citep[VI;][]{Jordan1999, blei2017variational, zhang2019advances} bring us the opportunity to exploit the problem in the new framework. VI converts the Bayesian inference into an optimization problem, and finds the best density in a pre-specified family to approximate the actual posterior distribution. To further improve computational efficiency for large scale problems, we adopt the stochastic variational inference framework by \cite{hoffman2013stochastic}. Using an unbiased estimate of the natural gradient during each iteration,  our algorithm is able to avoid costly numerical integration for the frequency parameter, and avoid scanning over all samples per iteration.

Under the same simulation as \cite{yuan2018near}, we show our approach only fails to recover the period  for $\sim 2$\% of the Miras, as opposed to their $\sim 9$\%. The absolute deviation error is remarkably reduced by 77\%.  Furthermore, the method of \cite{yuan2018near} costs hundreds of CPU hours, which is not possible to run without a computer cluster. In contrast, our method finishes within a few hours  on a personal computer. 

The benefit of the proposed model goes beyond fast and accurate period estimation. Firstly, existing methods only deliver a point estimate for the frequency (or period) without an uncertainty measurement. Our work is the first attempt to quantify and approximate the multimodal posterior uncertainty. This is achieved by allowing the variational distribution for frequency to be any continuous function over a compact interval. An analytical solution has been derived for its update in our work. Secondly, the PLR is traditionally determined by a two-step procedure: average magnitudes and periods are estimated for individual Miras separately; then the PLR is obtained via a least-squares fit. The proposed method is a more systematic way to estimate the PLR, which is the fundamental tool for distance measurements. By adopting Bayesian modeling, we directly harvest its estimate and uncertainty quantification as direct model products. 

The rest of the paper is organized as follows. Our proposed method is developed in Section~\ref{sec:proposedMethod}. Its computational algorithm via the stochastic variational inference is developed in Section~\ref{sec:stochasticVI}. We then conduct simulation studies to compare our method with existing methods in Section~\ref{sec:simulation}. Finally, in Section~\ref{sec:realdata}, the proposed method is applied to a real data set of M33 Miras \citep{yuan2018near}. Section~\ref{sec:downsampling} studies the robustness of the proposed method to downsampling. The Supplementary Material \citep{He2020} contains proofs of theoretical results and additional numerical results.

\section{The Hierarchical Model}
\label{sec:proposedMethod}

\begin{figure}[t]
	{\small
		\begin{align}
		y_{ibj}|s_{ibj} &\sim \sN(s_{ibj}, \sigma_{ibj}^2) \text{ for } i\in \mathcal{I},\ b\in\mathcal{B},\ j\in\mathcal{J}_{ib}, \label{eqn:hm:layer1}\\
		s_{ibj}& = m_{ib} +  \vb_{ibj}^T\vbeta_{ib}+h_{ibj}; \label{eqn:hm:layer1-2}\\
		m_{ib} | f_i, &\valpha_b, \gamma_{b}  \sim \sN( \vd_i^T\valpha_b ,  1/\gamma_{b})
		\text{ for }  i\in \mathcal{I},\ b\in\mathcal{B};\label{eqn:hm:layer2}\\
		\vbeta_{i}  |\mOmega 	& \sim \sN(\vzero, \mOmega^{-1})
		\text{ for }  i\in \mathcal{I};\label{eqn:hm:layer3}\\
		h_{ib}(\cdot) | f_i &\sim \mathcal{GP}( 0,  k_b(\cdot, \cdot) )
		\text{ for }  i\in \mathcal{I},\ b\in\mathcal{B};\label{eqn:hm:layer4}\\
		f_i &\sim \mathrm{Unif}(f_{\min}, f_{\max})
		\text{ for }  i\in \mathcal{I};\label{eqn:hm:layer5}\\
		\valpha_b  &\sim \sN\left(\bar{\valpha}_b,  (1/\bar{\delta})\mI\right)
		\text{ for }  b\in \mathcal{B};\label{eqn:hm:layer6}\\
		\gamma_b & \sim \text{Gamma}(\bar{\gamma}_b\bar{r}, \bar{r}) 
		\text{ for }  b\in \mathcal{B};\label{eqn:hm:layer7}\\
		\mOmega &\sim \text{Wishart}(\bar{\mOmega}/\bar{n}, \bar{n}).
		\label{eqn:hm:layer8}
		\end{align}}
	\vspace{-20pt}
	\caption{The hierarchical model for period estimation and PL relation construction.	\label{fig:hierarchicalModel}}
\end{figure}

Now we elucidate our Bayesian hierarchical model for  period estimation with multi-band Mira data.  Our dataset is a collection of $N$ Miras in a galaxy and the observations of each Mira  are collected with $B$ filters (bands) indexed by $b \in \{1,2,\cdots, B\}$.  For the $i$-th star $(i=1,\cdots, N)$, the data collected with the $b$-th filter is denoted as $\mathcal{D}_{ib} = \{(t_{ibj}, y_{ibj}, \sigma_{ibj})\}_{j=1}^{n_{ib}}$, where $y_{ibj}$ is the magnitude measured through the $b$-th filter at time $t_{ibj}$, and the corresponding measurement uncertainty is $\sigma_{ibj}$. As each Mira is observed through the $b$-th filter, all measurements for the $i$-th Mira are denoted as $\mathcal{D}_i = \cup_b \mathcal{D}_{ib}$. The period of the $i$-th Mira will be denoted as $p_i$, and its frequency is $f_i=1/p_i$. Note $p_i$ (or $f_i$) is one of our key quantities of interest.

\begin{figure}[t]
	\centering
	\resizebox{0.7\columnwidth}{!}{%
		\begin{tikzpicture}
		\tikzstyle{main}=[circle, minimum size = 10mm, thick, draw =black!80, node distance = 16mm]
		\tikzstyle{connect}=[-latex, thick]
		\tikzstyle{box}=[rectangle, draw=black!100]
		\node[main] (theta)  {$f_i$ };
		\node[main] (alphab) [below=of theta] { $\valpha_b$};
		\node[main, fill = black!10, left=of alphab] (alphabar) [] {$\bar{\valpha}_b$ };
		\node[main, , fill = black!10, below=of alphab, yshift=5mm](deltabar){$\bar{\delta}$};
		\node[main] (mib) [right=of theta] {$m_{ib}$};
		\node[main] (gammab) [below=of mib] {$\gamma_b$ };
		\node[main, fill = black!10] (gammabbar) [right=of gammab] {$\bar{\gamma}_b$ };
		\node[main, fill = black!10] (rbar) [below=of gammab, yshift=5mm] {$\bar{r}$ };
		\node[main] (beta) [above=of mib] {$\vbeta_{ib}$ };
		\node[main] (omega) [left=of beta,xshift = -28mm] {$\mOmega$ };	
		\node[main, fill = black!10] (omegabar) [left=of omega,yshift=10mm] {$\bar{\mOmega}$ };	
		\node[main, fill = black!10] (nbar) [left=of omega,yshift=-10mm] {$\bar{n}$ };	
		\node[main, fill = black!10] (yibj) [right=of mib] {$y_{ibj}$ };
		\node[main] (hibj) [above=of yibj] {$h_{ibj}$ };
		\path (alphabar) edge [connect] (alphab)
		(deltabar) edge [connect] (alphab)
		(gammabbar) edge [connect] (gammab)
		(rbar) edge [connect] (gammab)
		(theta) edge [connect] (mib)
		(alphab) edge [connect] (mib)
		(gammab) edge [connect] (mib)
		(hibj) edge [connect] (yibj)
		(mib) edge [connect] (yibj)
		(beta) edge [connect] (yibj)
		(omegabar) edge [connect] (omega)
		(nbar) edge [connect] (omega)
		(omega) edge [connect] (beta);
		\node[rectangle, inner sep=3.5mm, draw=black!100, fit =   (alphab)(gammab)(gammabbar)(alphabar),yshift=-1mm] {};
		\node[rectangle, inner sep=3.5mm, label=below:$b\in\mathcal{B}$,
		yshift=5mm, xshift=0mm,
		fit =   (alphab)(gammab)(gammabbar)(alphabar)] {};
		\node[rectangle, inner sep=4.5mm, draw=black!100, fit =   (hibj)(yibj),xshift=-3mm,yshift=0.5mm] {};
		\node[rectangle, inner sep=3.5mm, fit =   (yibj),
		label = below:$j\in \mathcal{J}_{ib}$, yshift=45mm, xshift = -6mm] {};
		\node[rectangle, inner sep=7mm,draw=black!100, fit=  (beta)(mib) (yibj)] {};
		\node[rectangle, inner sep=9mm, draw=black!100, fit =  (beta)(theta) (mib) (yibj)] {};
		\node[rectangle, inner sep=5mm, label=below:$i\in\mathcal{I}$, fit =  (theta), xshift = -8mm, yshift=45mm] {};
		\node[rectangle, inner sep=5mm, label=below:$b\in\mathcal{B}$, fit =  (mib), xshift = -6mm, yshift=45mm] {};
		\end{tikzpicture}}
	\caption{The hierarchical model diagram.	\label{fig:hierarchicalModelDiagram}}
\end{figure}
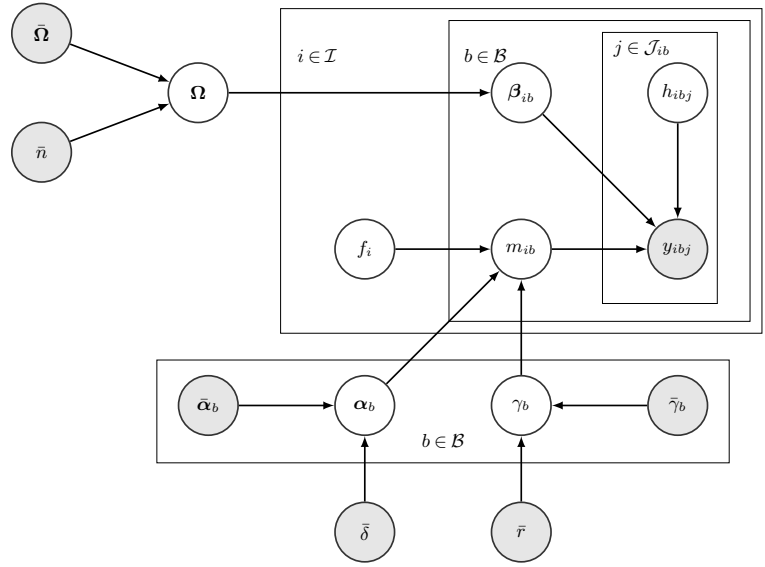

In the rest of the paper, our whole  collection of  Mira data will be denoted as $\mathcal{D} = \cup_i \mathcal{D}_i$.  The following  index sets will be also used: $\mathcal{I} = \{1,\cdots, N\}$,  $\mathcal{B} = \{1,\cdots, B\}$, and  $\mathcal{J}_{ib} = \{1,\cdots, n_{ib}\}$. Our  model is  based on the work of \cite{he2016period} and \cite{yuan2018near}, where each Mira is modeled separately.  We jointly model the whole collection $\sD$ of Miras, and simultaneously estimate the PLR. The overall  model is listed in Figure~\ref{fig:hierarchicalModel} and its diagram plotted in Figure~\ref{fig:hierarchicalModelDiagram}. This is a hierarchical Bayesian model, where Eqn.\ \eqref{eqn:hm:layer1}--\eqref{eqn:hm:layer1-2} constitute the level-1, Eqn.\ \eqref{eqn:hm:layer2}--\eqref{eqn:hm:layer5} constitute the level-2, and Eqn.\ \eqref{eqn:hm:layer6}--\eqref{eqn:hm:layer8} constitute the level-3 of the model. The following subsections explain the model components in details.

\subsection{The light curve decomposition}
\begin{figure}[t]
	\centering
	\includegraphics[width = 0.7\textwidth]{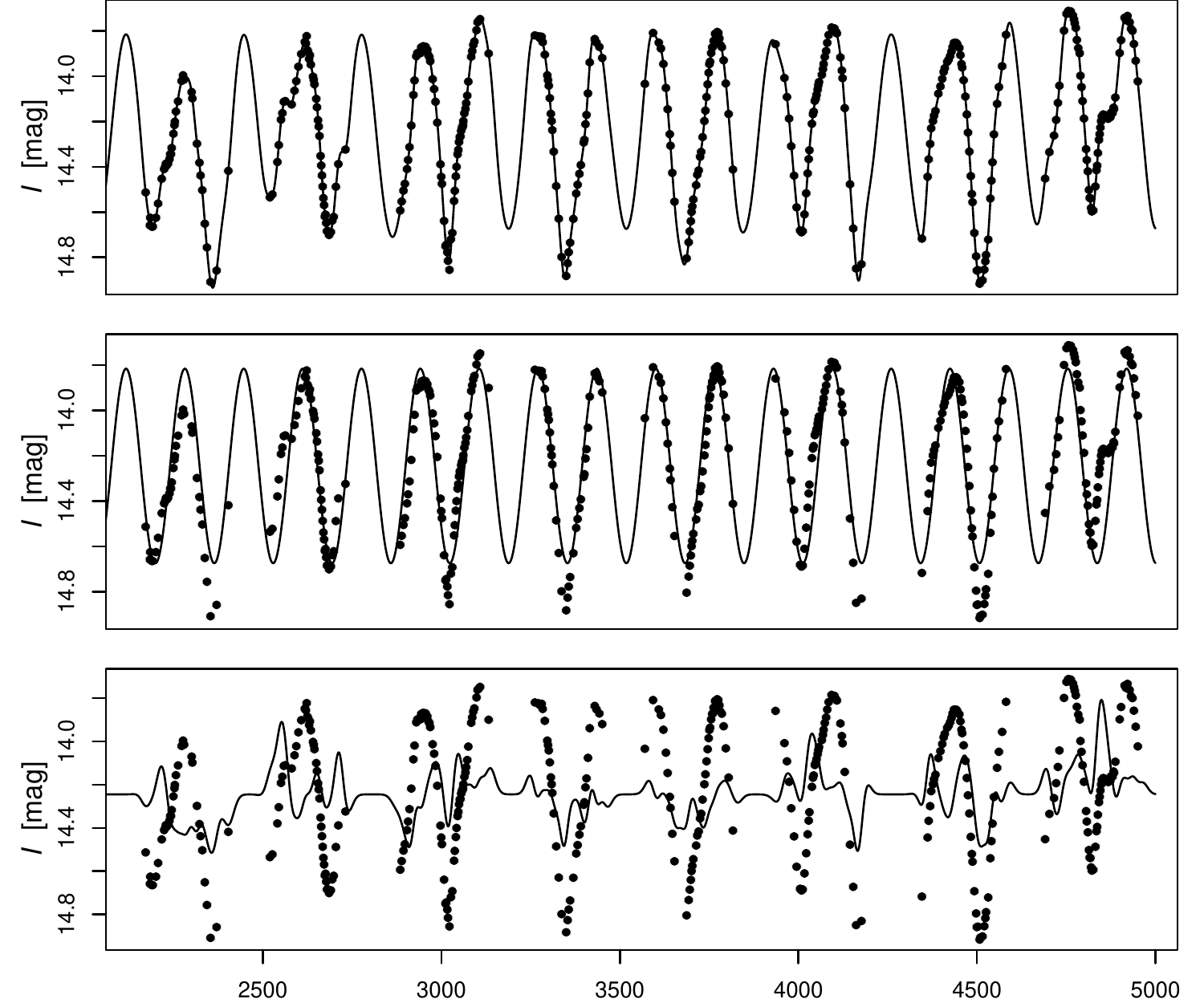}
	\caption{An example of Mira light curve decomposition. The solid curve in the upper panel is the fitted light curve. The solid curve in the middle panel is the exact periodic component. The lower panel is the non-periodic function.}
	\label{fig:firstMiraDecompose}
\end{figure}

Denote $\mathcal{D}_{ib} = \{(t_{ibj}, y_{ibj}, \sigma_{ibj})\}_{j=1}^{n_{ib}}$ as the observed light curve  for the $i$-th Mira with the $b$-th filter. The magnitude $y_{ibj}$ is decomposed as signal plus noise terms
$y_{ibj} = s_{ib} (t_{ibj}) + \sigma_{ibj}\epsilon_{ibj},$
where $s_{ib}(\cdot)$ is the signal as a function of time, and $\epsilon_{ibj}$'s are standard Gaussian noise. In other words, the observed light curve is noisy observation of   $s_{ib} (t)$ at sparse time points.  Following \cite{he2016period}, this signal function  is further decomposed  as
\begin{equation}
s_{ib} (t) = m_{ib} + \vb(f_i\cdot  t)^T\vbeta_{ib}+ h_{ib}(f_i\cdot t), \label{eqn:model:decompositionNew}
\end{equation}
where $m_{ib}$ is the average magnitude,  $\vb(u) =  (\cos(2\pi u), \sin(2\pi  u))^T$ is the sinusoidal basis vector and $\vbeta_{ib} = (\beta_{ib1}, \beta_{ib2})^T$  is the corresponding coefficient, and $h_{ib}(f_i\cdot t)$ is a smooth function that captures the non-periodic variations in Mira magnitudes.
Note the sinusoidal function $\vb(f_i\cdot  t)^T\vbeta_{ib}$ has a period of $p_i=1/f_i$. Figure~\ref{fig:firstMiraDecompose} illustrates  the decomposition~\eqref{eqn:model:decompositionNew} for the $I$-band light curve in Figure~\ref{fig:firstMira}. In the upper panel, the solid curve represents the fitted signal $s_{ib}(t)$. The middle panel shows the fitted periodic component $m_{ib} + \vb(f_i\cdot  t)^T\vbeta_{ib}$, and 
the lower panel is for the non-periodic component $m_{ib} + h_{ib}(f_i \cdot t ) $.

With simplified notations $s_{ibj} = s_{ib}(t_{ibj})$, $\vb_{ibj} = \vb(f_i \cdot t_{ibj})$ and $h_{ibj} = h_{ib}(f_i \cdot t_{ibj})$, Eqn.~\eqref{eqn:model:decompositionNew} constitutes the level-1 of the hierarchical model in Figure~\ref{fig:hierarchicalModel}. This top level models the individual observed light curve $\mathcal{D}_{ib}$ separately. Our joint modeling comes in by introducing the PLR model component in the next subsection.

\begin{remark}
	The Generalized Lomb-Scargle method \citep{zechmeister2009generalised} and its multi-band version only have the average magnitude and the sinusoidal component in the decomposition~\eqref{eqn:model:decompositionNew}. The additional term  $h_{ib}(\cdot)$ is introduced by \cite{he2016period} and \cite{yuan2018near} to account for the additional variability of Mira light curves. All of these methods model each Mira separately, without the hierarchical information introduced in the following subsections.
\end{remark}

\subsection{The Period-Luminosity relation}~\label{sec:plr}

For a specific band $b\in\mathcal{B}$, the average magnitude $m_{ib}$ and the period $p_i$ follow the PLR for all $i\in\sI$. As in Figure~\ref{fig:lmcPLR}, for each band, the PLR can be fitted by a quadratic polynomial in  $\log_{10} (p)$. 
Since our model is parameterized by frequency, the following equivalent PLR (noting $\log_{10} p_i = -\log_{10} f_i $) is employed,
\begin{equation} \label{eqn:modelPLR}
m_{ib} = \alpha_{b3}\log^2_{10} f_i + \alpha_{b2}\log_{10} f_i
+ \alpha_{b1}  + e_{ib} = \valpha_b^T\vd_i + e_{ib}\,.
\end{equation}
In the above, $\vd_i = (1,  \log_{10}(f_i), \log^2_{10}(f_i))^T$ is the PLR basis vector for the $i$-th Mira, $\valpha_b = (\alpha_{b1}, \alpha_{b2}, \alpha_{b3})^T$ is the PLR coefficient for the $b$-th band, and the PLR residual $e_{ib}$ is assumed to follow $\sN(0, 1/\gamma_{b})$ with precision $\gamma_b$. In summary, given  $f_i, \valpha_b, \gamma_{b}$,  the average magnitude  is dictated by  $m_{ib} | f_i, \valpha_b, \gamma_{b} \sim  \sN(\valpha_b^T\vd_i, 1/\gamma_{b}).$ In addition, we place on $f_i$ a uniform prior over a compact interval $[f_{\min}, f_{\max}]$, i.e., $f_i \sim \mathrm{Unif}(f_{\min}, f_{\max})$. The compact interval $[f_{\min}, f_{\max}]$ can be selected as the search range of interest. For most Miras, their frequency lies inside the interval $[1/1000, 1/100]$. These constitutes parts of the level-2 of our model in Figure~\ref{fig:hierarchicalModel}.We remark that while the level-1 of the model treats each light curve  separately, the level-2 dictates the common a-priori information shared by individual Mira parameters.

In our proposed model, the coefficient $\valpha_b$ and the precision $\gamma_b$ are also unknown parameters with conjugate prior
$\valpha_b \sim \sN(\bar{\valpha}_b, (1/\bar{\delta}) \mI)$, $\gamma_{b}\sim \text{Gamma}(\bar{\gamma}_b\bar{r}, \bar{r})$, where  $\bar{\valpha}_b$, $\bar{\gamma}_b$, $\bar{\delta}$, $\bar{r}$ are fixed hyper-parameters.  These become Eqn.~\eqref{eqn:hm:layer6}--\eqref{eqn:hm:layer7}, and part of level-3 of our model in Figure~\ref{fig:hierarchicalModel}.  Notice that the level-3 parameters $\valpha_b, \gamma_{b}$ and $\mOmega$ (to be introduced in the next subsection) govern the probability distribution of the Mira samples.

\begin{remark}
	Eqn.~\eqref{eqn:hm:layer2} plays a key role in our hierarchical model. It governs the relation between $m_{ib}$ and $f_i$, and  discourages implausible parameter values. This model component is the key to reduce the aliasing effect and improves the period estimation accuracy. Moreover, a direct product of our proposed method is the PLR coefficients $\valpha_b$'s, which is one of the important pursuits for  modern astronomical surveys.
\end{remark}

\subsection{The periodic component}

For the $i$-th star,  let $\vbeta_{i} = (\vbeta_{i1}^T,\vbeta_{i2}^T,\cdots, \vbeta_{iB}^T)^T$ be the collection of periodic coefficients across all bands. This vector has  multivariate Gaussian prior $\vbeta_{i} | \mOmega \sim \sN(\vzero, \mOmega^{-1})$ in Eqn.~\eqref{eqn:hm:layer3}  of Figure~\ref{fig:hierarchicalModel}.  The precision matrix $\mOmega$ is also  unknown  with a Wishart prior $\mOmega \sim \text{Wishart}(\bar{\mOmega}/\bar{n}, \bar{n})$ in Eqn.~\eqref{eqn:hm:layer8}, for fixed hyper-parameters $\bar{\mOmega}$ and $\bar{n}$.

The multivariate Gaussian prior~\eqref{eqn:hm:layer3} on $\vbeta_{i}$ empowers us the ability to modeling the correlation between the periodic components across all bands. \cite{long2016estimating} and \cite{yuan2018near} have noticed both periodic amplitude and phase  are linearly correlated across bands. The prior distribution as given in~\eqref{eqn:hm:layer3} on $\vbeta_i$ allows us to take this correlation into account. Nevertheless, the assumption imposed by Eqn.~\eqref{eqn:hm:layer3} is weaker than   the explicit regularization of \cite{long2016estimating}.

\subsection{The non-periodic component} \label{sec:proposedMethod:nonpara}

Recall the decomposition~\eqref{eqn:model:decompositionNew} has an additional term $h_{ib}(\cdot)$ accounting for the  non-periodic variation.  For each $j\in\mathcal{J}_{ib}$, we denote $u_{ibj} = f_i\cdot t_{ibj}$ as the phase, which is the frequency $f_i$ multiplied by the time points $t_{ibj}$. Note $h_{ib}(\cdot)$  is a function of $u_{ibj}$'s. For the a specific band $b\in\mathcal{B}$,  the random function $h_{i b}(\cdot)$ is modeled by  a Gaussian process \citep[GP,][]{Rasmussen2005} with zero mean and a kernel function $k_{b}(\cdot, \cdot)$.  Under the GP assumption,  the vector $\vh_{ib} = \left(h_{ib}(u_{ib1}),\cdots, h_{ib}(u_{ibn_{ib}})\right)^T$ follows a multivariate Gaussian distribution with zero mean and covariance matrix  $\mH_{ib} = \big[ k_{b}(u_{ibp},\, u_{ibq})  \big]_{p,q=1}^{n_{ib}}$. The Eqn.~\eqref{eqn:hm:layer4} in Figure~\ref{fig:hierarchicalModel} is exactly this GP component.   Note  the $b$-th band data of all Miras share a common kernel $k_b(\cdot,\cdot)$. In this work, we employed the following kernel function, $$k_{b}(u, u') = \tau_{b1} \cdot \exp\big\{-  (u-u')^2 /\tau_{b2} \big\} +  \tau_{b3}\cdot \mathrm{I}(u=u'),$$ where $\mathrm{I}(\cdot)$ is the indicator function, and  $\vtau_b = (\tau_{b1}, \tau_{b2}, \tau_{b3})^T$ is the kernel parameter for the $b$-th band.  The first term of the above kernel is the squared exponential kernel with variance parameter $\tau_{b1}$ and length-scale parameter $\tau_{b2}$. The second term is a nugget term accounting for additional variability. Our adoption of the squared exponential kernel follows a common practice in the kernel machine field, though other kernel functions can also be used. The Gaussian process curve fitting is generally not sensitive to the choice of kernel.

In this formulation, for each band, the non-period components of light curves of all Miras are assumed to be realizations of the same Gaussian process. This allows Miras with sparse observations or relatively low data quality to ``borrow strength" from one another when fitting individual light curves. Letting all Miras share the same set of kernel parameters has the additional advantage of avoiding the intensive optimization of kernel parameters for individual light curves, and thereby greatly speeding up computation.

\begin{remark}
	In the work of \cite{he2016period} and \cite{yuan2018near},  the term $h_{ib}(\cdot)$ is chosen as a function of time  $t$, and the kernel parameter is distinct for different bands and different Miras. In contrast, our $h_{ib}(\cdot)$ is  a function of phase $u = f_i\cdot t$, and the kernel parameter is shared by all Miras in a given band. 
 Our choice of using $h_{ib}(f_i\cdot t)$ makes the function argument of the non-periodic component consistent with the periodic component $\vb(f_i\cdot  t)^T\vbeta_{ib}$ as shown in~\eqref{eqn:model:decompositionNew}.
    This choice is supported by two numerical results presented in Section~\ref{sec:non-periodic} of the Supplementary Material \citep{He2020}.

\end{remark}

\subsection{The joint likelihood}

All of the above leads to our  model listed in Figure~\ref{fig:hierarchicalModel}, which has a few unknown parameters. In the rest of the paper, to simplify the notation, the PLR coefficients for all bands are collectively denoted as $\valpha = \{\valpha_{1},\cdots, \valpha_{B}\}$, and all precision parameters as $\vgamma = (\gamma_1, \cdots, \gamma_B)^T$. For the $i$-th Mira, its average magnitudes across all bands are denoted by  $\vm_i = (m_{i1}, \cdots, m_{iB})^T$. For the whole collection of Miras, all their average magnitude parameters are denoted by $\vm = \{\vm_1, \cdots, \vm_N  \}$. In addition, we set  $\vf = \{f_1, \cdots, f_N\}$ for the frequency parameters of all Miras and $\vbeta = \{\vbeta_{1}, \cdots, \vbeta_{N}\}$ for all periodic coefficients.

Given the hierarchical model in Figure~\ref{fig:hierarchicalModel},  Bayesian inference requires computing the posterior of parameters, which is proportional to the joint distribution of all parameters. For the data $\sD_i$ of the  $i$-th Mira, its related joint likelihood (of Eqn.~\eqref{eqn:hm:layer1} through~\eqref{eqn:hm:layer5}) is
\begin{align*}
&\prod_{b=1}^{B}\prod_{j=1}^{n_{ib}} \exp\Big\{-\frac{1}{2\sigma_{ibj}^2}
(y_{ibj} - m_{ib} - \vb_{ibj}^T\vbeta_{ib} - h_{ibj})^2 \Big\} 
\times\prod_{b=1}^{B} \vert \mH_{ib}\vert^{-1/2} 
\exp\Big\{-\frac{1}{2}\vh_{ib}^T \mH_{ib}^{-1} \vh_{ib} \Big\}\\
&\qquad\qquad \times \vert \mOmega\vert^{1/2} \exp\Big\{-\frac{1}{2} \vbeta_{i}^T\mOmega\vbeta_{i}\Big\}  \times\prod_{b=1}^{B} \gamma_b^{1/2}\exp\Big\{-\frac{\gamma_b}{2} (m_{ib} - \vd_i^T\valpha_{b})^2
\Big\}.
\end{align*}
The   Gaussian process component $\vh_{ib}$ can be integrated out. In fact, given $m_{ib}, \vbeta_{ib}, f_i$, the vector $\vy_{ib} = (y_{ib1}, \cdots, y_{ibn_{ib} })^T$ follows a Gaussian distribution with mean vector $m_{ib}\vone + \mB_{ib}\vbeta_{ib}$ and covariance matrix $\mSigma_{ib} = \mH_{ib} + \text{diag}(\sigma_{ib1}^2,\cdots, \sigma_{ibn_{ib}}^2)$, where $\vone\in\bbR^{n_{ib}}$ is a vector of ones and $\mB_{ib} = \big(\vb_{ib1},\cdots, \vb_{ibn_{ib}}\big)^T\in\bbR^{n_{ib}\times 2}$. Consequently, the likelihood for the observation of the $i$-th Mira simplifies to 
\begin{align}
L_{i} = 	&\prod_{b=1}^{B}|\mSigma_{ib}|^{-1/2}\exp\Big\{-\frac{1}{2}
(\vy_{ib} - m_{ib} \vone- \mB_{ib}\vbeta_{ib})^T \mSigma_{ib}^{-1}
(\vy_{ib} - m_{ib} \vone- \mB_{ib}\vbeta_{ib})  \Big\}\nonumber \\
&\qquad \times \vert \mOmega\vert^{1/2} \exp\Big\{-\frac{1}{2} \vbeta_{i}^T\mOmega\vbeta_{i}\Big\}  \times\prod_{b=1}^{B} \gamma_b^{1/2}\exp\Big\{-\frac{\gamma_b}{2} (m_{ib} - \vd_i^T\valpha_{b})^2
\Big\}. \label{eqn:likelihood:ithmira}
\end{align}
Given the whole dataset $\sD$, the overall joint distribution (of Eqn.~\eqref{eqn:hm:layer1} through~\eqref{eqn:hm:layer8}) is
\begin{align}
p(\vm, \vbeta, \vf, \mOmega, \vgamma, \valpha, \sD) 
= & \Big(\prod_{i=1}^{N} L_{i} \Big)
\times \exp\Big\{ -\frac{1}{2} \langle\bar{n}\bar{\mOmega}^{-1}, \mOmega\rangle
+ \frac{\bar{n}-2B-1}{2} \log\det \mOmega
\Big\} \nonumber\\
&\quad\times \prod_{b=1}^{B}  \exp\Big\{-\bar{r}\gamma_{b} + (\bar{r}\bar{\gamma}_b -1) \log \gamma_{b} -\frac{\bar{\delta}}{2} \Vert \valpha_b - \bar{\valpha}_b\Vert^2 \Big\}.
\label{eqn:model:joint}
\end{align}
The right hand side is the product of the likelihoods for all observations and  the priors of all related parameters. We know the posterior $ p(\vm, \vbeta, \vf, \mOmega, \vgamma, \valpha|\, \sD) $ is proportional to the above distribution up to an unknown constant. Traditionally, inference for this hierarchical Bayesian model is carried out by the Monte Carlo  Markov Chain (MCMC).  However, it is well known that the traditional MCMC lacks scalability to a large data set.   To make it computationally feasible to apply our model to real Mira light-curve datasets, we will develop an efficient algorithm based on the stochastic variational inference  in the next section. 

\begin{remark}
	Before fitting the model, we need to specify the hyper-parameters $\bar{\valpha}_b$, $\bar{\delta}$, $\bar{\gamma}_b$, $\bar{r}$, $\bar{\mOmega}$, $\bar{n}$. We can set relative small values for $\bar{r}$, $\bar{n}$, and large values for $\bar{\delta}$. This makes the  priors less informative. The other hyper-parameters ($\bar{\mOmega}$ and $\bar{\delta}$) can be easily specified based on information from existing high-quality astronomical surveys.  This is applicable to the first and second order parameters $\bar{\alpha}_{b2}$, $\bar{\alpha}_{b3}$ of PLRs, as the shape of PLRs is the same for Miras in different galaxies. The only exception is the intercept terms $\bar{\alpha}_{b1}$ for the PLR prior, as the distance for a new dataset is unknown.  Fortunately, for a new survey, we only need to set a roughly reliable value of  $\bar{\alpha}_{b1}$. For this purpose, we can apply multi-band generalized Lomb-Scargle (MGLS) method to a small subset of the data, and then use a robust method (e.g. minimal absolute deviation) to get a fitted value of $\bar{\alpha}_{b1}$. 

It is known in the literature that Bayesian hierarchical model may perform poorly when vague priors are used. \cite{gelman2006prior} observed that choice of ``non-informative" prior distributions on hierarchical variance parameters may make a big impact on the inference, and improper choice may lead to inaccurate posterior distributions.  In our numerical studies, by setting non-extreme values $\bar{r} = 1$, $\bar{n} = 1$ for the variance parameters, our prior distributions do not have the adverse shrinkage effect pointed out in  \cite{gelman2006prior}  and our model shows accurate estimation results. Moreover, we have sufficient data so that the posterior distribution is dominated by the information provided by the data.
\end{remark}

\begin{remark}
	We also need to specify a value for the kernel parameter $\vtau$. This is  achieved by applying the MGLS method to a small subset of the dataset. With the estimated frequency from MGLS, we fit sinusoidal model to the light curves and obtain the sinusoidal residuals. After that, we fit a Gaussian process to the residual and obtain maximum likelihood estimates of the kernel parameters.
\end{remark}

\section{Main Algorithm}
\label{sec:stochasticVI}

Variational inference \citep{Jordan1999,wainwright2008graphical} aims at approximating the true posterior by finding an optimal density function $q(\cdot) $ among a specific distribution family $\mathcal{Q}$. This $q(\cdot) $ is then employed for subsequent inference.  To find the optimal approximation, the commonly used strategy  is to  minimize  the Kullback-Leibler divergence (KL divergence) between $q(\cdot)$ and the actual posterior distribution $p(\vm, \vbeta, \vf, \mOmega, \vgamma, \valpha|\, \sD) $. In our context, we are trying to solve 
\begin{equation} \label{eqn:VIproblem}
\min_{q(\cdot)\in\mathcal{Q}} \int q(\valpha, \vgamma, \mOmega, \vf, \vm, \vbeta)
\log\frac{q(\valpha, \vgamma, \mOmega, \vf, \vm, \vbeta)}{p(\vm, \vbeta, \vf, \mOmega, \vgamma, \valpha|\, \sD) }\intd \vm\intd \vbeta \intd \vf \intd \mOmega \intd \vgamma \intd \valpha
\end{equation}
for $q(\cdot)$ among a specific distribution family $\mathcal{Q}$. Equivalently, the variational inference aims at maximizing the evidence lower bound  (ELBO) $\underline{L} (\sD ):= \Expect_{q}\big\{\log\frac{p(\vm, \vbeta, \vf, \mOmega, \vgamma, \valpha, \sD)}{q(\vm, \vbeta, \vf, \mOmega, \vgamma, \valpha)} \big\}$, where the expectation is taken with respect to $q(\vm, \vbeta, \vf, \mOmega, \vgamma, \valpha)$. By Jensen's inequality, it holds that
\begin{align*}
\log L(\sD) &=  
\log\Big[\int  p(\vm, \vbeta, \vf, \mOmega, \vgamma, \valpha, \sD)\, \intd \vm\intd \vbeta \intd \vf
\intd \mOmega \intd \vgamma \intd \valpha\Big] \\
&\ge 
\int q(\vm, \vbeta, \vf, \mOmega, \vgamma, \valpha) \log\frac{p(\vm, \vbeta, \vf, \mOmega, \vgamma, \valpha, \sD)}{q(\vm, \vbeta, \vf, \mOmega, \vgamma, \valpha)} 
\, \intd \vm\intd \vbeta \intd \vf \intd \mOmega \intd \vgamma \intd \valpha =   \underline{L} (\sD ) \, .
\end{align*}
The term $\log L(\sD)$ is the marginal likelihood for the dataset, and it is interpreted as the model evidence.  The above implies  the ELBO $\underline{L} (\sD )$ acts as a lower bound for $\log L(\sD)$.

The variational inference problem~\eqref{eqn:VIproblem} requires specifying the distribution family $\mathcal{Q}$. This  family should be large enough to approximate the true posterior accurately, while not too complicated for  ease of computation. The mean field family \citep{blei2017variational} is a popular choice, for which the distribution $q(\cdot)$ factors as a product of distributions for each parameter. In this work, we build a  structured  mean-field variational family $\mathcal{Q}$. The densities inside $\mathcal{Q}$  have the following  form,
\begin{align}
q(\vm, \vbeta, \vf, \mOmega, \vgamma, \valpha)
=&\prod_{i=1}^N  q(\vm_i, \vbeta_i, f_i) \times q(\valpha) q(\vgamma) \times q(\mOmega) \nonumber\\
= &\prod_{i=1}^N  \left[q(\vm_i,  \vbeta_i| f_i) q(f_i)\right] \times 
\prod_{b=1}^{B} \left[q(\valpha_b) q(\gamma_b)  \right] \times q(\mOmega)\, . \label{eqn:factoredQfun}
\end{align}
The  distribution specification for each factor of Eqn.~\eqref{eqn:factoredQfun} is listed in Table~\ref{tbl:simpleMeanField}.  Most of them belong to the exponential family.  Section~\ref{sec:canonicalReview} in the Supplementary Material \citep{He2020} provides a review on these exponential family distributions and our employed notions for their canonical parameters. 

For most parameters (e.g. $\valpha_{b}, \gamma_{b}, \mOmega$) of our  model,  their full conditional distributions belong to the exponential distribution family.  Their variational distributions are assumed to be  their conjugate counterparts.  For example, from the joint distribution \eqref{eqn:model:joint}, we can  find the full conditional for $\gamma_b$ is a Gamma distribution.  Correspondingly, the variational distribution $q(\gamma_b)$ is assumed to be a Gamma distribution. 

The variational distribution $q(\vm_i,  \vbeta_i, f_i)  =  q(\vm_i,  \vbeta_i | f_i) q(f_i)$ of each Mira is constructed with special consideration.  Recall  the actual posterior  is  multimodal in $f_i$. To facilitate uncertainty quantification of the estimated frequency, the variational distribution $q(f_i)$ is allowed to be any continuous density on the compact interval   $[f_{\mathrm{min}} , f_{\mathrm{max}} ]$.  Besides, for convenience, the distribution $q(\vm_i, \vbeta_i | f_i)$ is assummed to be Gaussian and depends on the  frequency $f_i$. This conditional dependence is necessary, because the  amplitude and phase of the fitted sinusoidal curve vary with $f_i$.

\begin{table}
	\centering
	\caption{Notation for the mean field distributions with their canonical parameters\\} \label{tbl:simpleMeanField}
	\begin{tabular}{|c|c|c||c|c|c|}
		\hline
		\multicolumn{3}{|c|}{Local}  & \multicolumn{3}{|c|}{Global} \\
		\hline
		Factor	& Distribution &  Parameter  &		Factor		& Distribution &  Parameter  \\
		\hline
		$q(\vm_i, \vbeta_i | f_i)$& Gaussian & $\eta_{i}^{\theta} = (\eta_{i}^{\theta1}, 
		\eta_{i}^{\theta 2})$ & 	$q(\valpha_b)$ &  Gaussian  & 
		$\eta_{b}^{\alpha} = (\eta_{b}^{\alpha1}, \eta_{b}^{\alpha2})$   \\
		$q(f_i)$ & Free Form &  None &	$q(\gamma_b)$ &  Gamma  & 
		$\eta_{b}^{\gamma} = (\eta_{b}^{\gamma 1}, \eta_b^{\gamma 2})$\\
		& & &
		$q(\mOmega)$ &  Wishart  & $\eta^{\mOmega} = (\eta^{\mOmega 1}, \eta^{\mOmega 2})$  \\		
		\hline
	\end{tabular}
\end{table}

After specifying $\mathcal{Q}$, the model inference  is completely framed as an optimization problem. Our target is maximizing the evidence lower bound, i.e. $\max_{\ q(\cdot) \in\mathcal{Q}}\,  \underline{L} (\sD )$, with $\mathcal{Q}$ determined in~\eqref{eqn:factoredQfun}. The optimal solution $q(\vm, \vbeta, \vf, \mOmega, \vgamma, \valpha)$ will then be used for model inference, in place of the actual  posterior  $p(\vm, \vbeta, \vf, \mOmega, \vgamma, \valpha|\, \sD)$. This optimization is typically carried out by the coordinate ascent algorithm \citep{blei2017variational}, where one factor in \eqref{eqn:factoredQfun} gets updated with the others fixed during each round of iteration. Though this updating strategy improves computational speed over MCMC in lots of cases,  it is still not fast enough for our problem at hand. For efficient computation, we will  exploit our hierarchical model structure under the framework of stochastic variational inference \cite[SVI,][]{hoffman2013stochastic}.  According to the SVI scheme, the unknown parameters can be divided into two groups:  local parameters and global parameters.  The local parameters are specific to each Mira.  For the $i$-th Mira,  its own local parameters include the average magnitude $\vm_i$,  the periodic coefficient $\vbeta_i$, and its frequency $f_i$. On the other hand, the global parameters govern all Mira samples. These global parameters  include the PLR coefficients  $\valpha$,  the PLR residual precision vector $\vgamma$, and the precision matrix $\mOmega$ for the sinusoidal coefficients.  By exploiting the model structure, SVI becomes much more efficient than the classical coordinate ascent algorithm. In fact, conditional on the global parameters, the local parameters of each Mira can be updated independently of each other.  On the other hand,  the global parameters can be updated based on mini-batch samples. The details to update local and global parameters will be provided in Section~\ref{sec:algorithm:local} and  Section~\ref{sec:algorithm:global}, respectively.

\subsection{Update local parameters}
\label{sec:algorithm:local}
In this subsection, we  derive analytic formulae to  update the local variational distribution $q(\vm_i,  \vbeta_i, f_i)  =  q(\vm_i,  \vbeta_i | f_i) q(f_i)$ for each $i\in\sI$, 
while the global $q(\valpha)$, $q(\mOmega)$ and $q(\vgamma)$ are fixed. 
For one specific $i\in\sI$, we need to maximize the ELBO with respect to  $q(\vm_i,  \vbeta_i, f_i)$. Note the ELBO  can be factored as
\begin{align}
\underline{L} (\sD )=& \Expect_{q}\Big\{
\log\frac{p(\vm, \vbeta, \vf, \mOmega, \vgamma, \valpha, \sD)}{q(\vm, \vbeta, \vf, \mOmega, \vgamma, \valpha)} \Big\} \nonumber \\
=& \sum_{j=1}^{N} \Expect_{q}\Big\{
\log\frac{p(\vm_j, \vbeta_j, f_j | \mOmega, \vgamma, \valpha, \sD_j)}{q(\vm_j, \vbeta_j, f_j)} \Big\} 
+\Expect_{q}\Big\{
\log\frac{\prod_{b=1}^{B} \left[p(\valpha_b) p(\gamma_b)  \right] \times p(\mOmega)}{
	\prod_{b=1}^{B} \left[q(\valpha_b) q(\gamma_b)  \right] \times q(\mOmega)} \Big\}  \nonumber\\
=& \Expect_{q}\Big\{
\log\frac{p(\vm_i, \vbeta_i, f_i | \mOmega, \vgamma, \valpha, \sD_i)}{q(\vm_i, \vbeta_i, f_i)} \Big\} +C\,. \label{eqn:mirai:elbo}
\end{align}
In the above, the second equation uses the conditional independence between observations, 
and the last  term $C$ is  a constant irrelevant to the update of
$q(\vm_i, \vbeta_i, f_i) $. According to the last line~\eqref{eqn:mirai:elbo}, we only need to consider
the data $\sD_i$ of the $i$-th Mira. This means given the global
$q(\valpha)$, $q(\mOmega)$ and $q(\vgamma)$, the local
variational distribution $q(\vm_i, \vbeta_i, f_i)$ can be updated independently of the other Miras.

The average magnitude $\vm_i$  and the sinusoidal coefficient $\vbeta_i$ can be conveniently updated as a whole.  For this purpose, we introduce a few notations. Let $\vtheta_{ib} = (m_{ib}, \vbeta_{ib}^T)^T$ be the parameter for the $b$-th band. Then, we arrange  $\vtheta_{ib}$'s of all bands into a vector as $\vtheta_{i} = (\vtheta_{i1}^T,\cdots, \vtheta_{iB}^T)^T\in\bbR^{3B}$.  Correspondingly, two index sets can be defined as $\mathcal{K} = \{1,4,7,\cdots, 3B-2\}$ and its complementary $\mathcal{L} = \{1,2,\cdots, 3B\}\backslash\mathcal{K}$. The index set $\mathcal{K}$  records the locations of $m_{ib}$'s in the vector $\vtheta_i$, while $\mathcal{L}$ corresponds to the locations of  $\vbeta_{ib}$'s inside $\vtheta_i$.

Recall our target of updating the local $q(\vm_i,  \vbeta_i, f_i)  =  q(\vm_i,  \vbeta_i | f_i) q(f_i)$ with fixed $q(\valpha)$, $q(\mOmega)$ and $q(\vgamma)$. With the definition of $\vtheta_{i}$, it is equivalent to state the update in terms of $q(\vtheta_i, f_i) = q(\vtheta_i|f_i) q(f_i)$. According to our distribution specification in Table~\ref{tbl:simpleMeanField}, $q(f_i)$ is allowed to be any continuous density on a compact interval   $[f_{\mathrm{min}}, f_{\mathrm{max}}]$, while $q(\vtheta_i|f_i) \propto \exp\Big\{ \big\langle \eta^{\theta1}_i, \vtheta_i \vtheta_i^T\big\rangle+ \big\langle \eta^{\theta2}_i, \vtheta_i \big\rangle \Big\}$ is assumed to be  Gaussian with canonical parameters $\eta^{\theta1}_i\in\bbR^{3B\times 3B}$ and $\eta^{\theta2}_i\in\bbR^{3B}$. Note the value of $\eta^{\theta1}_i$ and $\eta^{\theta2}_i$ depends on the frequency $f_i$. With these notations, our goal boils down to finding the optimal $q(f_i)$, and for each $f_i$ computing the optimal $\eta^{\theta1}_i,\eta^{\theta2}_i$. The optimal solutions should maximize the ELBO~\eqref{eqn:mirai:elbo}, which is now equivalently expressed as
\begin{align}
\underline{L} (\sD ) =& \Expect_{q}\Big\{
\log\frac{p(\vtheta_i, f_i | \mOmega, \vgamma, \valpha, \sD_i)}{q(\vtheta_i, f_i)} \Big\} +C\,. \label{eqn:mirai:elbo2}
\end{align}

This equivalent ELBO involves the conditional likelihood $p(\vtheta_i, f_i | \mOmega, \vgamma, \valpha, \sD_i)$, which is proportional to  the  likelihood in~\eqref{eqn:likelihood:ithmira}. To convert the expression in a compact form, we need to define a few more notations. According to~\eqref{eqn:hm:layer2}  and~\eqref{eqn:hm:layer3}  of our  model, the conditional  distribution of $\vtheta_{i}$ given $(f_i, \valpha, \vgamma, \mOmega)$ is  $\sN(\vtheta_0, \mTheta^{-1})$, with its mean vector $\vtheta_0$ and precision matrix $\mTheta$  defined as follows.   The sub-vectors of $\vtheta_0$ indexed by $\sK$ and $\sL$ are $\vtheta_0[\sK] = (\vd_i^T\valpha_{1}, \cdots, \vd_i^T\valpha_{B})^T$ and $\vtheta_0[\sL] = \vzero$, respectively. In fact, $\vtheta_0[\sK]$ contains the a-priori mean of $\vm_i$ in~\eqref{eqn:hm:layer2}, while  $\vtheta_0[\sL]$ is the a-priori mean of $\vbeta_{i}$ in~\eqref{eqn:hm:layer3}.  Similarly, the sub-matrices of  $\mTheta$  indexed by $\sK$ and $\sL$ are 
$$
\mTheta[\sK,\sK] = \diag(\gamma_1,\cdots, \gamma_B), \quad \mTheta[\sL,\sL] = \mOmega,  \quad \text{and}\quad \mTheta[\sK,\sL] = \big(\mTheta[\sL,\sK] \big)^T= \vzero\,,
$$
respectively. In the above, for example, $ \mTheta[\sK,\sL] $ is the submatrix of $\mTheta$ with rows indexed by $\sK$, and columns indexed by $\sL$. All of the above is simply writing~\eqref{eqn:hm:layer2} and~\eqref{eqn:hm:layer3} of all bands in a compact form, i.e., $\vtheta_{i} | (f_i, \valpha, \vgamma, \mOmega) \sim \sN(\vtheta_0, \mTheta^{-1})$. In addition, for the $b$-th band of the $i$-th Mira, define a basis matrix $\mC_{ib}  = (\vone, \mB_{ib}) \in\bbR^{n_{ib} \times 3}$. Then, by rewriting~\eqref{eqn:likelihood:ithmira} with these new notations,  it follows that
\begin{align*}
\log p(  \vtheta_{i},f_i|   \mOmega, \vgamma,\valpha, \sD_i) =&
-\frac{1}{2}\sum_{b=1}^{B} (\vy_{ib} - \mC_{ib}\vtheta_{ib})^T \mSigma_{ib}^{-1}  (\vy_{ib} - \mC_{ib}\vtheta_{ib})  - \frac{1}{2}\log\det \mSigma_{ib}\\
&\quad \qquad -\frac{1}{2} (\vtheta_{i} - \vtheta_0)^T\mTheta(\vtheta_{i} - \vtheta_0) +C',
\end{align*}
with some  irrelevant constants $C'$. The above expression will be plugged into~\eqref{eqn:mirai:elbo2} to find the optimal solution.  Our updating formulae for the local parameter is summarized by the following theorem, whose proof is provided in Section~\ref{sec:proof:theorem1} of the Supplementary Material \citep{He2020}.

\begin{theorem} \label{thm:localupdate}
	Suppose the data $\sD_i$ of the $i$-th Mira and  $q(\valpha)$, $q(\vgamma)$, and $q(\mOmega)$ are fixed. Then, for each $f_i$, the optimal $q(\vtheta_{i} | f_i)$ maximizing ELBO  $\underline{L} (\sD )$ has canonical parameters
	\begin{equation}\label{update:local1:theta}
	\eta^{\theta 1}_{i} =\Expect_{q(\vgamma)}\Expect_{q(\mOmega)}(\xi^{\theta 1}_{i}), \quad
	\text{and}\quad 
	\eta^{\theta 2}_{i} = \Expect_{q(\valpha)}(\xi^{\theta 2}_{i}).
	\end{equation}
	In the above,
	$\xi^{\theta 1}_i = -\frac{1}{2}\times
	\diag\left( \mC_{i1}^T\mSigma_{i1}^{-1}\mC_{i1} , \cdots,
	\mC_{iB}^T\mSigma_{iB}^{-1}\mC_{iB}\right) 
	-\frac{1}{2}\mTheta$ and
	$$ 	\xi^{\theta 2}_i = \left(
	(\mC_{i1}^T\mSigma_{i1}^{-1}\vy_{i1})^T, 
	\cdots, 	(\mC_{iB}^T\mSigma_{iB}^{-1}\vy_{iB})^T \right)^T + \mTheta\vtheta_0\,.
	$$
	Moreover, the optimal $q(f_i)$ is given by
	\begin{equation} \label{update:local1:freqMain}
	q(f_i) \propto \exp\Big\{  -\frac{1}{2} g(f_i)   -\frac{1}{2} \log \det (-\eta_i^{\theta1})\Big\},
	\end{equation}
	with  
	\begin{align*}
	g(f_i)=    \frac{1}{2} (\eta_i^{\theta 2})^T (\eta_i^{\theta 1})^{-1}
\eta_{i}^{\theta 2} +
\left\langle\Expect_{q(\mOmega)}\Expect_{q(\vgamma)}\mTheta ,\ \Expect_{q(\valpha)} \vtheta_{0} \vtheta_{0}^T\right\rangle+ \sum_{b=1}^{B} \left[  \vy_{ib}^T\mSigma_{ib}^{-1}\vy_{ib} + 
\log\det \mSigma_{ib}\right].
	\end{align*}
\end{theorem}

This theorem develops analytical formulae to update $q(\vtheta_i, f_i)$ for the $i$-th Mira. At each fixed $f_i$, the canonical parameter for $q(\vtheta_i | f_i)$ is directly available via~\eqref{update:local1:theta}. Notice the basis matrix $\mC_{ib}$, the covariance matrix $\mSigma_{ib}$ as well as the a-priori mean vector $\vtheta_{0}$ all implicitly depend on the frequency $f_i$. Besides, the optimal density $q(f_i)$ is computed by~\eqref{update:local1:freqMain}. In our implementation,  $q(f_i)$  is discretized and evaluated on a dense grid of $[f_{\min}, f_{\max}]$. The grid is gaped by a small value of $\Delta f$.  After that $q(f_i)$ is normalized to become a proper density. The final algorithm for updating $q(\vtheta_i, f_i) = q(\vtheta_i| f_i) q(f_i)$ is summarized in  Algorithm~\ref{alg:updateOneSample}. In our numerical studies, the computation is performed on $500$ equally spaced points on the interval  $[1/1000, 1/100]$.

\begin{algorithm}[t]
	\caption{Updating Local Parameters} \label{alg:updateOneSample}
	\begin{flushleft}
	\textbf{Input}: $f_{\text{min}}, f_{\text{max}}, \Delta f$. The dataset $\sD_i$.\\
\textbf{Output}: $q(\vtheta_i, f_i)$.
	\end{flushleft}
	\begin{algorithmic}[1]
		\State $f_i \leftarrow f_{\text{min}}$
		\While{$f_i\le f_{\text{max}}$}
		\State Update $q(\vtheta_i | f_i)$ by (\ref{update:local1:theta}).
		\State Compute $q(f_i)$ by (\ref{update:local1:freqMain}).
		\State $f_i \leftarrow f_i + \Delta f$
		\EndWhile
		\State Normalize $q(f_i)$ to become a proper density. 
	\end{algorithmic}
\end{algorithm}

\subsection{Update global parameters}
\label{sec:algorithm:global}

Now, we address   updating  the global  $q(\valpha), q(\mOmega)$  and $q(\vgamma)$,  with fixed local  variational distributions $q(\vm_i, \vbeta_{i}, f_i)$ for all $i\in\mathcal{I}$.  We firstly develop the update of $q(\mOmega)$ in detail. The way to update $q(\valpha)$ and $q(\vgamma)$ is derived similarly. 

When $q(\valpha)$, $q(\vgamma)$ and all  $q(\vm_i, \vbeta_{i}, f_i)$'s are fixed, optimizing the ELBO with respect to $q(\mOmega)$ is equivalent to maximizing $\Expect_q \log \left[\frac{p(\mOmega | \valpha, \vgamma, \vbeta, \vm, \vf, \sD)}{q(\mOmega)} \right]$.  This is because
\begin{align}
\underline{L} (\sD )=& \Expect_{q}\Big\{
\log\frac{p(\vm, \vbeta, \vf, \mOmega, \vgamma, \valpha, \sD)}{q(\vm, \vbeta, \vf, \mOmega, \vgamma, \valpha)} \Big\} \nonumber \\
=& \Expect_q \log \left[\frac{p(\mOmega | \valpha, \vgamma, \vbeta, \vm, \vf, \sD)}{q(\mOmega)} \right]
+\Expect_q \log \left[\frac{p( \valpha, \vgamma, \vbeta, \vm, \vf, \sD)}{q(\vm, \vbeta, \vf, \mOmega, \vgamma, \valpha)/q(\mOmega)} \right] \nonumber\\
=& \Expect_q \log \left[\frac{p(\mOmega | \valpha, \vgamma, \vbeta, \vm, \vf, \sD)}{q(\mOmega)} \right]+C''\,, \label{eqn:omega:elbo}
\end{align}
where $C''$ is some constant invariant to the change of $q(\mOmega)$. In the above, the expectation $\Expect_q(\cdot)$ is taken with respect to  $q\in\mathcal{Q}$ in~\eqref{eqn:factoredQfun} at the current iteration. From the joint distribution~(\ref{eqn:model:joint}),  we can find the full conditional for $\mOmega$ as
\begin{align*}
p(\mOmega | \valpha, \vgamma, \vbeta, \vm, \vf, \sD) 
\propto \exp\Big\{-\frac{1}{2} \Big\langle 
\bar{n}\bar{\mOmega}^{-1} + \sum_{i=1}^{N} \vbeta_{i} \vbeta_{i}^T,
\mOmega \Big\rangle + \frac{N+\bar{n}-2B-1}{2} \log|\mOmega|
\Big\}\,.
\end{align*}
This full conditional is a Wishart distribution with  canonical parameter $\xi^{\mOmega1} = -(\bar{n}\bar{\mOmega}^{-1} + \sum_{i=1}^{N} \vbeta_{i} \vbeta_{i}^T)/2$ and $\xi^{\mOmega 2} = (N+\bar{n}-2B-1)/2$.  In addition, note $q(\mOmega)\propto \exp\big\{\langle \eta^{\mOmega1}, \mOmega\rangle + \eta^{\mOmega2} \log|\mOmega| \big\}$  is also  Wishart  with canonical parameters $\eta^{\mOmega1}$ and $\eta^{\mOmega2}$. It can be verified  the optimal  $q(\mOmega)$ maximizing~\eqref{eqn:omega:elbo} has  canonical parameters 
\begin{equation} \label{eqn:optimal:etaomeag1}
\eta^{\mOmega1} = \Expect_q(\xi^{\mOmega1}) =
-\frac{1}{2}\Big[\bar{n}\bar{\mOmega}^{-1} + \sum_{i=1}^{N} \Expect_{q(\vm_i, \vbeta_{i}, f_i)}
\left(\vbeta_{i} \vbeta_{i}^T\right)\Big]\,,
\end{equation}
and $\eta^{\mOmega 2} = \xi^{\mOmega 2}$.  
Although it is  straightforward  to compute $\eta^{\mOmega 2} = \xi^{\mOmega 2}=(N+\bar{n}-2B-1)/2$, the involved computation in~\eqref{eqn:optimal:etaomeag1} is  expensive for a large dataset. In particular, the expectation with respect to $q(\vm_i, \vbeta_i, f_i)$ does not have a close form expression. It can only be carried out by costly numerical integration. Moreover, this  has to be done for all of the $N$ observations during each algorithm iteration, which is especially troublesome when $N$ is large.

Fortunately, the intensive computation can be avoided by  SVI \citep{hoffman2013stochastic}, which only requires an unbiased estimation of natural gradient.  For the canonical parameter  $\eta^{\mOmega1}$ of  $q(\mOmega)$, its  natural gradient of the ELBO is  $\Expect_q(\xi^{\mOmega1}) - \eta^{\mOmega1}$. It is the difference between the expected parameters of the full conditional and the parameters of the current $q(\mOmega)$. The gradient ascent algorithm would update $q(\mOmega)$  by $ \eta^{\mOmega1} \leftarrow \eta^{\mOmega1} + \kappa_t \left[\Expect_{q}(\xi^{\mOmega1}) - \eta^{\mOmega1}\right]$ with some step size $\kappa_t$ at the $t$-th iteration. In the context of  stochastic gradient algorithm \citep{ghadimi2013stochastic}, it is plausible to employ an unbiased estimation of the natural gradient via sampling.  For this purpose, at the $t$-th iteration, we sample a subset of indices $\widetilde{\sI}_t= \{i_1, i_2, \cdots, i_I\}$ from $\mathcal{I} = \{1,\cdots, N\}$ without replacement and with equal probability. The cardinality of the set is  $|\widetilde{\sI}_t| = I$ for some fixed $I$.  This corresponds to the mini-batch strategy, and the batch size $I$ is usually set to be $1,4,8,16$, etc. Given the set $\widetilde{\sI}_t$ and for each $j\in\widetilde{\sI}_t$, we further sample a frequency value $\tilde{f}_{j}$ from the current variational distribution $q(f_{j})$. Then, an unbiased estimate of the natural gradient $\Expect_q(\xi^{\mOmega1}) - \eta^{\mOmega1}$ is
$$
-\frac{1}{2}\Big[\bar{n}\bar{\mOmega}^{-1} + \frac{N}{I}\sum_{j\in\widetilde{\sI}_t} \Expect_{q(\vm_{j}, \vbeta_{j} | \tilde{f}_{j})}
\Big(\vbeta_{j} \vbeta_{j}^T\Big)\Big] - \eta^{\mOmega1}.
$$
Based on the unbiased natural gradient estimate, the update for $q(\mOmega)$ becomes
\begin{align}
\eta^{\mOmega1} &\leftarrow \eta^{\mOmega1} + \kappa_t
\Big(-\frac{1}{2}\Big[\bar{n}\bar{\mOmega}^{-1} + \frac{N}{I}\sum_{j\in\widetilde{\sI}_t}
\Expect_{q(\vm_{j}, \vbeta_{j} | \tilde{f}_{j})}
\Big(\vbeta_{j} \vbeta_{j}^T\Big)\Big]
- \eta^{\mOmega1}\Big), \label{eqn:updateOmega1} \\
\eta^{\mOmega 2} &\leftarrow (N+\bar{n}-2B-1)/2. \label{eqn:updateOmega2} 
\end{align}
Notice we only need to compute with $I$ samples in the udpate~\eqref{eqn:updateOmega1}.  In addition, the expectation with respect to 
$q(\vm_j, \vbeta_{j}| \tilde{f}_j)$ is straightforward because it is a Gaussian distribution.  The optimal value for $\eta^{\mOmega 2}$ is directly provided in \eqref{eqn:updateOmega2},  and $\eta^{\mOmega 2}$  keeps fixed at this value during the iteration. These features make the algorithm  scale easily to a large collection of  observations.

Under the same principle, we can find the full conditional for
$\gamma_{b}$ as
$q(\gamma_{b} |  \valpha, \mOmega, \vbeta, \vm, \vf, \sD) \propto
\exp\big\{\xi_b^{\gamma 1} \gamma_{b}
+ \xi_b^{\gamma 2} \log \gamma_{b}
\big\}$,
which is gamma distribution with canonical parameter 
$\xi_b^{\gamma 1} = -\bar{r} - \sum_{i=1}^{N} (m_{ib} - \vd(f_i)^T\valpha_{b})^2/2$
and $\xi_b^{\gamma 2} = N/2+\bar{\gamma}_b\bar{r}- 1$. Note 
$\vd(f) = (1,  \log_{10}(f), \log^2_{10}(f))^T$ is the PLR basis.
With the sampled subset of data and the sampled frequencies,
the SVI algorithm updates $q(\gamma_b)$ by
\begin{align}
\xi_b^{\gamma 1} &\leftarrow  \xi_b^{\gamma 1} + \kappa_t \Big( -\bar{r} - \frac{N}{I}\sum_{j\in\widetilde{\sI}_t}
\Expect_{q(\vm_j, \vbeta_{j}| \tilde{f}_j)} \Expect_{q(\valpha_{b})}
\left[ m_{jb}  - \vd(\tilde{f}_j)^T \valpha_{b} \right]^2/2 \Big), \label{eqn:updateGamma1} \\
\xi_b^{\gamma 2} &\leftarrow N/2+\bar{\gamma}_b\bar{r}- 1. \label{eqn:updateGamma2} 
\end{align}
Similarly,  the full conditional for  $\valpha_b$ is $q(\valpha_b |  \vgamma, \mOmega, \vbeta, \vm, \vf, \sD) \propto \exp\Big\{
\left\langle \eta^{\alpha1}_b, \ \valpha_b\valpha_b\right\rangle + \left\langle \eta^{\alpha2}_b , \ \valpha_b \right\rangle \Big\}$.
This full conditional is multivariate Gaussian with canonical parameter 
$\eta^{\alpha1}_b =-\frac{\bar{\delta}}{2}\mI - \frac{\gamma_{b}}{2} \sum_{i=1}^N \vd(f_i) \vd^T(f_i)$, 
and $\eta^{\alpha2}_b = \bar{\delta} \bar{\alpha}_b + \gamma_{b}\sum_{i=1}^{N} m_{ib} \vd(f_i)$.
The SVI algorithm updates  $q(\valpha_{b})$ by 
\begin{align}
\eta^{\alpha1}_b &\leftarrow \eta^{\alpha1}_b  + \kappa_t\Big(-\frac{\bar{\delta}}{2}\mI - \frac{\gamma_{b}}{2} \times \frac{N}{I}\sum_{j\in\widetilde{\sI}_t } \vd(\tilde{f}_j)\vd(\tilde{f}_j)^T\Big), \label{eqn:updateAlpha1} \\
\eta^{\alpha2}_b &\leftarrow  \eta^{\alpha2}_b + \kappa_t \Big(\bar{\delta} \bar{\alpha}_b + \gamma_{b}\times 
\frac{N}{I}\sum_{j\in\widetilde{\sI}_t } \Expect_{q(\vm_j, \vbeta_{j}| \tilde{f}_j)} (m_{jb}) \times \vd(\tilde{f}_j)
\Big).
\label{eqn:updateAlpha2}
\end{align}

\begin{algorithm}[t]
	\caption{Main Algorithm} \label{alg:simpleMain}
	\begin{algorithmic}[1]
		\For{$t=1,2,\cdots$}
		\State Set step size $\kappa_t = (c_1 + t)^{-c_2}$.
		\State Sample  $\widetilde{\sI}_t= \{i_1, i_2, \cdots, i_I\}$ from $\sI = \{1,2,\cdots, N\}$ without replacement.
		\For{$j\in \widetilde{\sI}_t$}
		\State Update local  $q(\vtheta_j, f_j)$ by Algorithm~\ref{alg:updateOneSample}.
		\State Sample $\tilde{f}_{j}$ according to $q(f_{j})$.
		\EndFor
		\State Update $q(\mOmega)$ by (\ref{eqn:updateOmega1}) and  (\ref{eqn:updateOmega2}). 
		\For{$b=1,2,\cdots, B$}
		\State Update $q(\gamma_b)$ by (\ref{eqn:updateGamma1}) and  (\ref{eqn:updateGamma2}).
		\State Update $q(\valpha_b)$ by (\ref{eqn:updateAlpha1}) and  (\ref{eqn:updateAlpha2}).
		\EndFor
		\EndFor
	\end{algorithmic}
\end{algorithm}

The final algorithm is summarized in Algorithm~\ref{alg:simpleMain}. The algorithm chooses  the step size $\kappa_t = (c_1 + t)^{-c_2}$. At the start of each iteration,  a small subset of samples is selected and their local parameters get updated in Line 3--7 of Algorithm~\ref{alg:simpleMain}.  With the sampled and updated local variational distribution, the global parameters get updated by \eqref{eqn:updateOmega1}--\eqref{eqn:updateAlpha2}.    In our numerical studies, we set $c_1 \in[1000,2000]$ and $c_2\in[0.5,1]$. We also take mini-batch size $I=8$, and  find that 1000 iterations are enough for convergence.

\section{Simulation}
\label{sec:simulation}

We have conducted two simulation experiments to compare the performance of our method (SVI) with some existing methods.  These include the generalized Lomb-Scargle method \citep[GLS,][]{zechmeister2009generalised} and its multi-band version 
\citep[MGLS,][]{vanderplas2015periodograms}, the single-band semi-parametric model \citep[SP,][]{he2016period} and its multi-band extension \citep[MSP,][]{yuan2018near}.

\begin{table}
	\centering
	{\scriptsize
		\begin{minipage}{.52\linewidth}
			\caption{Performance for Simulation I}	\label{tab:sim_M33}
			\centering
			\begin{tabular}{r|ccccc}
				\hline\hline
				& GLS & MGLS & SP & MSP & SVI \\ 
				\hline
				ADE ($\times 10^{-4}$) & 5.98 & 3.97& 5.70 & 2.18 & \textbf{0.52} \\ 
				RR (\%) & 72.00& 82.44 & 78.58 & 91.26 & \textbf{98.04} \\ 
				\hline
			\end{tabular}
		\end{minipage}%
		\begin{minipage}{.48\linewidth}
			\centering
			\caption{SVI Confidence Set} \label{tab:sim_M33:confset}
			\begin{tabular}{r|cccc}
				\hline\hline
				& \multicolumn{4}{c}{Coverage} \\ 
				\hline
				Nomimal (\%) & 90 & 95 & 99 & 99.5  \\ 
				Actual (\%) & 78.5 & 86.2 & 95.3 & 97.8 \\ 
				\hline
			\end{tabular}
		\end{minipage} 
	}
\end{table}

\begin{figure}
	\centering
	\includegraphics[width =0.42 \textwidth]{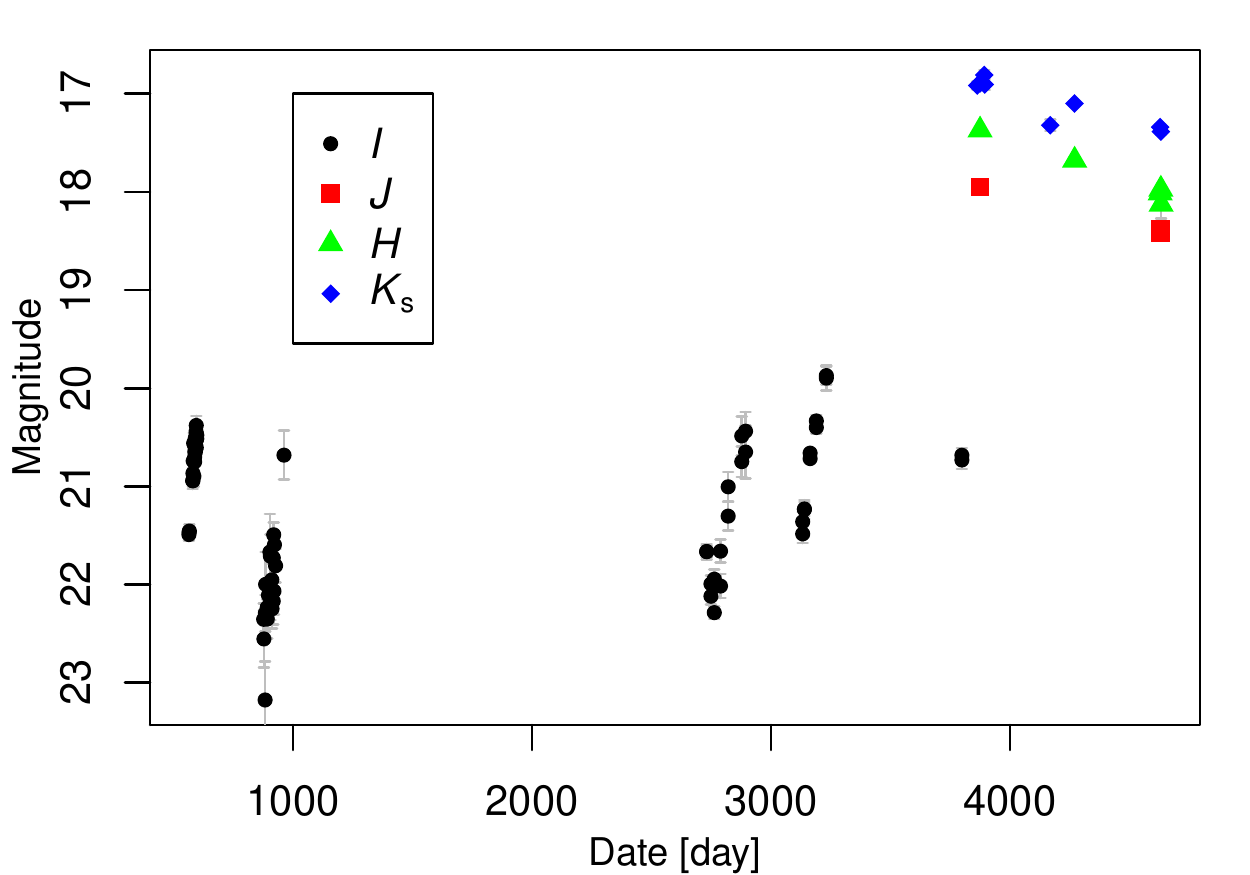}
	~\includegraphics[width =0.42 \textwidth]{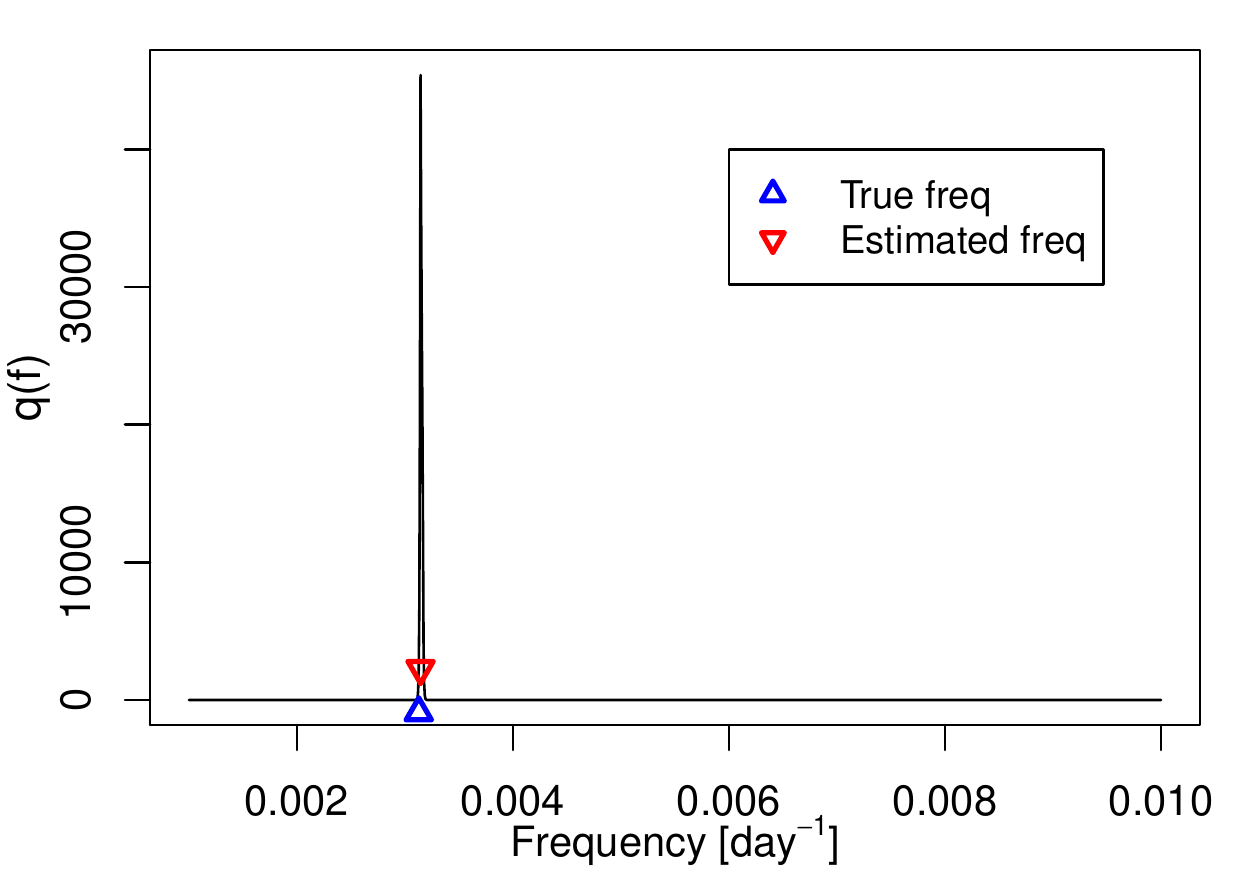}
	
	\includegraphics[width =0.42 \textwidth]{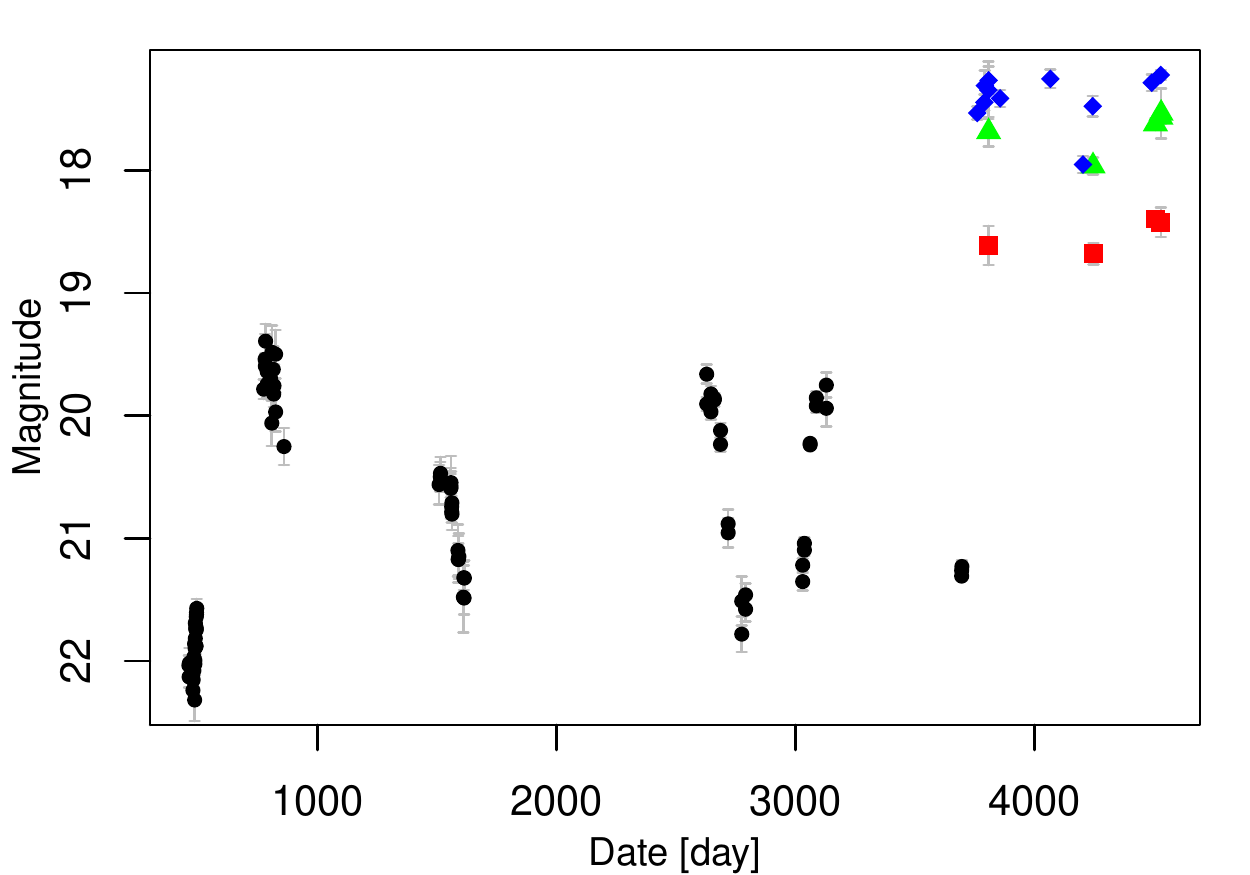}
	~\includegraphics[width =0.42\textwidth]{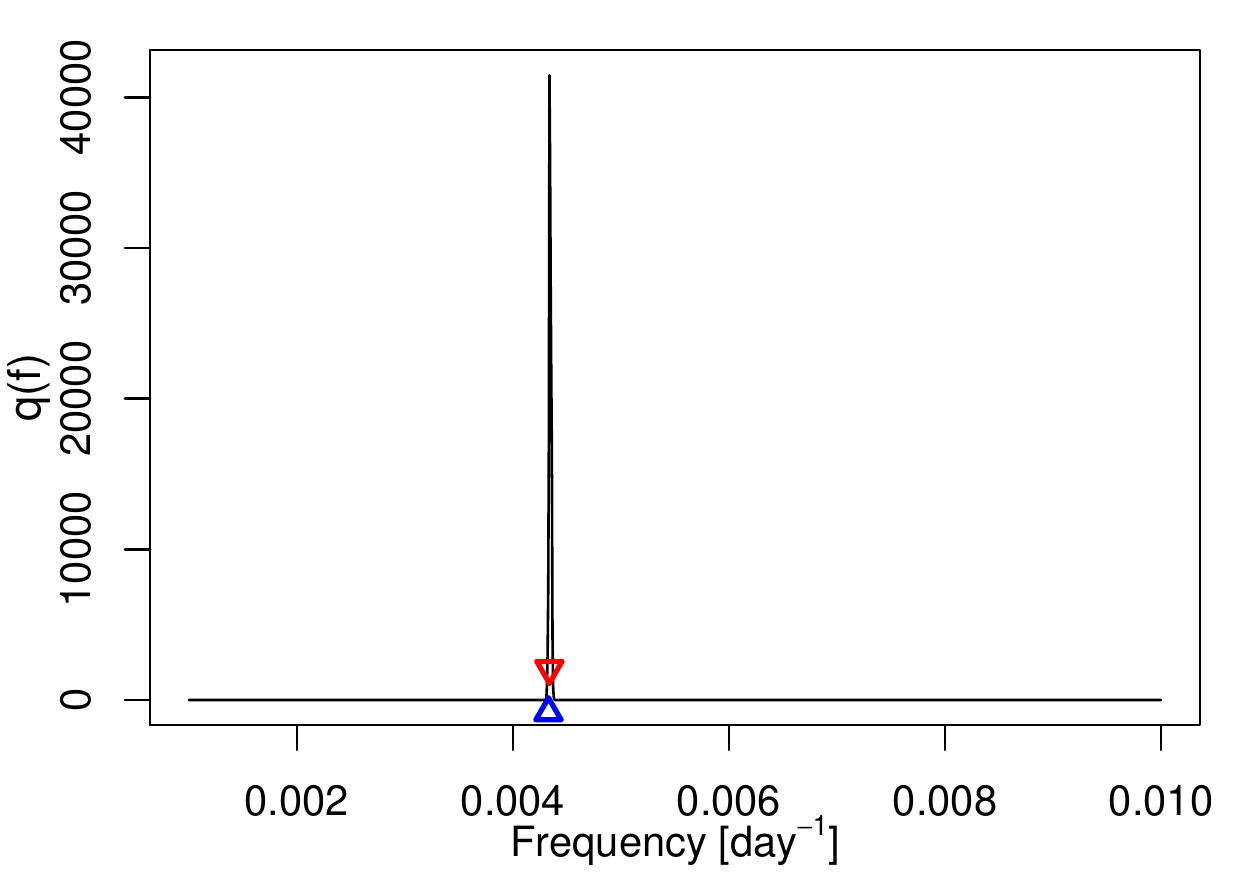}
	
	\includegraphics[width =0.42 \textwidth]{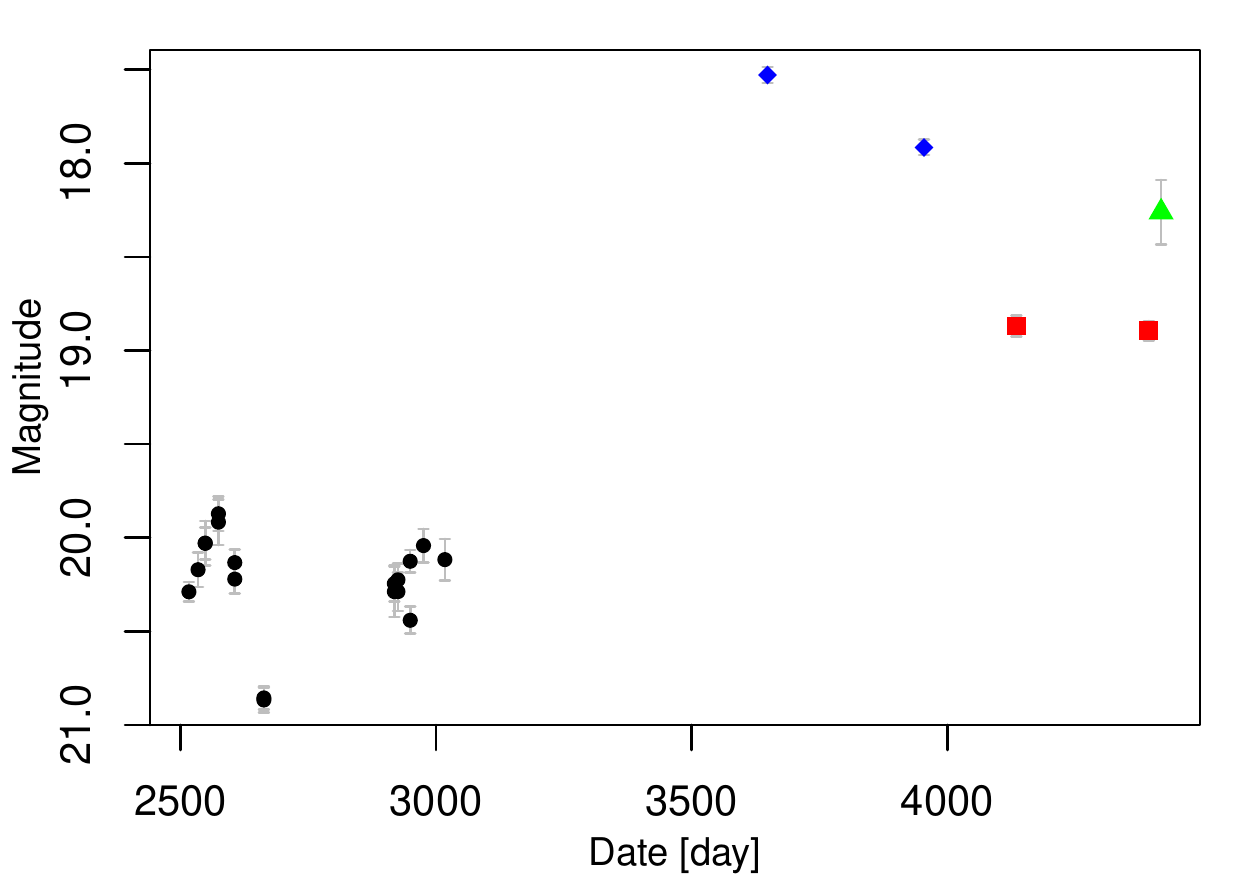}
	~\includegraphics[width =0.42 \textwidth]{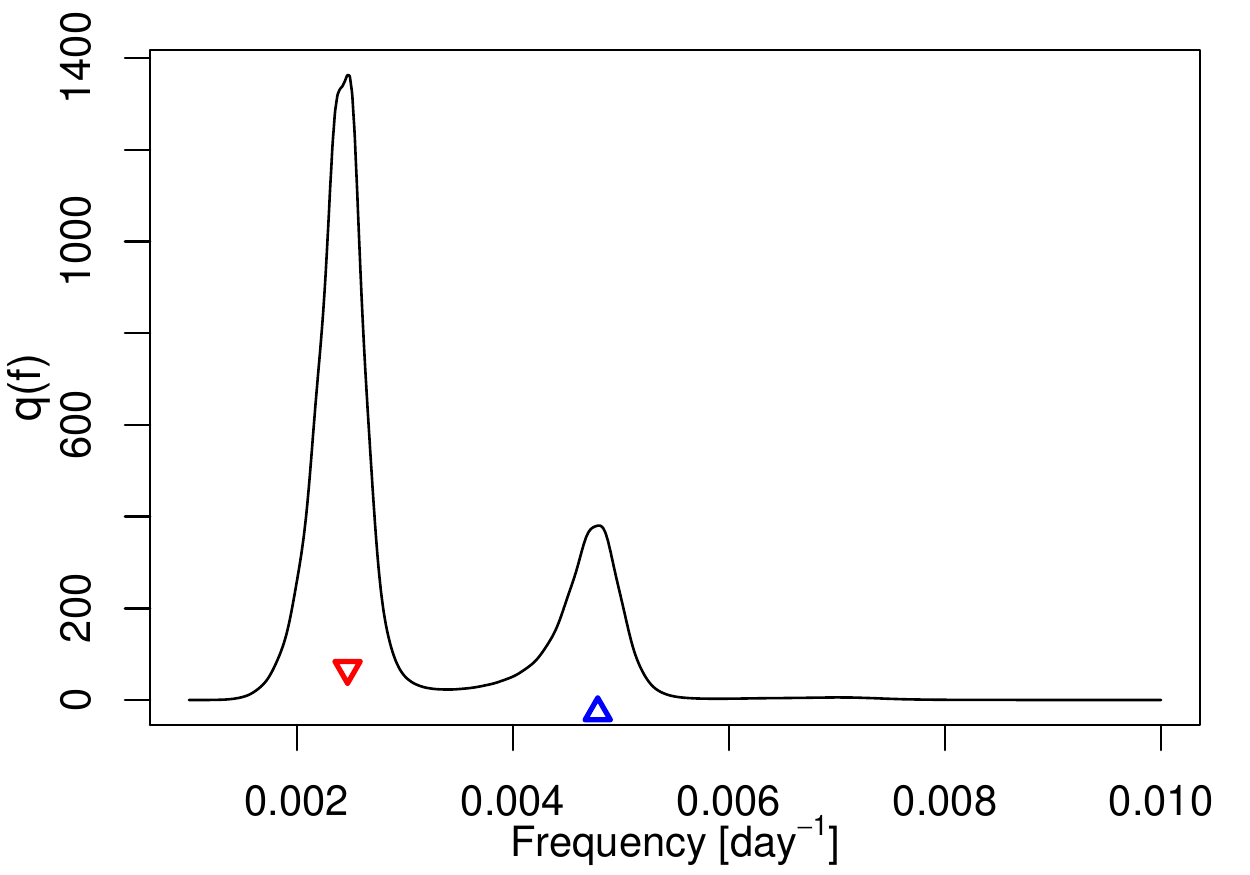}
	
	\includegraphics[width =0.42 \textwidth]{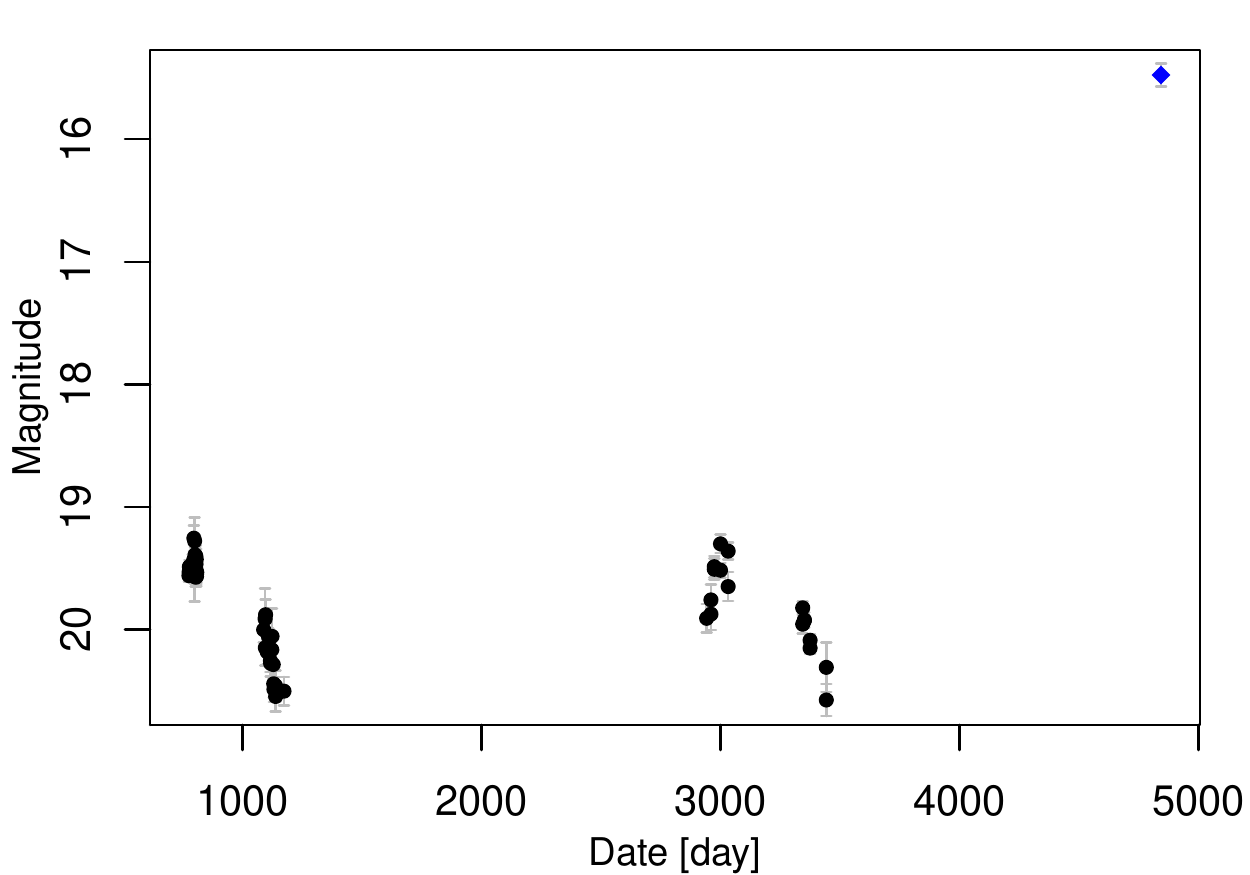}
	~\includegraphics[width =0.42 \textwidth]{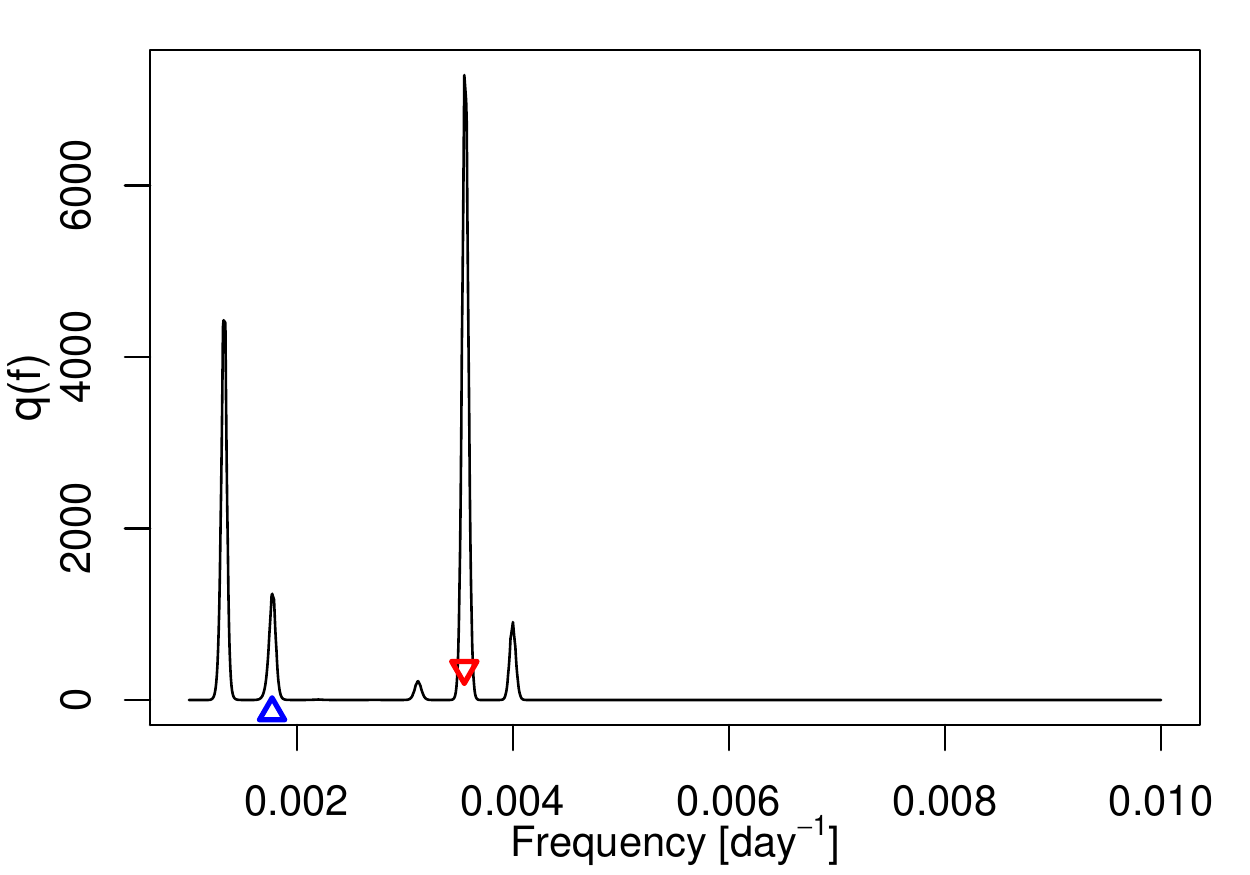}
	\caption{In the left column are four simulated Mira light curves from Simulation I. In the right column are the  estimated variational distributions $q(f_i)$ for the corresponding Mira in the same row.} \label{fig:simu1:lc}
\end{figure}

The first simulation uses the same simulated dataset used in \cite{yuan2018near}, where the light curves for 5,000 O-rich Miras were simulated in the $I$, $J$, $H$, and $K_s$ bands. The cadence and sampling quality of the light curves closely matched their actual survey of M33. Given the observation time points, the signal light curves were generated using the Mira template of \cite{yuan2017large}. This template is trained from the well-observed OGLE LMC data set. After this, the magnitude of the generated light curves were shifted by $+6.27$ mag to account for the relative distance between the LMC and M33. Finally, the light curve magnitude values were contaminated by realistic noise. Four simulated multi-band Mira light curves are shown in the left column of Figure~\ref{fig:simu1:lc}. 

The five methods GLS, MGSL, SP, MSP, and SVI are applied to this dataset. The single-band models (GLS and SP) are applied to the $I$-band dataset as it has the most data points. For our proposed method, we use maximum a posterior (MAP) as a point estimate for the frequency. In other words, for the $i$-th star, its estimated frequency is  $\hat{f}_i = \arg\max_{f_i} q(f_i)$ with the final estimated variational distribution $q(f_i)$.

The accuracy of period estimation is measured by two metrics: recovery rate and absolute deviation error. The recovery rate (RR) is the percentage of light curves with accurately estimated frequency.  Suppose for the $i$-th simulated Mira, $f_{i0}$ is its true frequency and $\hat{f}_{i}$ is the estimated frequency. The estimation is considered to be accurate if their absolute difference is less than a threshold value, i.e., if $|f_{i0} - \hat{f}_i| \le \lambda$. We choose $\lambda = 2.7\times 10^{-4}$ in this work,  which is the average half distance to the sidelobes in the frequency spectra \citep{he2016period}. Then, the recovery rate (RR) is computed as $\mathrm{RR} = \frac{1}{N} \sum_{i=1}^{N} \mathrm{I}(|f_{i0} - \hat{f}_i| \le \lambda)$, where $\mathrm{I}(\cdot)$ is the indicator function. The other employed metric is the absolute deviation error (ADE), computed by $\mathrm{ADE} =\frac{1}{N} \sum_{i=1}^{N} |f_i - \hat{f}_i | $. 

For all methods, their recovery rate and ADE on this dataset are reported in Table~\ref{tab:sim_M33}.  Notice the single-band GLS method has worst performance in both metrics. By exploiting more observational information, the multi-band  methods (MGLS and MSP) increase the RR by 10\% and reduces ADE by more than 40\%, compared with their  single-band counterparts (GLS and SP). It is also notable that  SP (or MSP) outperforms GLS (or MGLS resp.).  Recall the SP and MSP have modeled the sinusoidal residuals by Gaussian process.  The improvement demonstrates the effectiveness of modeling the quasi-periodic signal of Mira light curves.  Lastly, we note our SVI method has the best performance. The proposed SVI has  remarkably estimated almost all samples accurately. It has a recovery rate of $98.20\%$ and ADE accuracy of $0.52\times 10^{-4}$.

In the right column of  Figure~\ref{fig:simu1:lc}, we plot  the fitted variational distribution $q(f_i)$ for each simulated Mira light curve shown in the left column of the same row.  The red reversed triangles mark the estimated frequencies $\hat{f}_i$, and the blue triangles mark the true frequencies. The first two rows demonstrate the cases when the true frequencies are correctly identified. The last two rows show the cases when the MAP $\hat{f}_i$ fails to estimate the true frequency. However, in these cases when the signal is weak, our proposed method still recovers the multi-modal posterior of the frequency, and the true frequency lies around one of the posterior modes.

The plots of $q(f_i)$ illustrate the power of this method to quantify the frequency estimation uncertainty. To further assess the uncertainty quantification from $q(f_i)$, we construct a $1-\alpha$ level confidence set for each Mira in the simulated dataset.  The confidence set $\sC_i = \{f_i: q(f_i) > c_i\}$ is constructed with some $c_i$  such that $\int_{\sC_i}  q(f_i)\intd f_i = 1-\alpha$. Note each $\sC_i$ may be composed of a few short intervals due to the multimodal nature of $q(f_i)$.  For different value of  $\alpha$, the confidence set is computed, and the actual coverage rate is  evaluated over the simulated dataset.  The empirical results  are reported in Table~\ref{tab:sim_M33:confset}. For example, when the nominal coverage rate is $95\%$ (with $\alpha=5\%$), the actual coverage rate is only $86.2\%$. It indicates the confidence sets are not wide enough and  have an under-coverage issue.  This problem is  inherent  for mean filed variational inference which underestimates  estimation uncertainty \citep{wang2019frequentist}. 

Taking a closer look of the coverage rate issue, we found that, in most cases when the confidence set misses the true frequency, the true frequency resides very close to the confidence set; it is rare that the confidence set completely misses the true frequency. More precisely, let $\mathcal{C}_i$ be the 99\% confidence set for the $i$ sample. Let $\lambda = 2.7\times 10^{-4}$ be the threshold to determine the recovery rate (RR) as used in the paper. We say that the set $\mathcal{C}_i$ ``entirely misses'' the true frequency, if $\text{dist}(f_{0i},\mathcal{C}_i)>\lambda$; and it ``almost covers" the true frequency, if otherwise. In other words, the $\mathcal{C}_i$ is considered as completely failed if the distance of the true frequency $f_{i0}$ to the set $\mathcal{C}_i$ is larger than $\lambda$. Out of the total 5,000 simulated samples, for 234 samples (4.68\%) the 99\% confidence set fails to cover the corresponding the true frequencies; among them, for only 19 (0.38\%) samples, the confidence set entirely misses the true frequency, and for 215 samples (4.3\%), the confidence set almost covers the true frequency.

To further assess the performance of the proposed method, we conducted another, more comprehensive simulation study where 90 datasets with distinct characteristics were generated by considering 90 different combinations of sampling cadence patterns, different observation noise levels, and various number of sampling points. Some of these datasets are more  challenging than others with higher level of data sparsity and noise.  Our proposed method still delivers the best performance in terms of  the two metrics (RR and ADE) for each dataset. Details of this simulation study are presented in Section~\ref{sec:suppsimu} of the Supplementary Material \citep{He2020}.

\begin{figure}[t]
	\centering
	\begin{subfigure}[b]{0.32\textwidth}
		\includegraphics[width =\textwidth]{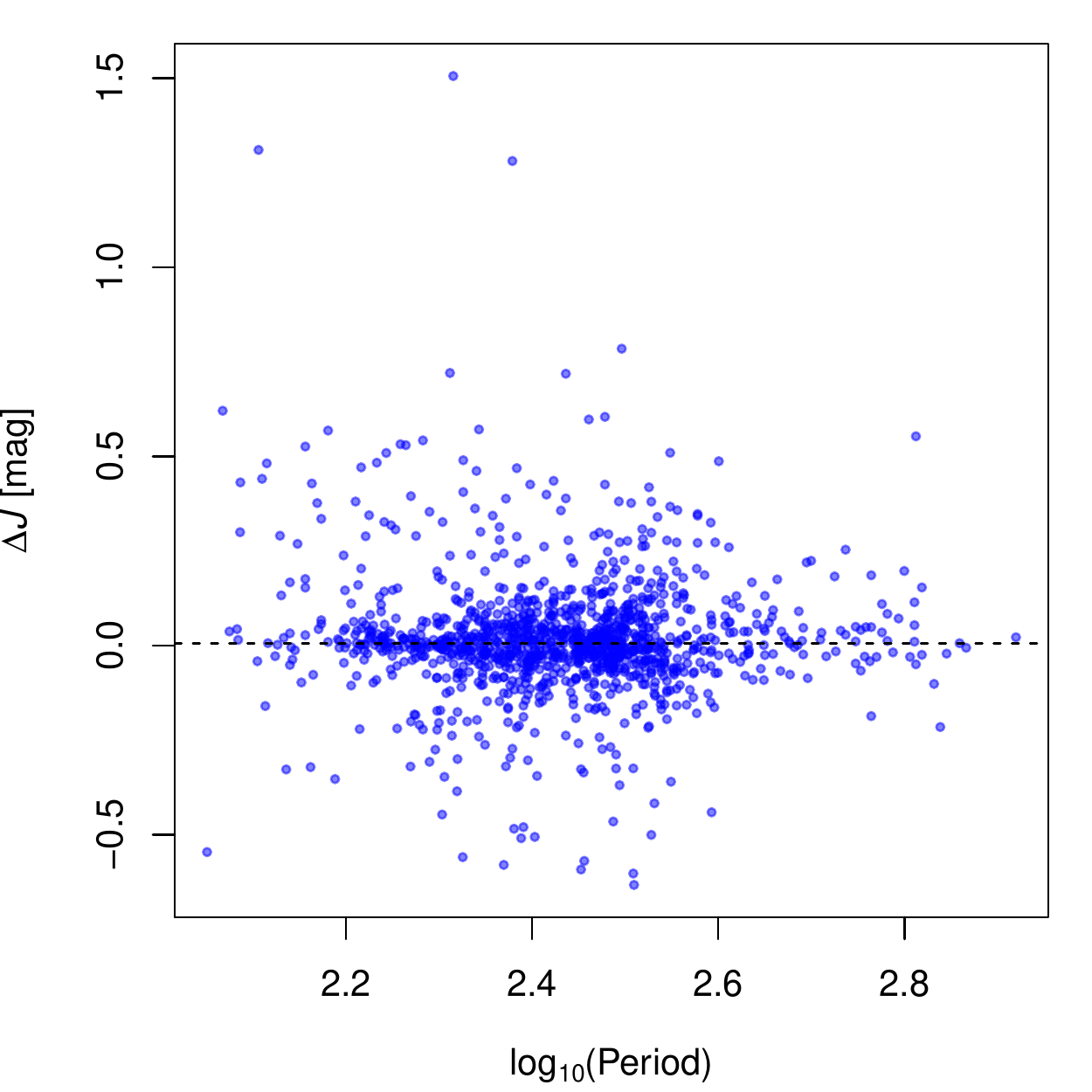}
		\caption{}
		\label{fig:real:comparePM2}
	\end{subfigure}~
	\begin{subfigure}[b]{0.32\textwidth}
		\includegraphics[width =\textwidth]{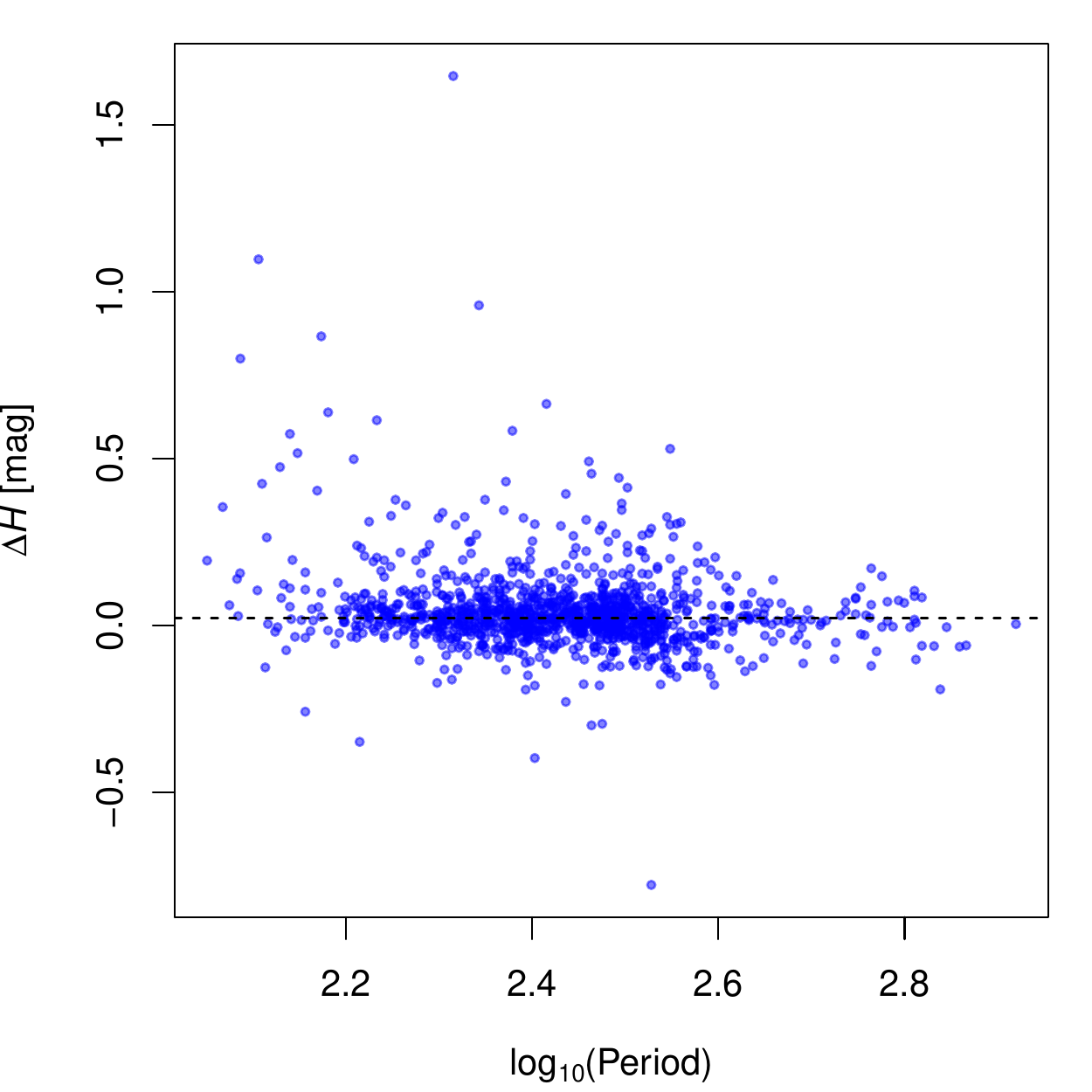}
		\caption{}
		\label{fig:real:comparePM3}
	\end{subfigure}~
	\begin{subfigure}[b]{0.32\textwidth}
		\includegraphics[width =\textwidth]{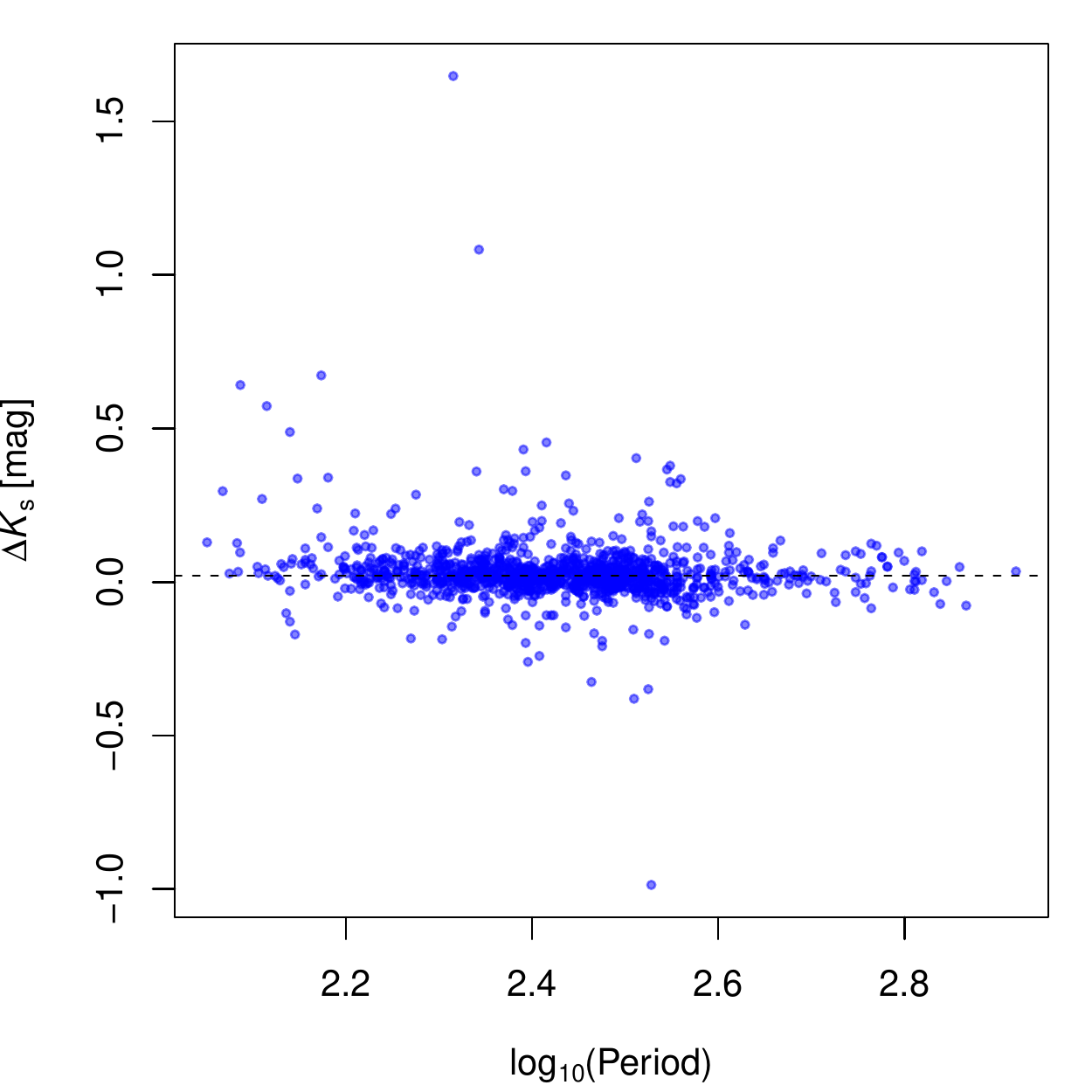}
		\caption{}
		\label{fig:real:comparePM4}
	\end{subfigure}
	\caption{
		The horizontal axis is the logarithm of SVI period, and the vertical axis is the difference of average magnitude between the results of SVI and  \cite{yuan2018near}.  Left, center and right panels show $J-$, $H$ and $K_s$ magnitudes, respectively. The horizontal dashed lines indicate the average difference between the two methods.	\label{fig:real:comparePM}  }
\end{figure}

\section{The M33 Dataset}
\label{sec:realdata}

In this section we analyze the dataset of M33 Miras from \cite{yuan2018near}. The authors collected images and photometric measurements from several sources \citep{javadi2015uk,macri2001direct,pellerin2011m} to generate light curves in four bands ($I$, $J$, $H$, and $K_s$) with a median number of observations of 68, 5, 6, and 11, respectively. \cite{yuan2018near} identified 1,781 candidate Miras and classified 1,265 of them as O-rich, 88 as C-rich, and 428 as unknown, respectively.  The O-rich Miras were used to estimate distance moduli (logarithms of the distance) for M33. From the $J$-, $H$- and $K_s$-band PLRs, they obtained values of $24.82\pm 0.06$ mag, $24.82\pm 0.06$  mag, and $24.75 \pm 0.06$  mag, respectively. Earlier works based on Cepheid variables in the same galaxy yielded estimates of its distance modulus as $24.65\pm 0.12$ mag \citep{Lucas2001extragalactic}, $24.53\pm 0.11$ mag \citep{scowcroft2009effect}, and $24.62 \pm 0.07$  mag  \citep{gieren2013araucaria}. Detached eclipsing binaries resulted in an estimate of $24.92\pm 0.12$  mag \citep{bonanos2006first}, while RR Lyrae variables provided $24.67\pm 0.08$  mag  \citep{sarajedini2006rr}.

\begin{figure}
\begin{minipage}[t]{0.46\linewidth}
	\includegraphics[width =\textwidth]{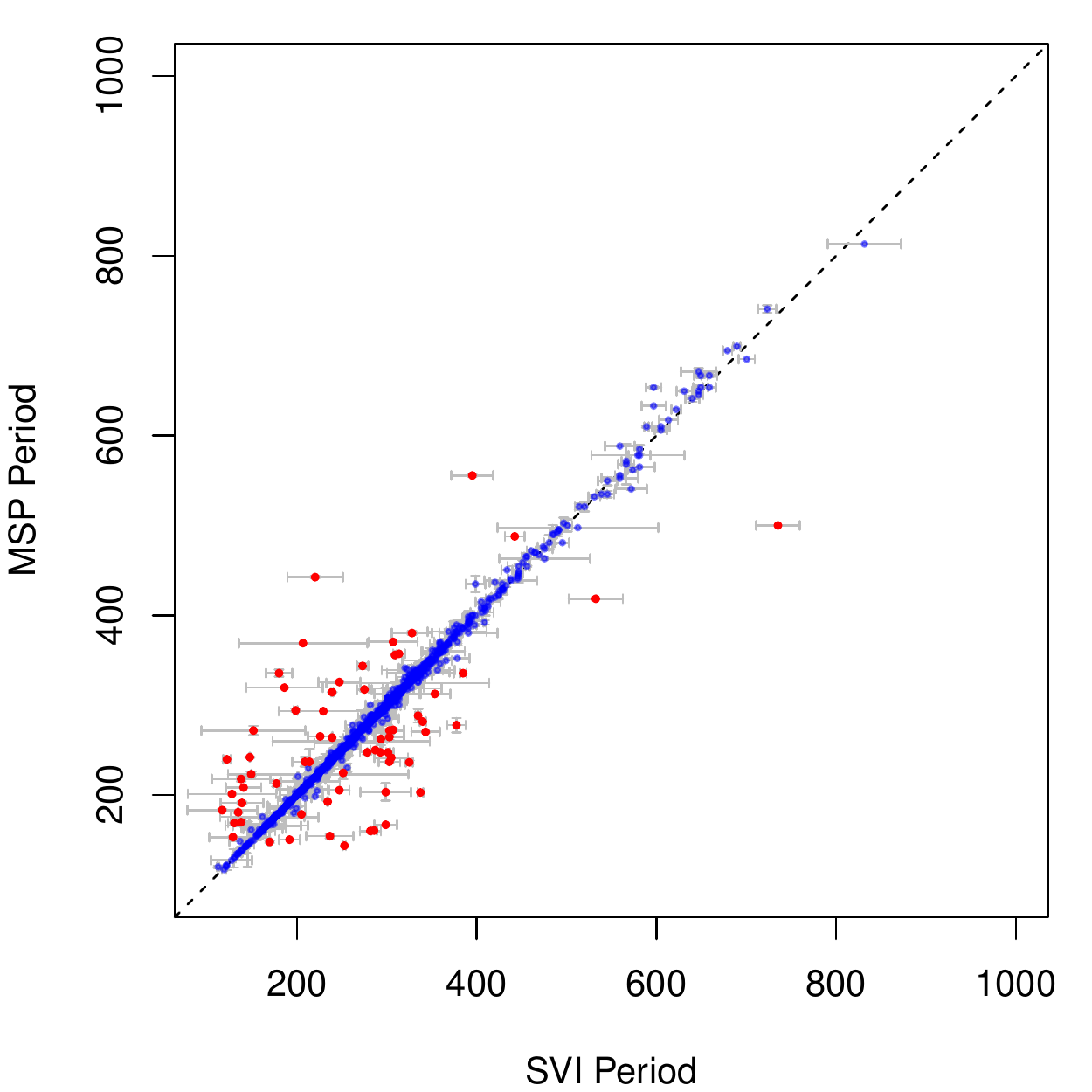}
		\caption{ Comparison of the period estimated by SVI and  by \cite{yuan2018near}.  The red points are Miras with inconsistent periods from the two methods. 	\label{fig:real:comparePM1}}
\end{minipage}\hfill
\begin{minipage}[t]{0.52\linewidth}
	\includegraphics[width =\textwidth]{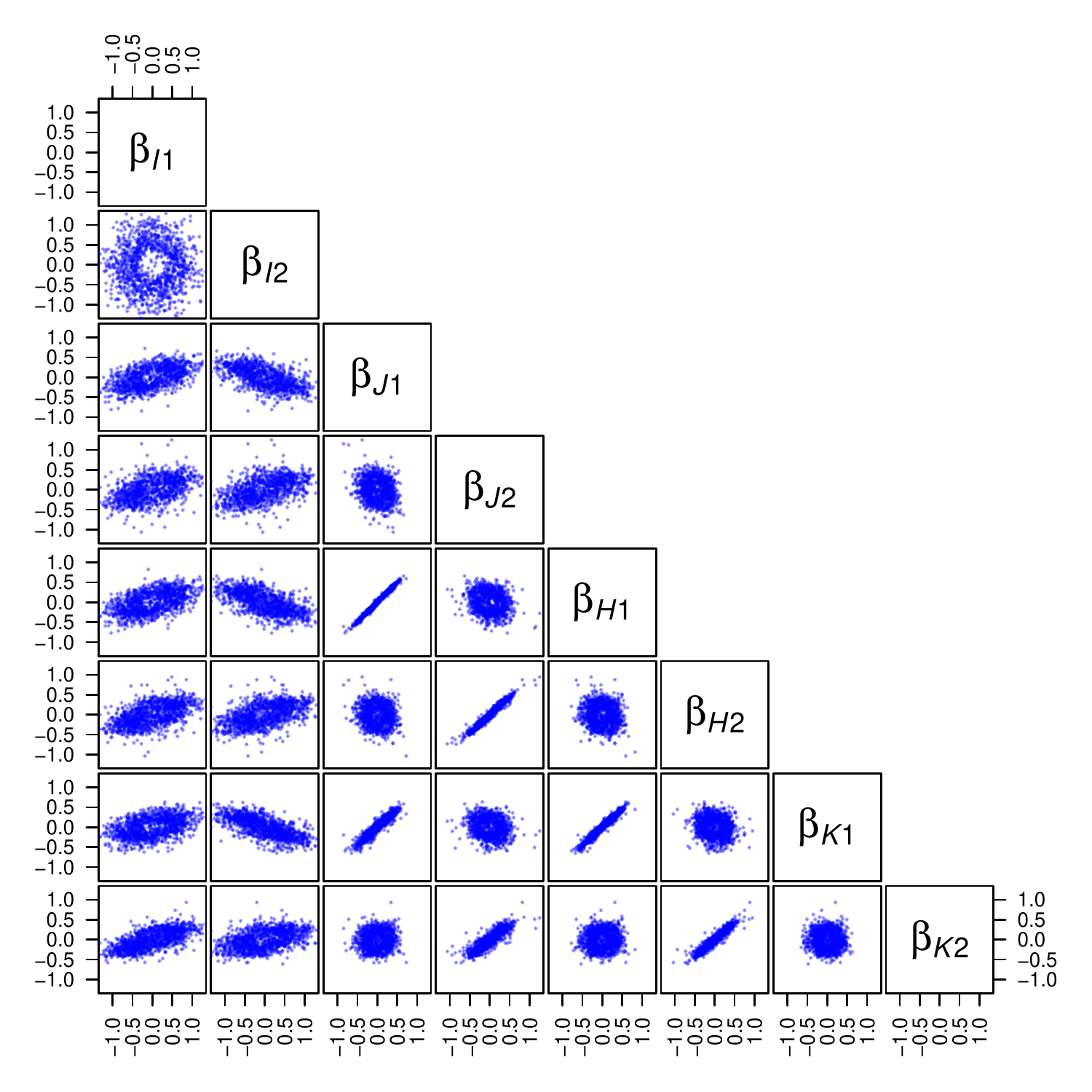}
	\caption{Scatterplot for the sinusoidal coefficients for M33 Miras. }
	\label{fig:real:betacorr}
\end{minipage}
\end{figure}
We carry out an analysis with our proposed method using the 1,265 candidate Miras classified as O-rich by \cite{yuan2018near}. Our method fits the light curves reasonably well, as can be visually assessed by plotting the MAP estimator; see Section~\ref{sec:visual} of the Supplementary Material for some randomly selected examples. We obtain the frequency and average magnitude for each Mira. The frequency of the $i$-th Mira is directly  estimated  by $\hat{f}_i = \arg\max_{f_i} q(f_i)$. The  uncertainty for its estimated period $\hat{p}_i = 1/\hat{f}_i$ is computed by $\sigma^2(\hat{p}_i) =\int (1 / f_i)^2 q(f_i)\mathrm{d} f_i -\big[\int (1 / f_i) q(f_i)\mathrm{d} f_i\big]^2$. The computation of $\hat{f}_i$ is straightforward and more reliable, as our model has built-in PLR  to suppress aliasing frequencies.  In contrast, the work of \citet{yuan2018near} involves an ad-hoc post-estimation correction.   For each Mira, they test both the primary and secondary peaks in their log-likelihood. Their final estimated frequency is selected as the one that best fits the overall PLR.

\begin{table}
	\centering
			\caption{$J$-, $H$- \& $K_s$-band PLR coefficients for candidate M33 O-rich Miras}
\resizebox{1.05\columnwidth}{!}{%
	{\scriptsize
\label{tab:realdata1}
		\begin{tabular}{cc|ccc|cccc}
			\hline\hline
			galaxy & band &  \multicolumn{3}{c|}{linear (period$<400$d)}  & \multicolumn{4}{c}{quadratic}  \\
			& & $a_0$ & $a_1$ & $\sigma$ &  $a_0$ & $a_1$ & $a_2$  & $\sigma$ \\
			\hline
			\hline
			LMC (Yuan+17)&$  J$ & $12.67\pm 0.01$ & $-3.48\pm 0.09$& $0.15$ &
			$12.70\pm 0.01$ & $-3.49\pm 0.09$ & $-1.54\pm 0.23$ & $0.15$ \\
			M33 (Yuan+18) &$  J$&$18.92\pm 0.01$&$\dots$ & $0.25$  & 
			$18.94\pm 0.01$&$\dots$&$\dots$ & $0.25$ \\
			M33 (SVI) &$  J$&$18.97 \pm 0.01$&$-3.32\pm 0.06$ & $0.22$  & 
			$19.01\pm 0.01$ & $-3.36\pm0.05$ &  $-1.52\pm0.02$ & $0.23$ \\
			\hline
			LMC (Yuan+17)&$  H$&$11.91\pm 0.01$&$-3.64\pm 0.09$ & $0.16$  & 
			$11.96\pm 0.01$&$-3.59\pm 0.10$&$-3.40\pm 0.31$& $0.16$ \\
			M33 (Yuan+18) &$  H$&$18.17\pm 0.01$&$\dots$& $0.24$  & 
			$18.27\pm 0.01$&$\dots$&$\dots$& $0.26$ \\
			M33 (SVI) &$  H$&$18.24\pm 0.01$&$-3.59\pm 0.06$& $0.21$  & 
			$18.27\pm0.01$ & $-3.11\pm0.04$ & $-3.24\pm0.02$& $0.22$ \\
			\hline
			LMC (Yuan+17)&$K_s$&$11.56\pm 0.01$&$-3.77\pm 0.07$ & $0.12$  & 
			$11.59\pm 0.01$&$-3.77\pm 0.08$&$-2.23\pm 0.20$& $0.12$ \\
			M33 (Yuan+18) &$K_s$&$17.78\pm 0.01$&$\dots$ & $0.21$ & 
			$17.83\pm 0.01$&$\dots$&$\dots$& $0.22$ \\
			M33 (SVI) &$K_s$&$17.85\pm 0.01$&$-3.93\pm 0.06$ & $0.20$ & 
			$17.87\pm0.01$ & $-3.68\pm0.04$ & $-2.20\pm0.02$ & $0.20$ \\
			\hline
		\end{tabular}
	}}
\end{table}

A  comparison of  the periods estimated by our  method and the  MSP method is shown in Figure~\ref{fig:real:comparePM1}.   Most of the estimated periods from the two methods are consistent,  with only $5.8\%$ Miras (red points) deviating by more than 10\%.    We also compare the average $JHK_s$ magnitude between  our determinations and that of \citet{yuan2018near} in Figure~\ref{fig:real:comparePM2},~\ref{fig:real:comparePM3} and~\ref{fig:real:comparePM4}.  Our determination has slighter higher average magnitude values. Nevertheless, the two results are generally consistent, despite there are some outliers. The standard deviations of the difference are 0.16, 0.13, and 0.10 mag for the $J$, $H$, and $K_s$ bands, respectively.

\begin{table}
	\centering
		\caption{Derived Distance Moduli for M33\label{tab:realdata2}} 
				\resizebox{1.05\columnwidth}{!}{%
			{\scriptsize
		\begin{tabular}{cccccccc}
			\hline\hline
			band	& $\Delta a_0$ & $\Delta \bar{m} $ &  $\Delta A_\lambda$ & 
			$\Delta \mathrm{ct}$ & $ \Delta \mu$   & $\mu_{\mathrm{LMC} }$  & 
			$\mu_{\mathrm{M33}}$\\ 
			\hline
			$J$ & $6.311\pm 0.014$ & $-0.036\pm 0.035$  & $0.029\pm 0.008$ & $0.016 \pm 0.036$ & $6.320\pm 0.053$ & $18.493 \pm 0.048$& $24.813\pm 0.071$ \\
			$H$ & $6.312\pm 0.014$ & $-0.033\pm 0.029$  & $0.018\pm 0.005$ & $0.010 \pm 0.040$ & $6.307\pm 0.052$ & $18.493 \pm 0.048$& $24.800\pm 0.070$ \\
			$K_s$ & $6.288\pm 0.014$ & $-0.026\pm 0.024$  & $0.012\pm 0.003$ & $-0.007 \pm 0.032$ & $6.267\pm 0.042$ & $18.493 \pm 0.048$ & $24.760\pm 0.064$ \\
			\hline
		\end{tabular}
	}}
\end{table}

In addition, we compute the posterior mean of the sinusoidal coefficient $\vbeta_i$ for each Mira.  The scatter plot of the expected $\vbeta_i$'s (the expectation is taken with respect to $q(\vbeta_i)$)  is shown in Figure~\ref{fig:real:betacorr}. Notice that the coefficients for the $JHK_s$ bands have a tight correlation, and their relation with the $I$-band coefficients $\beta_{I1}, \beta_{I2}$ is also non-negligible. The correlation structure in the real data justifies our model  Eqn.~\eqref{eqn:hm:layer3} on $\vbeta_i$. By exploiting this  structure, our method is able to constrain the model degree of freedom and  estimate  parameters more accurately.

We note the doughnut shape of the posterior joint distribution of $\beta_{I1}, \beta_{I2}$ shown in Figure~\ref{fig:real:betacorr} does not suggest model misspecification. In fact, it is the Bayesian method in action. The posterior distribution needs not be consistent with the prior. In Bayesian statistics, it is expected that when there is enough data, the information provided in the data will ``wash out the prior."

Next, we  will compare our estimated PLRs with the results from \cite{yuan2017large} and \cite{yuan2018near}.  For each band, \cite{yuan2018near} fit  a quadratic PLR 
\begin{equation} \label{eqn:equiPLR}
M = a_0 + a_1 (\log_{10}(P) - 2.3) + a_2 (\log_{10}(P) - 2.3)^2
\end{equation}
to \textit{all} Miras; and a linear PLR  $M = a_0 + a_1 (\log_{10}(P) - 2.3)$ to  \textit{Miras with period less than 400 days}. Notice the above quadratic PLR is simply an equivalent reparameterization of our PLR in~\eqref{eqn:modelPLR}. In the work of \cite{yuan2018near},  the PLRs are firstly fitted to the LMC Miras \citep{yuan2017large}. After that, they fix the coefficients $a_1$ and $a_2$, and only fit the intercept $a_0$ to the M33 Miras. Table~\ref{tab:realdata1} presents their  results. The  coefficients for the LMC Miras are  in the rows of \texttt{LMC (Yuan+17)}, and  their results on the M33 Miras are  in the rows named \texttt{M33 (Yuan+18)}. 

\begin{figure}
	\centering
	\begin{subfigure}[b]{0.32\textwidth}
		\includegraphics[width=\textwidth]{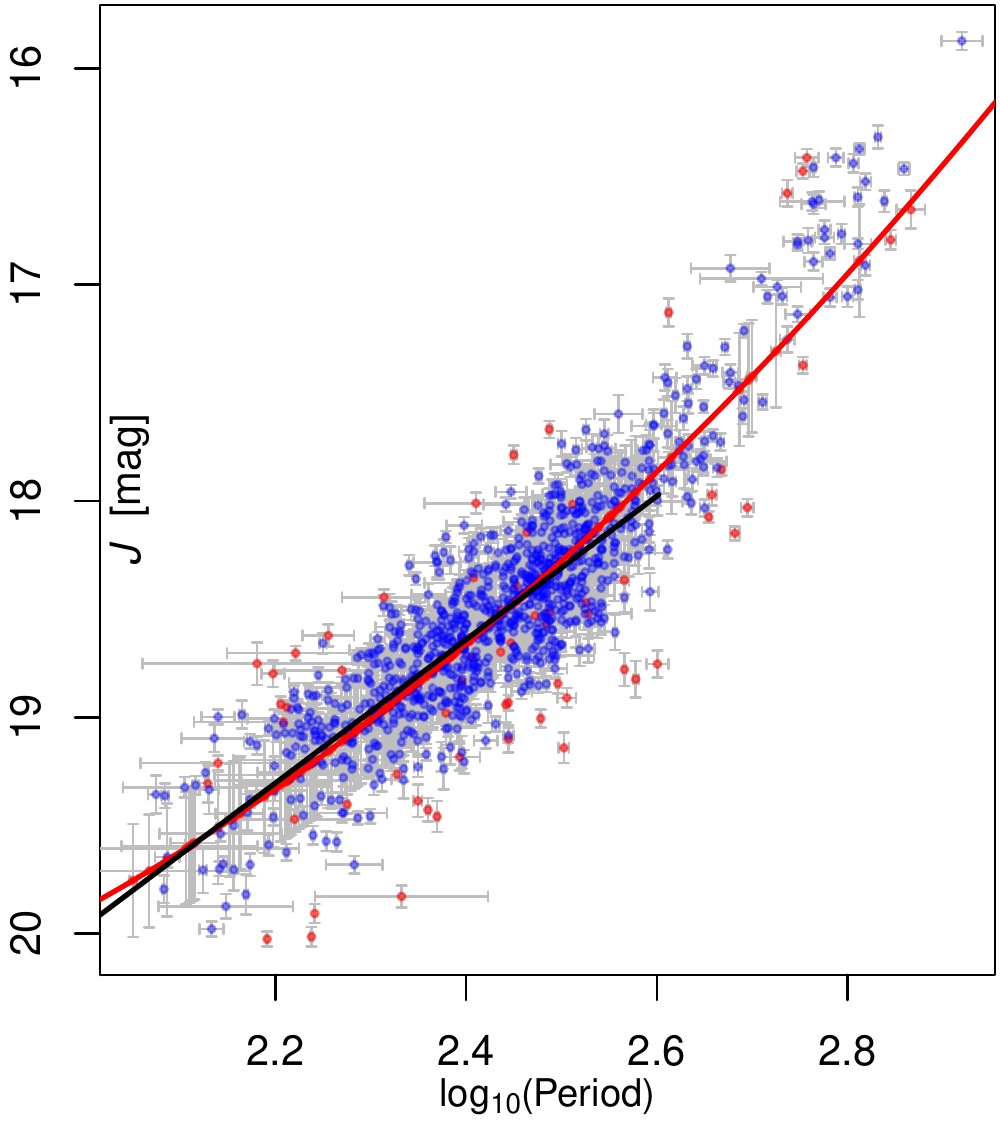}
	\end{subfigure}
	\begin{subfigure}[b]{0.32\textwidth}
		\includegraphics[width=\textwidth]{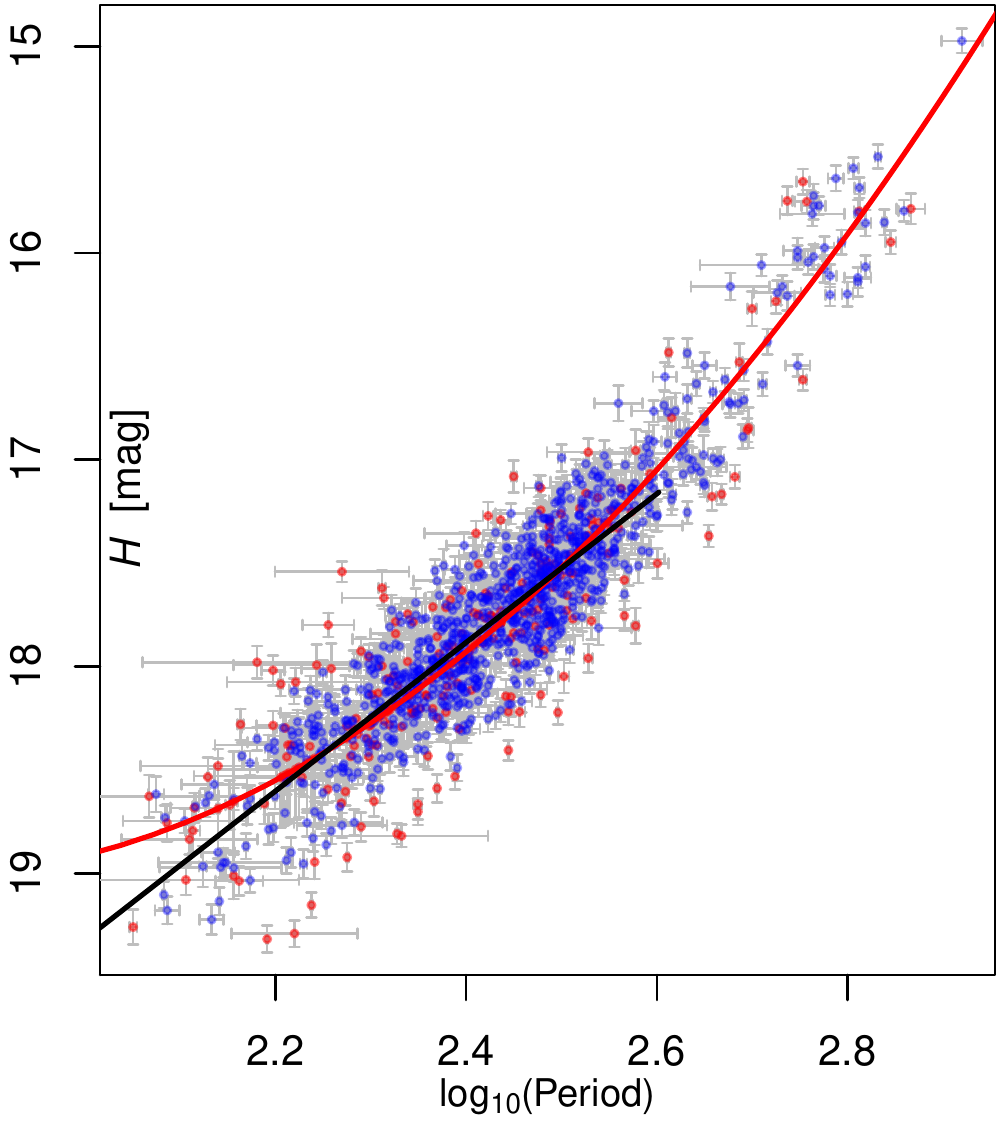}
	\end{subfigure}
	\begin{subfigure}[b]{0.32\textwidth}
		\includegraphics[width=\textwidth]{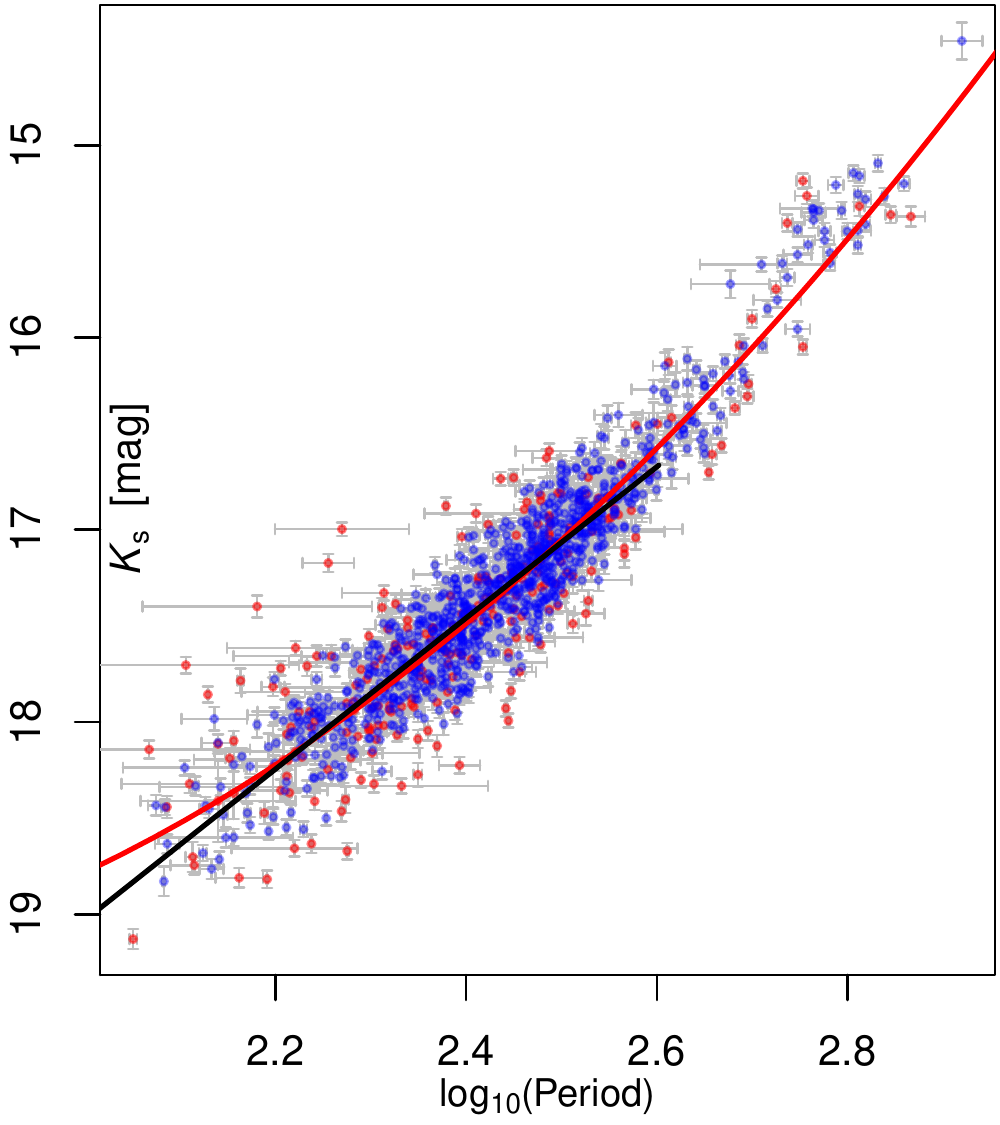}
	\end{subfigure}
	\caption{The PLRs of M33 Miras in the $J$ (left), $H$ (center) and $K_s$ (right) bands. The vertical axes shown the average magnitude in the respective band while the horizontal axes are the logarithm of the period. Red curves are the fitted PLRs from our SVI method. Black lines are fitted linear PLRs for Miras with period less than 400 days. }\label{fig:M33PLR}
\end{figure}

The quadratic PLRs from our models are plotted as red curves in Figure~\ref{fig:M33PLR}. In the figure, each point represents one Mira with our estimated average magnitude and period. For Miras with period less than 400 days, we also fit additional linear PLRs, which are shown as black lines. Our coefficients  are  presented in the rows named \texttt{M33 (SVI)} in Table~\ref{tab:realdata1}. For comparison, the reported coefficient values have been converted to respect the equivalent parameterization~\eqref{eqn:equiPLR}. Notice our estimated  $a_1$ and $a_2$ are consistent with the LMC results from \citet{yuan2017large} within the error budget. 

In Table~\ref{tab:realdata1}, the column $\sigma$ is the standard deviation of the  PLR residuals.  Smaller $\sigma$ means tighter PLRs. It is remarkable that our model achieves competitive performance as the M33 result of \cite{yuan2018near}, without their ad-hoc period correction tricks. This  verifies the quality the automatically estimated PLRs embedded in our model.

Lastly, we compute the distance modulus of the M33 galaxy. Denote $a_0^{\mathrm{LMC}}$ as the zero-order coefficient of the LMC quadratic  PLR from  \cite{yuan2017large}. Denote  $a_0^{\mathrm{M33}}$ as the zero-order coefficient of our M33 quadratic PLR . Then, up to a few corrections,  $\Delta a_0 = a_0^{\mathrm{M33}} - a_0^{\mathrm{LMC}}$  is equal to the difference in distance moduli of the two galaxies. These corrections include: conversion to flux-averaged magnitude $\Delta \bar{m}$, correction for the different interstellar extinction $\Delta A_\lambda$ towards LMC and M33; and the color term bias $\Delta \mathrm{ct}$ of the photometric calibration. We note the flux correction $\Delta \bar{m}$ is necessary because the average magnitude was computed in units of flux for the LMC Miras. This correction is carried out as follows: We fit the signal light curve $s_{ib}(t)$ in~\eqref{eqn:model:decompositionNew} for each Mira over the whole observational time domain. The signal $s_{ib}(t)$ is in the unit of magnitude and can be converted to flux via $s'_{ib}(t) = 10^{-0.4\cdot s_{ib}(t)}$. The average value $l_{ib}$ of $s'_{ib}(t)$ is computed, and it is converted back to magnitude by $m_{ib}' = -2.5\log_{10}(l_{ib})$. Note some irrelevant constants have been dropped during these conversions. Then, the correction value $\Delta \bar{m}$ for each band is the average  difference $m_{ib}'  - m_{ib}$ from all Miras.

After making these  corrections,  we obtain the \textit{relative} distance modulus $\Delta \mu = \Delta a_0 +\Delta \bar{m} +\Delta A_\lambda + \Delta \mathrm{ct}$ between the LMC and M33. We adopt a distance modulus $\mu_{\mathrm{LMC}} = 18.493\pm 0.048$ for LMC \citep{pietrzynski2013}. Thus, the distance modulus for M33 is $\mu_{\mathrm{M33}} = \mu_{\mathrm{LMC}}  +\Delta \mu$.
This procedure is applied separately to the $J$-, $H$- and $K_s$-band data and summarized in Table~\ref{tab:realdata2}. From the respective PLRs, our analysis yields distance moduli of $24.813\pm 0.071$~mag, $24.800\pm 0.070$~mag, and $24.760\pm 0.064$~mag, respectively. These are also consistent with earlier estimates.

\section{Effect of downsampling}~\label{sec:downsampling}
In Section~\ref{sec:realdata}, we showed that our proposed SVI method produces distance estimates and uncertainties which are consistent with other methods without resorting to ad hoc period estimate corrections. This section studies how robust these estimates are to downsample the light curves, e.g., how do distance estimates change as the number of photometric measurements and/or light curve goes down. 

We conducted an experiment of downsampling the actual M33 data set used in Section~\ref{sec:realdata}. We compared our SVI method with two other multi-band methods, MGLS and MSP, without using any ad hoc period estimate corrections.  To assess the reliability and quality of the estimated PLR for distance estimation in downsampling, we measured the changes in the PLR intercept and in the PLR scattering between the downsampled estimates and the full sample estimates.

In each replication of the experiment, $\tilde{N} = 50$ or  $\tilde{N} = 100$ Miras are randomly drawn from the original $N=1,265$  candidate  O-rich Miras. For each selected Mira,   we also put a restriction $\bar{n}_{b}$ over the number of photometric measurements for its $b$-th band  light curve ($b\in\{I,J,H,K_s\}$). In particular, for the $b$-th band light curve of the $i$-th selected Mira, we sample $\tilde{n}_{ib} = \min\{n_{ib}, \bar{n}_b\}$ photometric measurements  without replacement from the original $n_{ib}$ measurements. Six combinations of $\tilde{N}$ and $\bar{n}_b$ are considered as in Table~\ref{table:downsample}. 

\begin{table}[h]
	\centering
	\caption{Experiment settings for downsampling the actual M33 observations}
	\label{table:downsample}
	\renewcommand{\arraystretch}{1.25}
	\vskip.5em
    \begin{tabular}{c|c|c|c|c|c} 
    	\hline \hline
    	Setting & $\tilde{N}$ & $\bar{n}_{I}$  & $\bar{n}_{J}$ & $\bar{n}_{H}$ & $\bar{n}_{K_s}$\\
    	\hline
    	S1 & 50 & 10  & 5 & 5 & 5 \\
    	S2 & 50 & 20  & 10 & 10 & 10 \\
    	S3 & 50 & 30  & 15 & 15 & 15 \\
    	S4 & 100 & 10  & 5 & 5 & 5 \\
    	S5 & 100 & 20  & 10 & 10 & 10 \\
    	S6 & 100 & 30  & 15 & 15 & 15 \\
    	\hline
\end{tabular}
  \renewcommand{\arraystretch}{1}
\end{table}

\begin{figure}[hp]
	\centering
	\includegraphics[width =0.83 \textwidth]{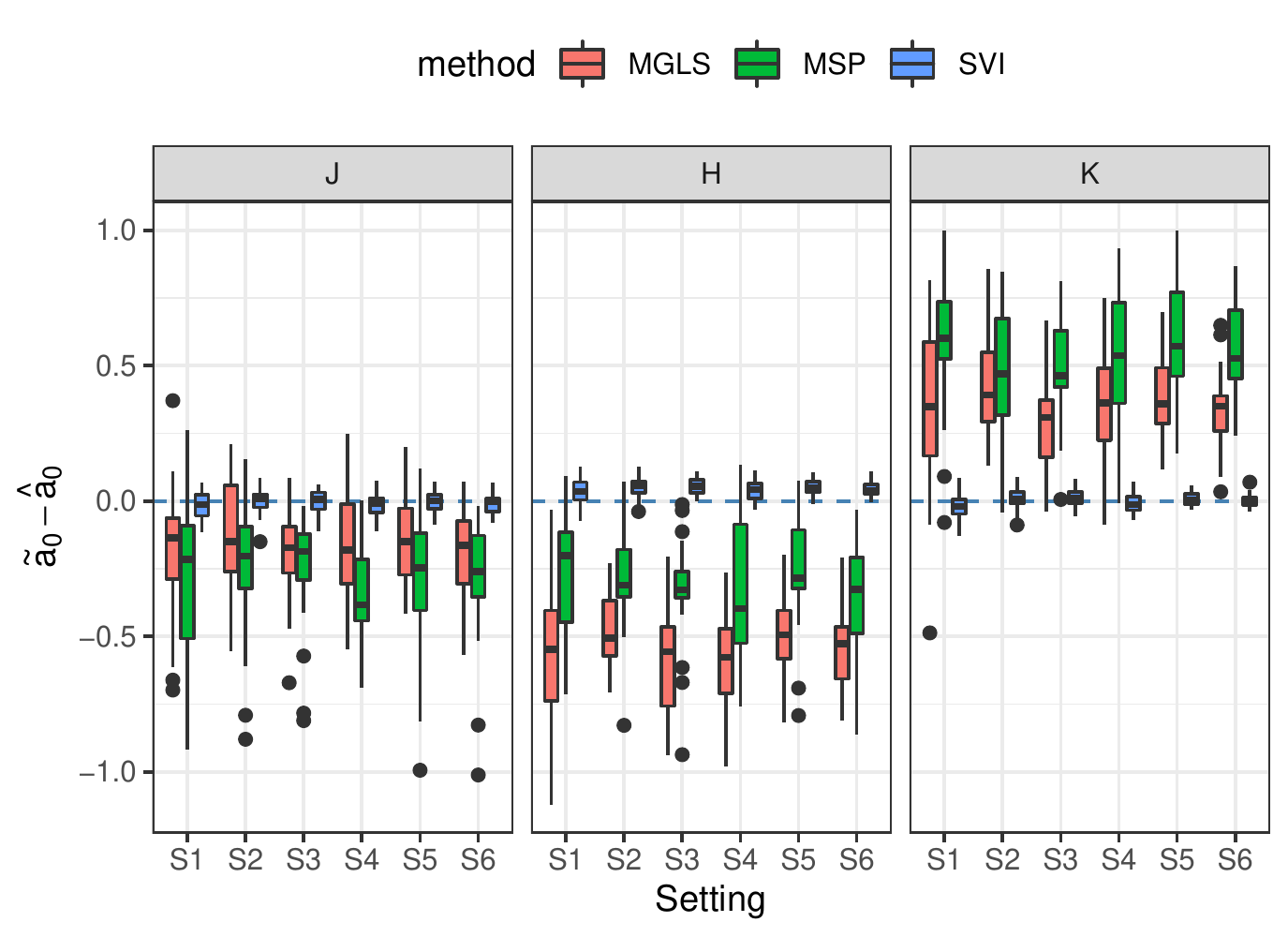}
	\includegraphics[width =0.83\textwidth]{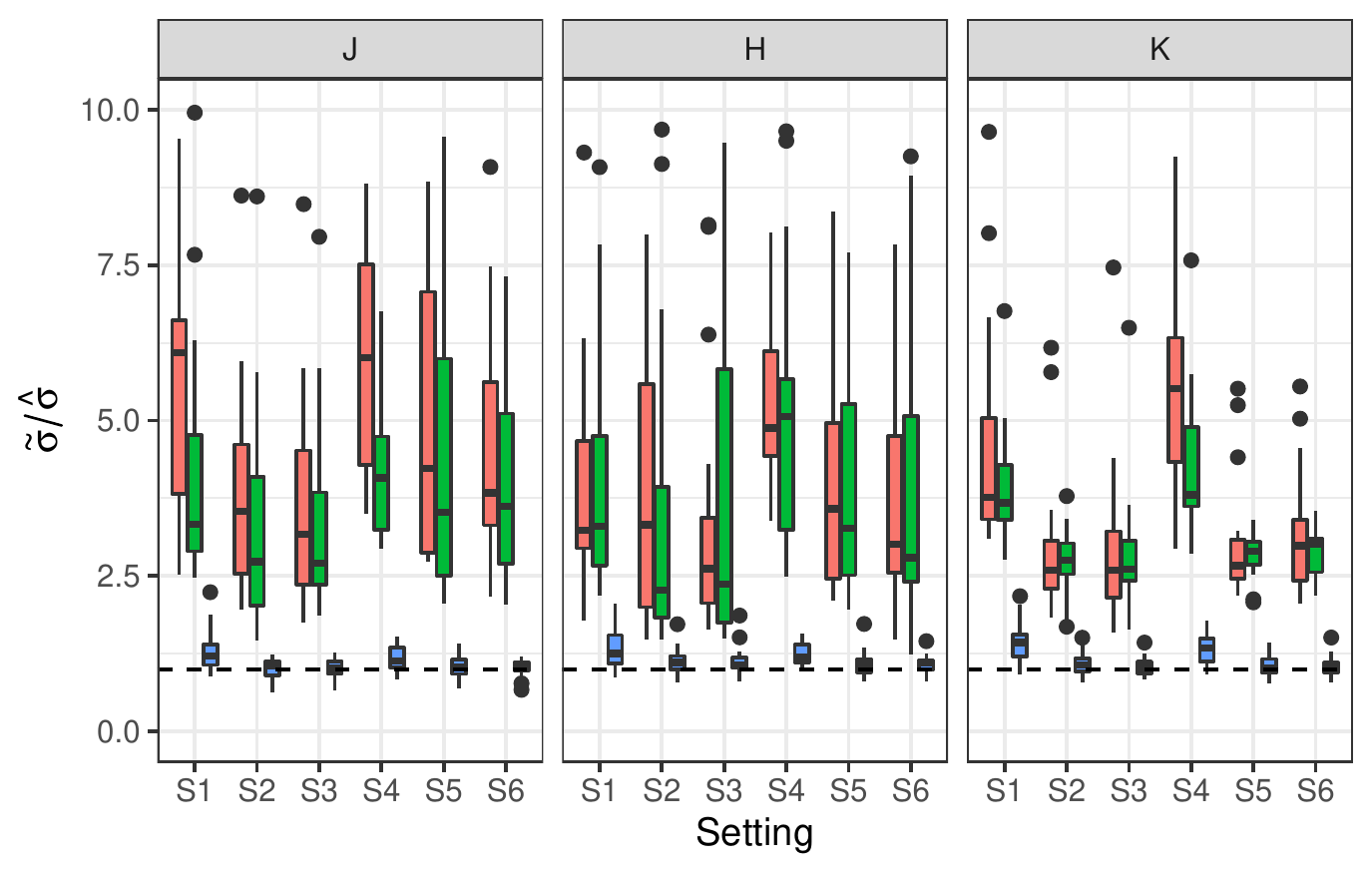}
	
	\caption{Method comparison via down-sampling the real M33 observation. The first row shows the difference between the downsampled PLR intercept and the full sample PLR intercept. The second row shows the standard deviation ratio of the downsampled PLR residuals and the full sample PLR residuals. The three columns correspond to the $J$ ,$H$ and $K_s$ bands, respectively. The result of 20 replications for MGLS, MSP and SVI are shown in the red, green and blue boxplots, respectively.} \label{fig:subsamplediff}
\end{figure}

Each setting in Table~\ref{table:downsample} is repeated 20 times.
In each replication of one setting, three multi-band methods (MGLS, MSP and  SVI) are applied  to estimate the period and the average magnitude for each down-sampled Mira.  
To fit a quadratic PLR for MGLS and MSP, we fix the coefficient $a_1$ and $a_2$ (estimated from LMC data) and only estimate the intercept term $a_0$. Suppose the intercept estimated from the subsample is denoted as $\tilde{a}_0$ and the full sample estimate is denoted $\hat{a}_0$. The first row of Figure~\ref{fig:subsamplediff} shows the boxplot of the difference $\tilde{a}_0 - \hat{a}_0$. The three panels in the first row correspond to the $J$, $H$ and $K_s$ bands, respectively; and the results for MGLS, MSP and SVI are colored as red, green, and blue, respectively. The intercept difference $\tilde{a}_0 - \hat{a}_0$ is closest to $0$ for our SVI method, and the variability is also small. 
As comparison, MGLS and MSP have much larger bias and variability than SVI. Compared with MGLS, the MSP method has  smaller bias in the $H$ band but larger bias in the $J$ and $K_s$ bands.

We also compare the scattering of the PLR residuals. Let $\tilde{\sigma}$ be the standard deviation of the PLR residuals for each method in each experiment replication. Also, let $\hat{\sigma}$  be the standard deviation of the PLR residuals obtained by our SVI method for the full sample. The second row of Figure~\ref{fig:subsamplediff} shows the boxplot of the ratio   $\tilde{\sigma}/\hat{\sigma}$ for the 20 replications of the three methods. The residual scattering ratio is only slightly above $1$ for our SVI method over the six scenarios of downsampling. As comparison, MGLS and MSP have substantially larger residual scattering ratio. The residual scattering of MSP is smaller than that of MGLS.

These results suggest that our SVI is much more robust to downsample Mira light curves than MGLS and MSP methods. It would be an interesting research topic to study the implication of these results for telescope cadence selection and the necessary density of time domain sampling. 

\section*{Acknowledgements}
He's Project 11801561 was supported by NSFC. L. Macri acknowledges support from the Mitchell Institute for Fundamental Physics and Astronomy at Texas A\&M University. Huang's work was partially supported by NSF grants DMS-1208952, IIS-1900990, and CCF-1956219. The authors would like to thank two anonymous reviewers for helpful comments which led to inclusion of Section~\ref{sec:downsampling} and other significant improvements of the paper. 

\begin{supplement}
		\textbf{Supplementary proof and results}. The  supplementary material contains a review of the notations of the exponential family used in the paper. It also contains the proof of Theorem~\ref{thm:localupdate} and additional numerical results.

	\textbf{Supplementary code and dataset}. The  code and datasets are released online at \url{https://github.com/shiyuanhe/sviperiodplr}. They serve to reproduce the simulation and real data analysis results in this work.
\end{supplement}

\clearpage

\clearpage

\begin{center}
\Large  \bf		Supplementary Materials to ``Simultaneous inference of periods and period-luminosity relations for Mira variable stars'' \par
\end{center}

\setcounter{section}{0}
\setcounter{figure}{0}
\setcounter{equation}{0}

\renewcommand{\theequation}{S.\arabic{equation}}
\renewcommand{\thefigure}{S.\arabic{figure}}
\renewcommand{\thesection}{S.\arabic{section}}

\vspace{35pt}

\section{Canonical Exponential Family}\label{sec:canonicalReview}
This section reviews the notation for the exponential family used in the paper.
The general form for the canonical exponential family is
$$
f(\vx|\eta) = h(\vx) \exp\{\eta^T t(\vx) - A(\eta)\}.
$$
where $\eta$ is the canonical parameter, $r(\vx)$ is the sufficient statistics,
and $A(\eta)$ is the log-partition function.
It is known that $\Expect (t(\vx)) = \nabla_{\eta} A(\eta)$.

For a $p$ dimensional multivariate Gaussian distribution with mean $\vmu$ and precision matrix $\mOmega$, its density is
\begin{align*}
	&\exp\Big\{-\frac{1}{2} (\vx - \vmu)^T\mOmega(\vx - \vmu) +\frac{1}{2}\log\det\mOmega+
	\frac{1}{2}\log 2\pi \Big\}	\\
	=& \exp\Big\{\langle-\frac{1}{2}\mOmega, \vx \vx^T  \rangle + \langle \mOmega\vmu, \vx\rangle  - \frac{1}{2}\vmu^T\mOmega\vmu +\frac{1}{2}\log\det\mOmega + \frac{1}{2}\log 2\pi \Big\}
\end{align*}
with $t(\vx) = (\vx\vx^T, \vx)$, $\eta = (\eta^1, \eta^2) =  (-\frac{1}{2}\mOmega, \mOmega\vmu)$ and 
$A(\eta) = \frac{1}{2}\vmu^T\mOmega\vmu -\frac{1}{2}\log\det\mOmega$. 
In addition, it holds that
\begin{align*}
	\Expect (\vx) = -\frac{1}{2} (\eta^1)^{-1}\eta^2,\; 	\Var (\vx) =  -\frac{1}{2} (\eta^1)^{-1}, \;
	\Expect (\vx\vx^T) = -\frac{1}{2} (\eta^1)^{-1} + \frac{1}{4} (\eta^1)^{-1}  \eta^2 (\eta^2)^T (\eta^1)^{-1} .  
\end{align*}
The expression for its entropy is
$$
-\int \log f(\vx|\eta) f(\vx|\eta)  \mathrm{d}\vx = \frac{1}{2} \log \det \left(2\pi e \mOmega^{-1}\right)
= -\frac{1}{2}\log \det \left(- \eta^{1}/(\pi e)\right)\,. 
$$

For gamma distribution with shape $\alpha$ and rate $\beta$, its density is
$$
\exp\Big\{-\beta x + (\alpha - 1) \ln x + \alpha \ln \beta - \ln \Gamma(\alpha) 
\Big\}
$$
with $t(x) = (x, \ln x)$ and canonical parameter 
$\eta = (\eta^1, \eta^2) = (-\beta, \alpha - 1)$. We also have 
$\Expect(x) = \alpha/\beta = -(\eta^2 + 1) / \eta^1$.

For $p$ dimensional Wishart distribution with $n (>p-1)$ degrees of freedom and 
scale matrix $\mV \succ 0$, its density is
$$
\exp\Big\{
-\frac{1}{2}\tr\left(\mV^{-1} \mX\right) +
\frac{n-p-1}{2} \log\det \mX -\frac{np}{2} \log 2 - 
\frac{n}{2} \log\det \mV - \log\Gamma_p(n/2)
\Big\}
$$
with canonical parameter $\eta = (\eta^1, \eta^2) = (-\frac{1}{2}\mV^{-1}, (n-p-1)/2)$. The mean is $\Expect \mX = n\mV = -0.5 \cdot(2\eta^2+p+1) \cdot (\eta^1)^{-1}$.

\section{Proof of Theorem~\ref{thm:localupdate}}\label{sec:proof:theorem1}
With fixed $q(\valpha), q(\mOmega), q(\vgamma)$,
updating $q(\vtheta_{i}, f_i ) = q(\vtheta_{i}| f_i ) q(f_i)$ to maximize ELBO 
in~\eqref{eqn:mirai:elbo2} can be  equivalently expressed by maximizing
\begin{align}
	&\int \log \frac{p(f_i, \vtheta_{i} | \valpha, \mOmega, \vgamma, \sD_i)}{q(f_i, \vtheta_{i})}
	\times q(f_i, \vtheta_{i}) q(\valpha) q(\mOmega) q(\vgamma)\ \intd\vtheta_{i} \intd f_i
	\intd \valpha \intd \mOmega \intd \vgamma \nonumber\\
	=& \int h(f_i) q(f_i) \intd f_i - \int \log q(f_i) q(f_i) \intd f_i\,,\label{proof:local1:qfKL}
\end{align}
where
\begin{align}
	h(f_i) =& \int \log \frac{p(f_i, \vtheta_{i} | \valpha, \mOmega, \vgamma, \sD_i)}{q(\vtheta_{i} | f_i)}
	\times q(\vtheta_{i} | f_i) q(\valpha) q(\mOmega) q(\vgamma)\ \intd\vtheta_{i} 
	\intd \valpha \intd \mOmega \intd \vgamma  \nonumber \\
	=&\Expect_{q(\vtheta_{i} | f_i)} \Expect_{q(\valpha)} \Expect_{q(\mOmega)} \Expect_{q(\vgamma)}\left[
	\log \frac{p(\vtheta_{i} |f_i,  \valpha, \mOmega, \vgamma,  \sD_i)}{q(\vtheta_{i} | f_i)} 
	+ \log p(f_i | \valpha, \mOmega, \vgamma)
	\right]\,. \label{proof:local1:qfKL:theta}
\end{align}
To find the optimal $q(\vtheta_{i}, f_i)$, we firstly find the optimal $q(\vtheta_{i} | f_i)$ at each fixed $f_i$. Then from equation~(\ref{proof:local1:qfKL}), $q(f_i)$ is readily available as $q(f_i) \propto \exp(h(f_i))$.

Notice
\begin{align*}
	\log p( f_i, \vtheta_{i}| \valpha, \mOmega, \vgamma, \sD_i) &\propto
	-\frac{1}{2}\sum_{b=1}^{B} (\vy_{ib} - \mC_{ib}\vtheta_{ib})^T \mSigma_{ib}^{-1}  (\vy_{ib} - \mC_{ib}\vtheta_{ib})  - \frac{1}{2}\log\det \mSigma_{ib}\\
	&\qquad \qquad -\frac{1}{2} (\vtheta_{i} - \vtheta_0)^T\mTheta(\vtheta_{i} - \vtheta_0) ,
\end{align*}
where the irrelevant constants have been dropped. Notice in the above, the basis matrix $\mC_{ib}$, the covariance matrix $\mSigma_{ib}$ as well as the prior mean vector $\vtheta_{0}$ all depend on the frequency $f_i$. The frequency has uniform prior. 
The  full conditional for $\vtheta_{i}$ is
\begin{align*}
	p( \vtheta_i|   f_{i},  \valpha,  \mOmega, \vgamma, \sD_i) \propto\exp
	& \Big\{ \sum_{b=1}^{B}
	\left\langle -\frac{1}{2} \mC_{ib}^T\mSigma_{ib}^{-1}\mC_{ib}, \vtheta_{ib}\vtheta_{ib}^T  \right\rangle
	+ \sum_{b=1}^{B}\left\langle \mC_{ib}^T\mSigma_{ib}^{-1}\vy_{ib},
	\vtheta_{ib}\right\rangle \Big\} \\
	&\times \exp\Big\{-\frac{1}{2} (\vtheta_{i} - \vtheta_0)^T\mTheta(\vtheta_{i} - \vtheta_0)\Big\} \\
	\propto\exp
	& \Big\{\left\langle \xi_i^{\theta 1}, \vtheta_{i}\vtheta_{i}^T  \right\rangle
	+ \left\langle  \xi_i^{\theta 2},	\vtheta_{i}\right\rangle \Big\}.
\end{align*}
This full conditional is Gaussian with canonical parameters
\begin{align*}
	\xi^{\theta 1}_i &= -\frac{1}{2}\times
	\diag\left( \mC_{i1}^T\mSigma_{i1}^{-1}\mC_{i1} , \cdots,
	\mC_{iB}^T\mSigma_{iB}^{-1}\mC_{iB}\right) 
	-\frac{1}{2}\mTheta\,,  \\
	\xi^{\theta 2}_i& = \left(
	(\mC_{i1}^T\mSigma_{i1}^{-1}\vy_{i1})^T, 
	\cdots, 	(\mC_{iB}^T\mSigma_{iB}^{-1}\vy_{iB})^T \right)^T + \mTheta\vtheta_0\,.
\end{align*}

For fixed $f_i$,  $q(\vtheta_{i} | f_i)$ is imposed as a Gaussian distribution with canonical 
parameter $(\eta_i^{\theta 1}, \eta_i^{\theta 2})$. 
Holding $q(\valpha), q(\vgamma), q(\mOmega)$ and $f_i$ fixed, the optimal $q(\vtheta_i | f_i)$ maximizing \eqref{proof:local1:qfKL:theta} is found by setting its canonical parameter as
$\eta^{\theta 1}_{i} = \Expect_{q(\valpha)}\Expect_{q(\vgamma)}\Expect_{q(\mOmega)}(\xi^{\theta1}_{i})$
and $\eta^{\theta 2}_{i} = \Expect_{q(\valpha)}\Expect_{q(\vgamma)}\Expect_{q(\mOmega)}(\xi^{\theta2}_{i})$. 
This is the updating equation~(\ref{update:local1:theta}).

In order to compute $q(f_i)\propto \exp(h(f_i))$,  we now need to evaluate $h(f_i)$, which is
$$
h(f_i) =\ \Expect_{q(\vtheta_{i} | f)}\Expect_{q(\valpha)} 
\Expect_{q(\vgamma)}
\log p( f_i, \vtheta_{i}| \vy_i, \valpha, \mOmega, \vgamma) 
- \Expect_{q(\vtheta_{i} | f)} q(\vtheta_{i} | f)\,.
$$
Then, with irrelevant constants dropped, we have
\begin{align}
	h(f_i) 
	\propto &\ \langle \eta_i^{\theta 1},\, \Expect_{q(\vtheta_{i} | f_i) } (\vtheta_{i}\vtheta_{i}^T)\rangle + 
	\langle \eta_i^{\theta 2},\, \Expect_{q(\vtheta_{i} | f_i) } (\vtheta_{i})\rangle - 
	\frac{1}{2} \langle\Expect_{q(\mOmega)}\Expect_{q(\vgamma)}\mTheta ,\, \Expect_{q(\valpha)} \vtheta_{0} \vtheta_{0}^T\rangle\nonumber\\
	&\qquad \qquad- \frac{1}{2} \sum_{b=1}^B 
	\left[ \vy_{ib}^T\mSigma_{ib}^{-1}\vy_{ib} + \log\det \mSigma_{ib} \right] 
	-\frac{1}{2} \ln \det (-\eta_i^{\theta1}/(\pi e)) . \label{proof:local1:g1}
\end{align}
The last term is the entropy of the Gaussian distribution $q(\vtheta_{i} | f_i)$. In addition, the first and second moments of $\vtheta_{i}$ can be evaluated,
\begin{align}
	&\langle \eta_i^{\theta 1}, \Expect_{q(\vtheta_{i} | f_i) } (\vtheta_{i}\vtheta_{i}^T)\rangle + 
	\langle \eta_i^{\theta 2},\Expect_{q(\vtheta_{i} | f_i) } (\vtheta_{i})\rangle \nonumber\\
	=&\langle \eta_i^{\theta 1}, -\frac{1}{2} (\eta_i^{\theta 1})^{-1} 
	+ \frac{1}{4}  (\eta_i^{\theta 1})^{-1} \eta_{i}^{\theta 2} (\eta_{i}^{\theta 2})^T(\eta_i^{\theta 1})^{-1} \rangle + 
	\langle \eta_i^{\theta 2}, -\frac{1}{2}  (\eta_i^{\theta 1})^{-1} \eta_{i}^{\theta 2}\rangle \nonumber\\
	=&-\frac{3B}{2} -\frac{1}{4} (\eta_i^{\theta 2})^T(\eta_i^{\theta 1})^{-1} \eta_{i}^{\theta 2}. \label{proof:local1:g2}
\end{align}
Now, (\ref{proof:local1:g1}) and (\ref{proof:local1:g2})  together imply (\ref{update:local1:freqMain}).

\section{Discussion of parameterization of the non-periodic component of a light curve}~\label{sec:non-periodic}

In our formulation of the non-periodic component of a light curve (Section~\ref{sec:proposedMethod:nonpara}), $h_{ib}(\cdot)$ is a function of phase $u = f_i\cdot t$. This differs from the work of \cite{he2016period} and \cite{yuan2018near}, where the term $h_{ib}(\cdot)$ is chosen as a function of time $t$. This section provides some numerical results supporting the new formulation.

Our choice of the model is guided by the goal of using the same kernel length-scale parameter $\tau_{b2}$
for all Miras so that the computationally expensive process of finding the optimal length-scale parameter individually for each Mira can be avoided.
Equation (10) reads
\begin{equation}
	s_{ib} (t) = m_{ib} + \vb(f_i\cdot  t)^T\vbeta_{ib}+ h_{ib}(f_i\cdot t), \label{eqn:model:decompositionNewSupp}
\end{equation}
where $\vb(u) =  (\cos(2\pi u), \sin(2\pi  u))^T$.
Note that the argument in the periodic component $\vb(f_i\cdot  t)^T\vbeta_{ib}$ is $f_i\cdot t$. Our choice of using $h_{ib}(f_i\cdot t)$ makes the function argument of the non-period component consistent with the period component. This choice is supported by two numerical results given below. 

\textit{First result.} We fit the Gaussian process model to the $I$- band light curves of OGLE MLC O-rich Miras under the two parameterizations ($h_{ib}(t)$ versus $h_{ib}(f_i t)$) of the non-period component. We used the kernel given in Section 2.4 of the paper and compared the optimal fitted length-scale parameters. Because these MLC light curves are densely sampled with high signal-to-noise ratio, we can fit the Gaussian process model directly to each Mira, without using the hierarchical model structure developed in our work (each Mira has its own length-scale parameter).

\begin{figure}[h]
	\centering
	\includegraphics[width =0.48 \textwidth]{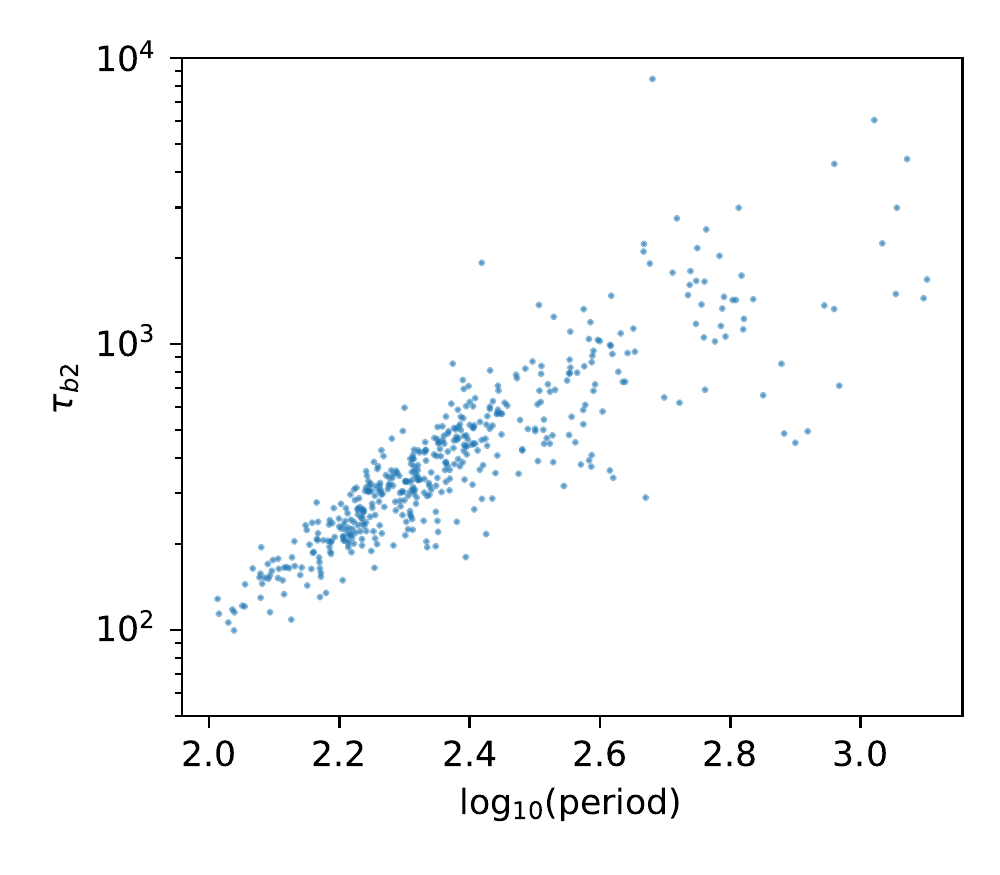}
	\includegraphics[width =0.48 \textwidth]{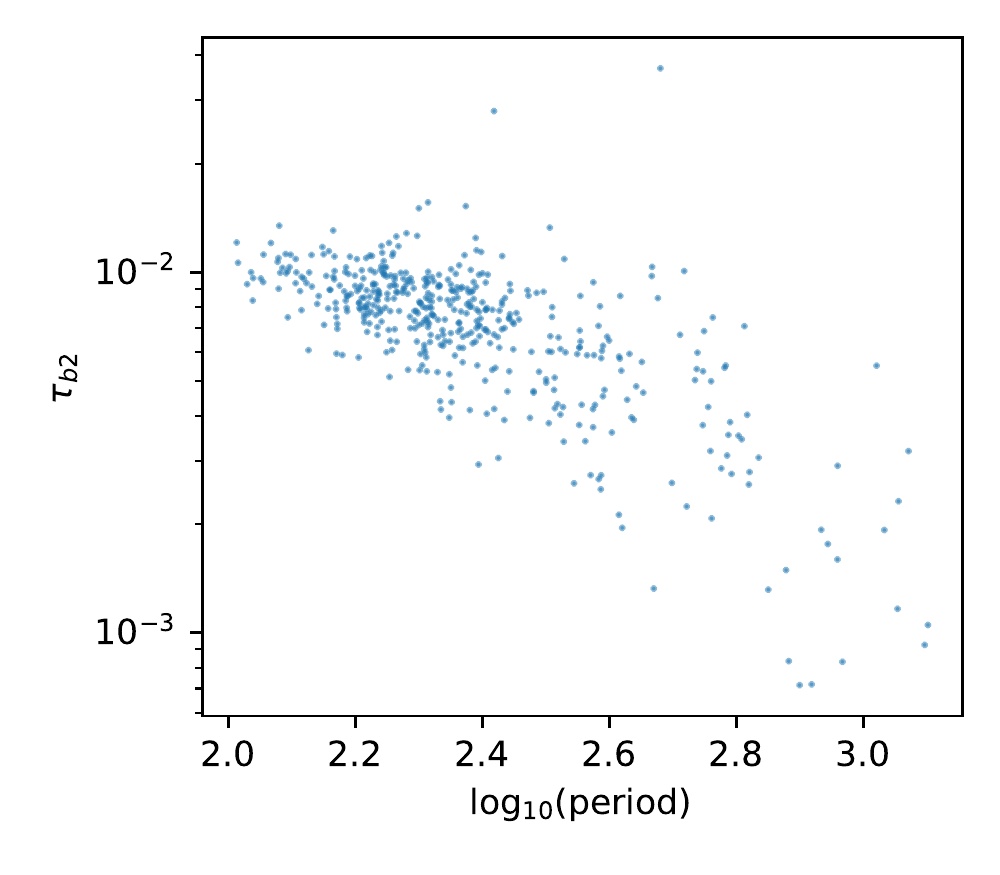}
	
	\caption{The fitted length-scale parameters $\tau_{b2}$ (on the $\log_{10}$ scale) versus the logarithm of the periods for O-rich LMC Mira $I$-band light curves. The model fitting is carried out independently to  469 OGLE LMC O-rich  Miras.  The left penal shows the fitted $\tau_{b2}$ of the unscaled model $h_{ib}(t)$. The right penal shows the fitted $\tau_{b2}$ of the scaled model $h_{ib}(f_i\cdot t)$. 
	} \label{fig:realdata:kernelTau}
\end{figure} 

\begin{figure}[h!]
	\centering
	\includegraphics[width =0.48 \textwidth]{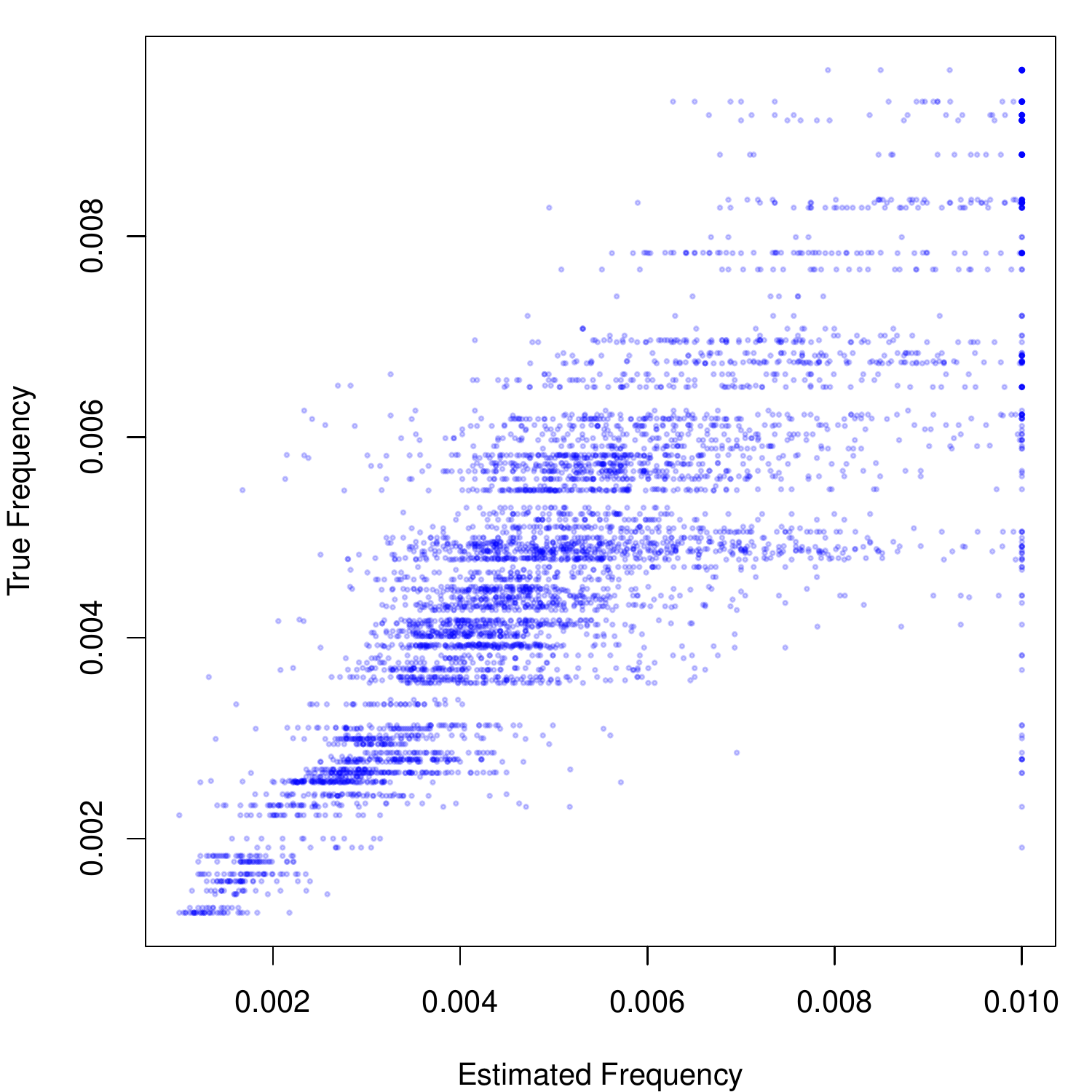}
	\includegraphics[width =0.48 \textwidth]{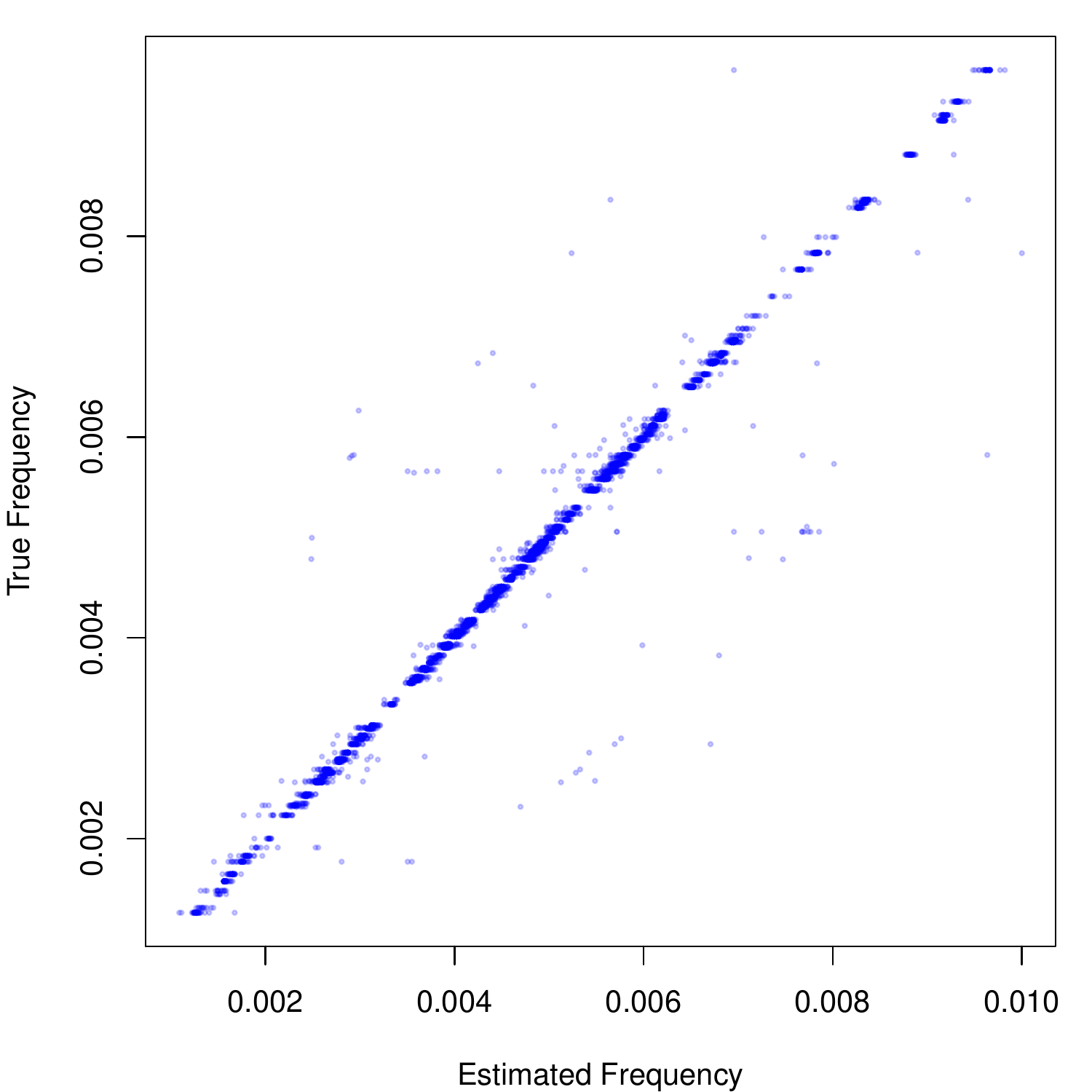}
	\caption{The frequency estimation comparison of using unscaled version $h_{ib}(t)$ (the left penal) versus the scaled version $h_{ib}(f_i\cdot t)$ (the right penal) in the Beyesian hierarchical model. In each penal, the horizontal axis is the estimated frequency, and the vertical axis is the true frequency. } \label{fig:simu:trueEst}
\end{figure} 

The left and right panels of Figure~\ref{fig:realdata:kernelTau} shows respectively the results of using the unscaled model $h_{ib}(t)$ and the scaled model $h_{ib}(f_i \cdot t)$. The vertical axis (on the $\log_{10}$ scale) is the optimal fitted $\tau_{b2}$, and the horizontal axis is the logarithm of the period. The left panel  shows that $\tau_{b2}$ varies in a large range, invalidating the idea of letting all Miras share the same length-scale parameter, while the right panel shows much smaller scattering of $\tau_{b2}$. The positive correlation between $\tau_{b2}$ and $\log_{10}(p)$ is evident on the left panel but much less so on the right panel. This result suggests that the scaled model $h_{ib}(f_i \cdot t)$ is preferred if we let all Miras share the same length-scale parameter in the Gaussian process model of the non-periodic component.

\textit{Second result.} 
Using the simulated data for 5,000 O-rich Miras presented in Section~4, we tested the unscaled version $h_{ib}(t)$ with the shared kernel parameter for all Miras, and keeping the rest of procedures the same in our Bayesian hierarchical model.
It turns out the model performance deteriorates considerably. More precisely, when $h_{ib}(t)$ is used instead of $h_{ib}(f_i\cdot t)$, the recovery rate (RR) as defined in the paper drops from $98.04\%$ to $28.04\%$, and the absolute deviation error (ADE) increases from $0.52\times 10^{-4}$ to $8.51\times 10^{-4}$. 

Figure~\ref{fig:simu:trueEst} depicts the estimation result. The left penal shows the result of using $h_{ib}(t)$, while the right penal presents the result of using $h_{ib}(f_i \cdot t)$. In each penal, the horizontal axis is the estimated frequency, and the vertical axis is the true frequency. It is obvious that the left panel shows a large scattering, while on the right panel most points reside on the diagonal. This result strongly supports our use of the scaled model $h_{ib}(f_i\cdot t)$.

\section{Additional Simulation Study}
\label{sec:suppsimu}

In addition to the simulation in the main text,  we have conducted a more extensive simulation. We generated 90 datasets of O-rich Mira light curves, containing  observations in the $I$ and $K_s$ bands. In a given dataset, each star has exactly $n_I$ and $n_K$ data points in the $I$ and $K_s$ bands, respectively.  The number of data points $(n_K, n_I)$ takes one of the following ten choices: $(5,5)$, $(5,10)$, $(5,20)$, $(5,30)$, $(10,10)$, $(10,20)$, $(10,30)$, $(20,20)$, $(20,30)$ and $(30,30)$.

Each dataset chooses one of three patterns (cadence) to generate time points. The observation time points are generated by the cadence of the M33 dataset (C1), the cadence of  the OGLE LMC dataset (C2), or simply by uniform distribution (C3). For the first cadence C1, the $K_s$-band data is obtained at a later time than the $I$-band data. For the second cadence C2, the two bands are observed roughly at the same time but have a seasonal pattern, such that there are no observations for a range of months each year when the Sun is too close to the galaxy. As for the last cadence, the two bands have similar uniform coverage over the whole observational time interval.

\begin{figure}[t]
	\centering
	\includegraphics[width = 0.65\textwidth]{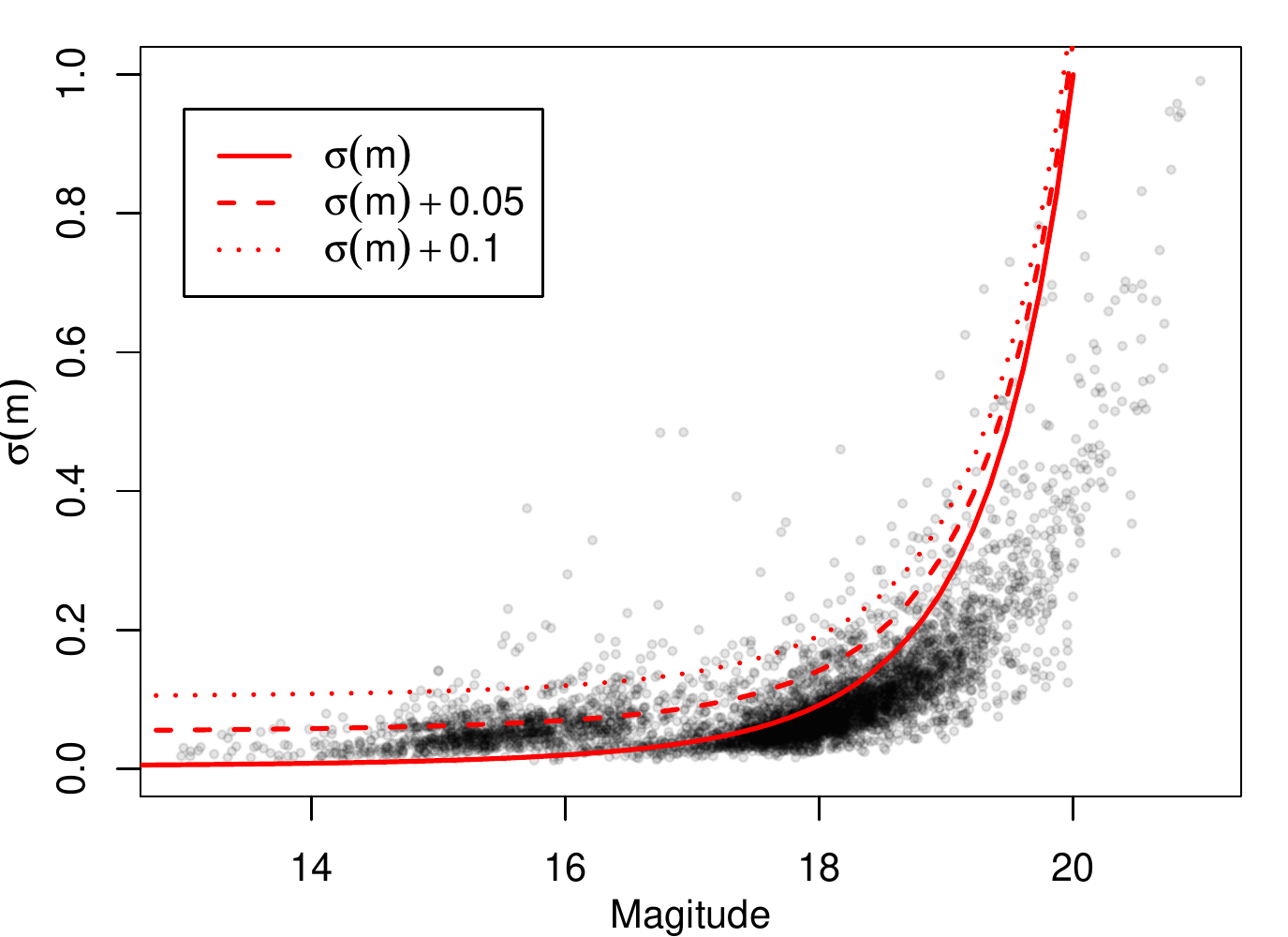}
	\caption{The magnitude noise relation for Simulation II. The three noise levels (N1, N2, and N3) are plotted as solid, dashed and dotted curves, respectively.  \label{fig:simu2:magnoise}}
\end{figure}

Given the observation time points, the light curve magnitudes are also generated from the template of \cite{yuan2017large}.  In addition, Gaussian noise is added to the magnitudes. According to real data from the actual survey, the noise level increases as the object becomes fainter (with larger magnitude values). A commonly-used function to fit the magnitude-noise relation is $\sigma(m) = \exp\left(a \cdot m^c - b \right)$. We adopted the values $a = 1.82\times 10^{-6}$,  $b = 5.84$ and $c = 5$ that closely match the OGLE survey. This is our first noise setting (N1). The other two settings increase the noise level by using $\sigma(m)+0.05$ (N2) and $\sigma(m) + 0.1$ (N3), respectively. Figure~\ref{fig:simu2:magnoise} illustrates the three noise levels. In the figure, the grey points represent the actual noise-mag relation from the OGLE survey. They are 20,000 randomly-selected observation points from OGLE. For our second simulation, the signal-to-noise ratio is lower than that of the OGLE survey, as the curves are slightly higher than the points. Therefore, it is more difficult to recover the correct periods.

\begin{figure}[h]
	\centering
	\includegraphics[width =0.95\textwidth]{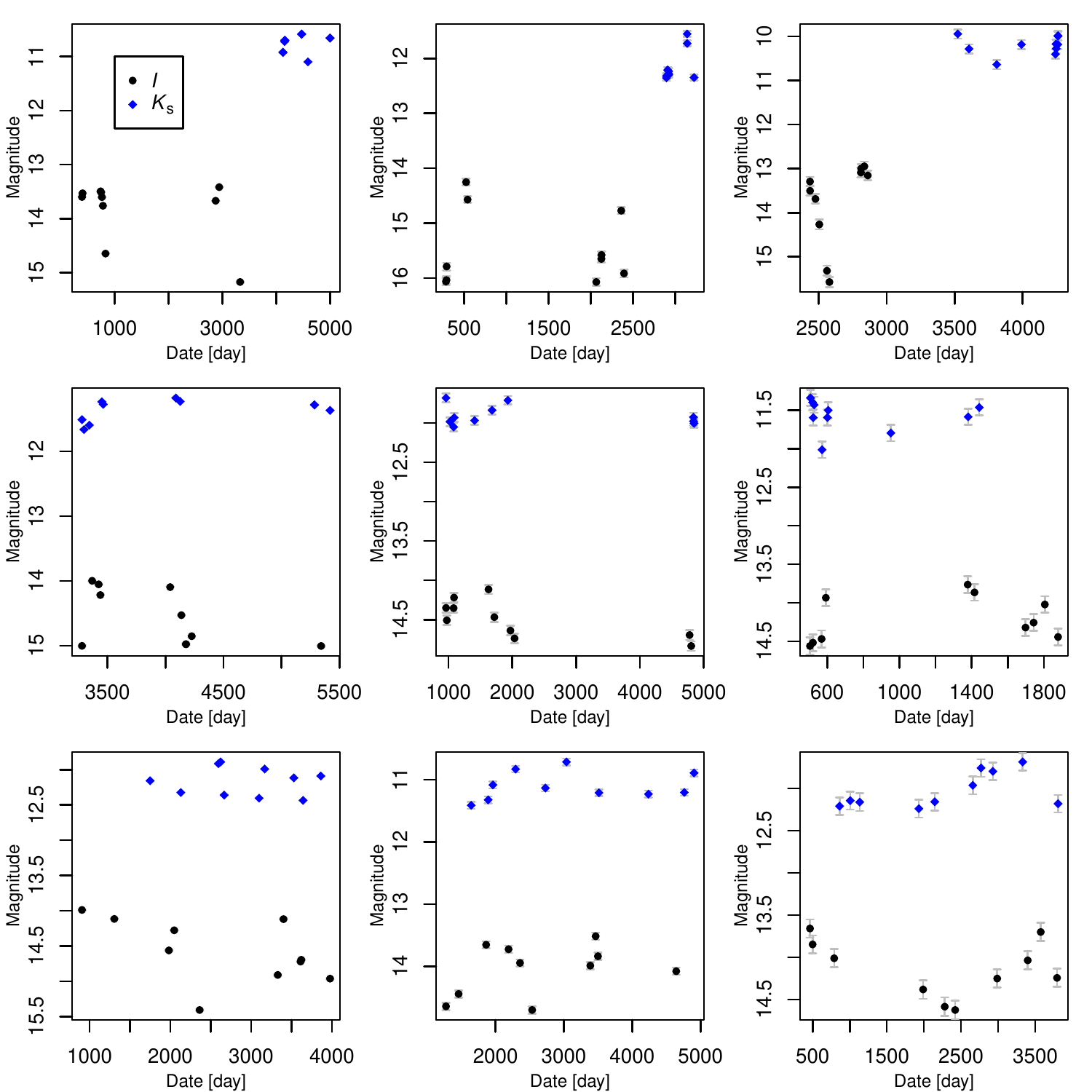}
	\caption{Randomly selected synthetic Mira data for Simulation 2. The first, second and third rows
		correspond to sampling cadence C1, C2 and C3, respectively. The first, second, and third columns
		correspond to noise level N1, N2 and N3, respectively.  Black points are $I$-band light curves, and blue squares
		are $K_s$-band light curves.} \label{fig:simu2:examples}
\end{figure}

In summary, we have 10 choices for the number of data points, 3 choices of sampling cadence, and 3 choices of noise level. By all  possible combinations, the above procedure produces 90 datasets. Figure~\ref{fig:simu2:examples} illustrates a few generated light curves for the choice of sample size $n_K=30$ and $n_I = 20$. The first, second and third rows correspond to sampling cadences C1, C2 and C3, respectively, while the first, second, and third columns correspond to noise levels N1, N2 and N3, respectively. The black points and blue squares denote $I$- and $K_s$-band light curves, respectively.

\begin{figure}[t]
	\centering
	\includegraphics[width =0.78 \textwidth]{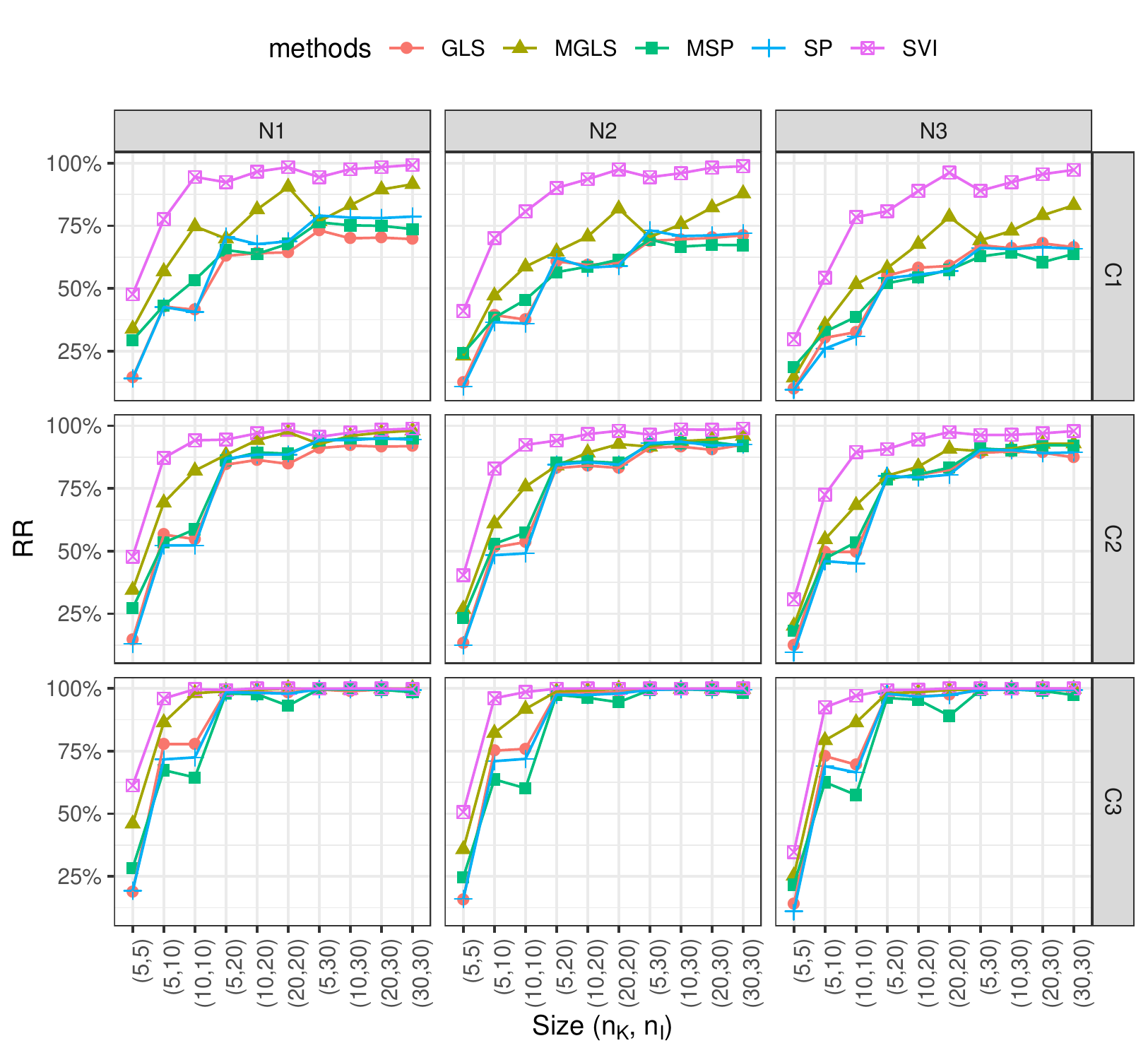}
	\caption{The recovery rate for Simulation II. } \label{fig:simu2:rr}
\end{figure}

\begin{figure}[t]
	\centering
	\includegraphics[width =0.78\textwidth]{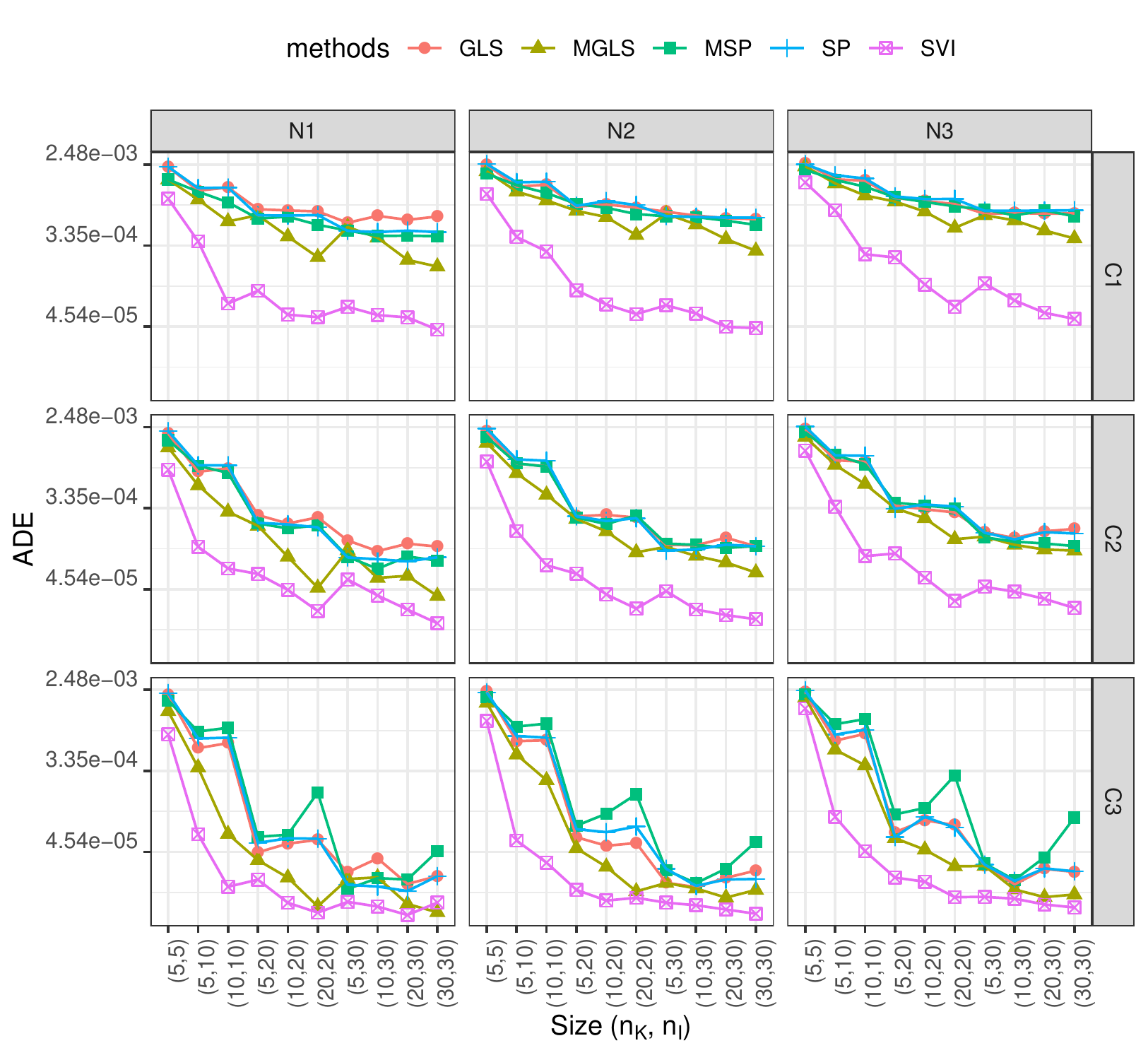}
	\caption{The absolute deviation error  for Simulation II. }
	\label{fig:simu2:msre}
\end{figure}

We apply the methods SP, MSP, GLS, MGLS and SVI to the simulated datasets. In these datasets, the single-band methods are applied to the $I$-band data. The same metrics as in the simulation Section~\ref{sec:simulation} of the paper, RR and ADE, are used for model comparison. Figure~\ref{fig:simu2:rr} shows the recovery rate for the 90 datasets. The horizontal axis represents the varying sample sizes $(n_K, n_I)$ for $K_s$- and $I$-band data. Its  columns correspond to different noise levels (N1--3 from left to right), and its rows for various sampling cadences (C1--3 from up to bottom). Similarly, Figure~\ref{fig:simu2:msre} shows the result for the ADE, with its vertical axes on the  logarithm scale. It is remarkable that the proposed method SVI has the best performance, and  consistently estimates period with higher accuracy for all of the datasets. For this simulation, the single-band methods SP and GLS have a similar performance. However, MGLS outperforms MSP in both metrics.  This  suggests the MSP might not be robust for a dataset with slightly different configurations from its model setting.

\section{Visual assessment of light curve fitting for Miras in M33}~\label{sec:visual}
\begin{figure}[p]
	\centering
	\includegraphics[width =0.48 \textwidth]{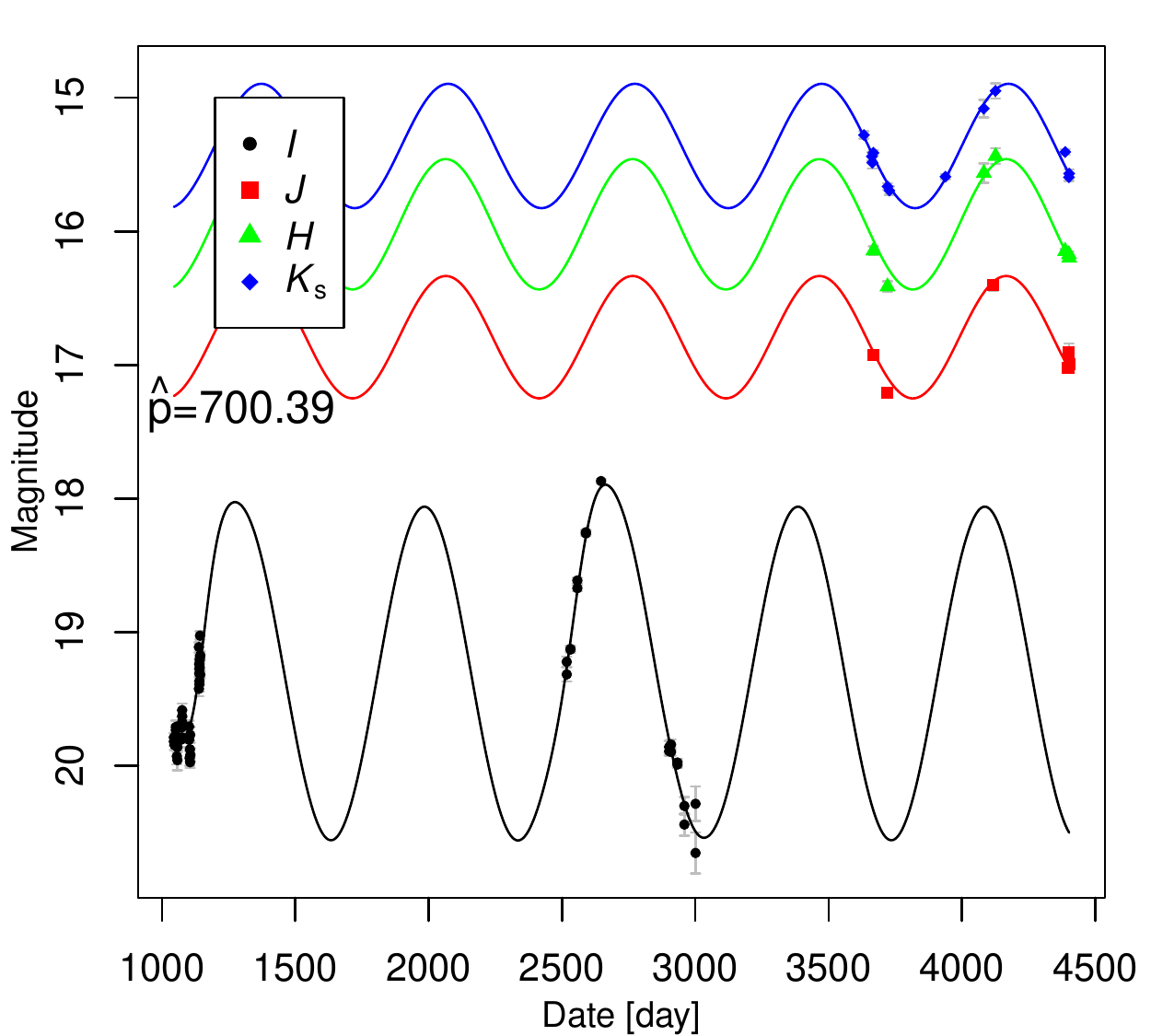}
	\includegraphics[width =0.48 \textwidth]{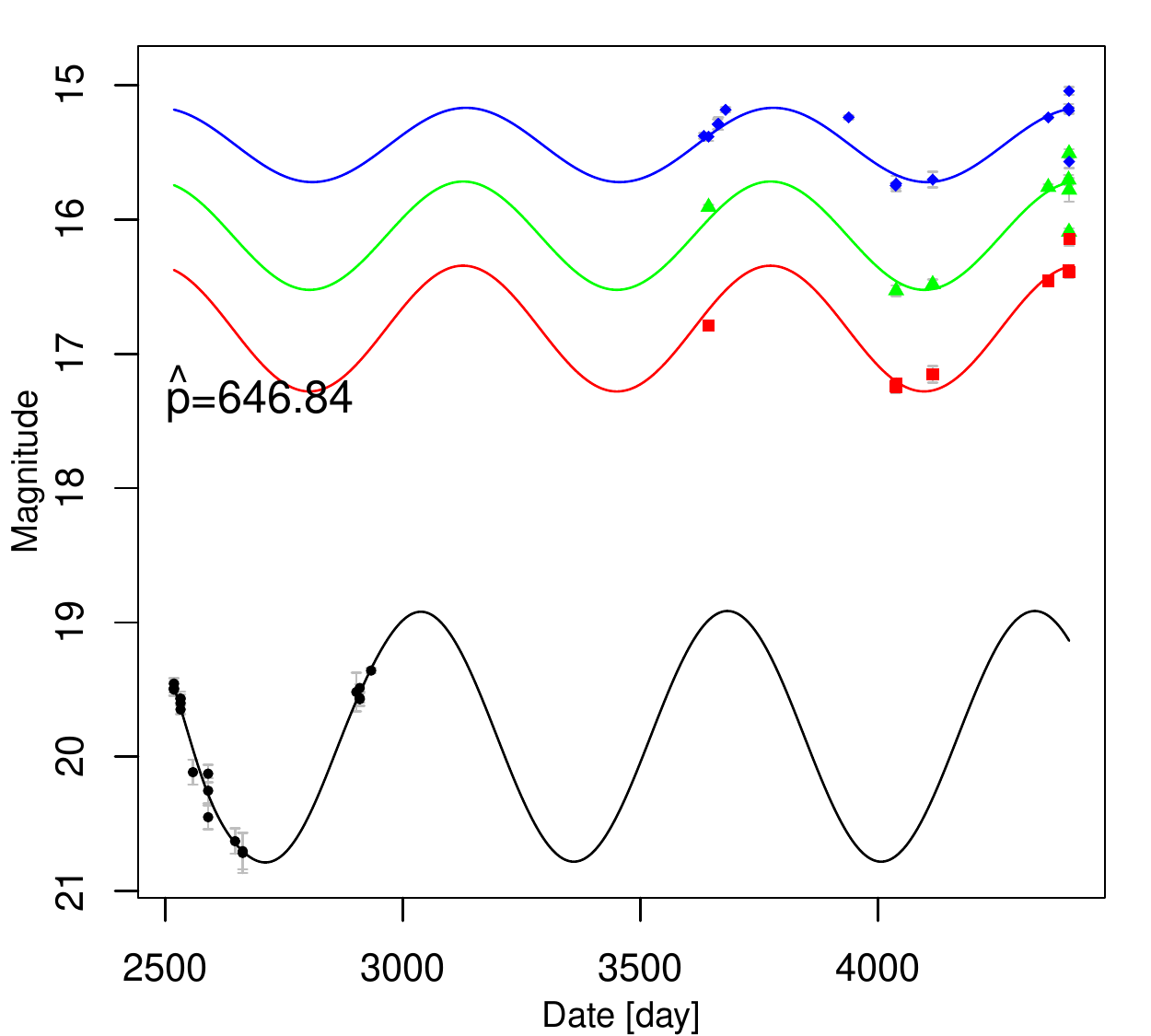}
	\includegraphics[width =0.48 \textwidth]{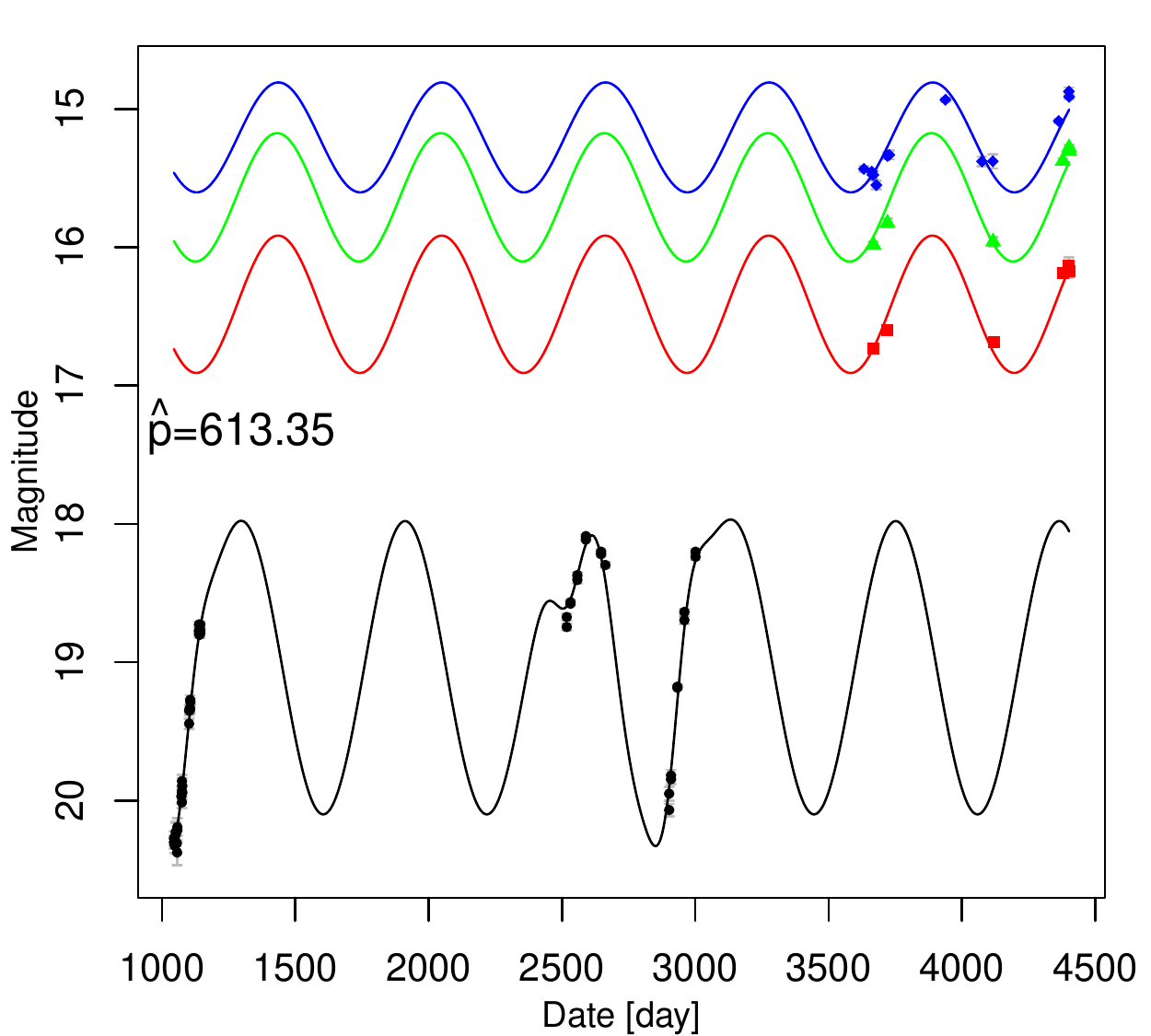}
	\includegraphics[width =0.48 \textwidth]{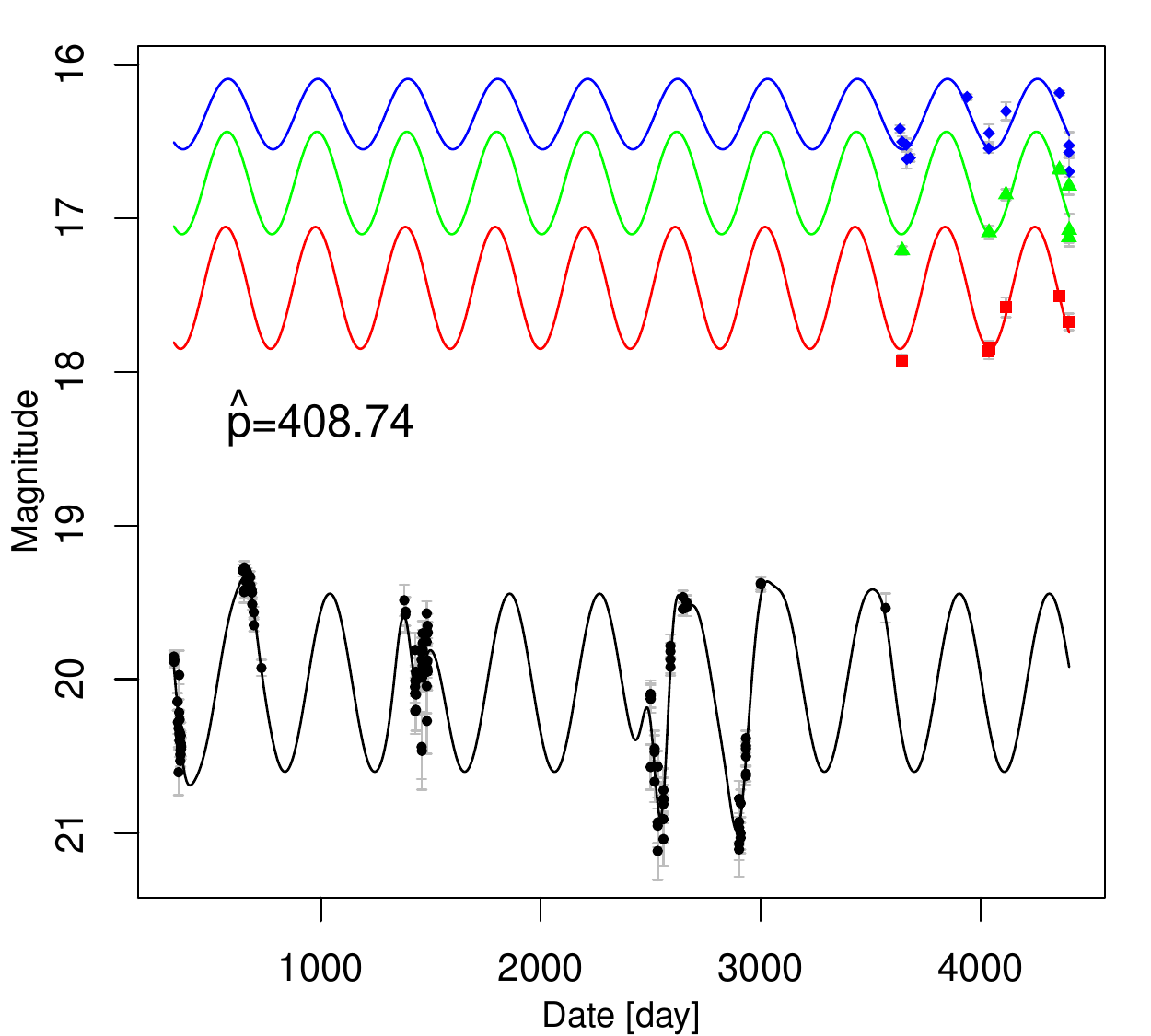}
	\includegraphics[width =0.48 \textwidth]{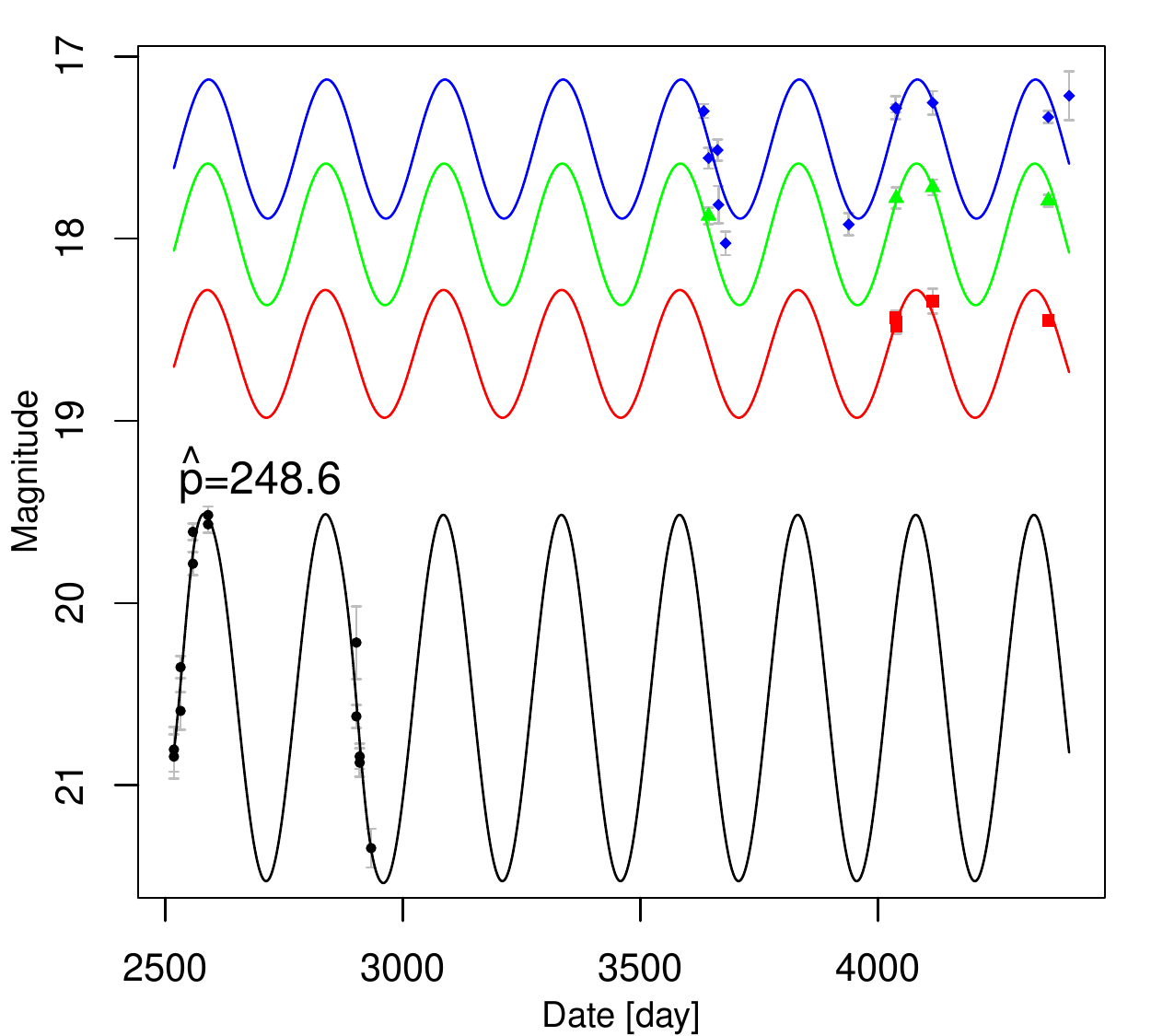}
	\includegraphics[width =0.48\textwidth]{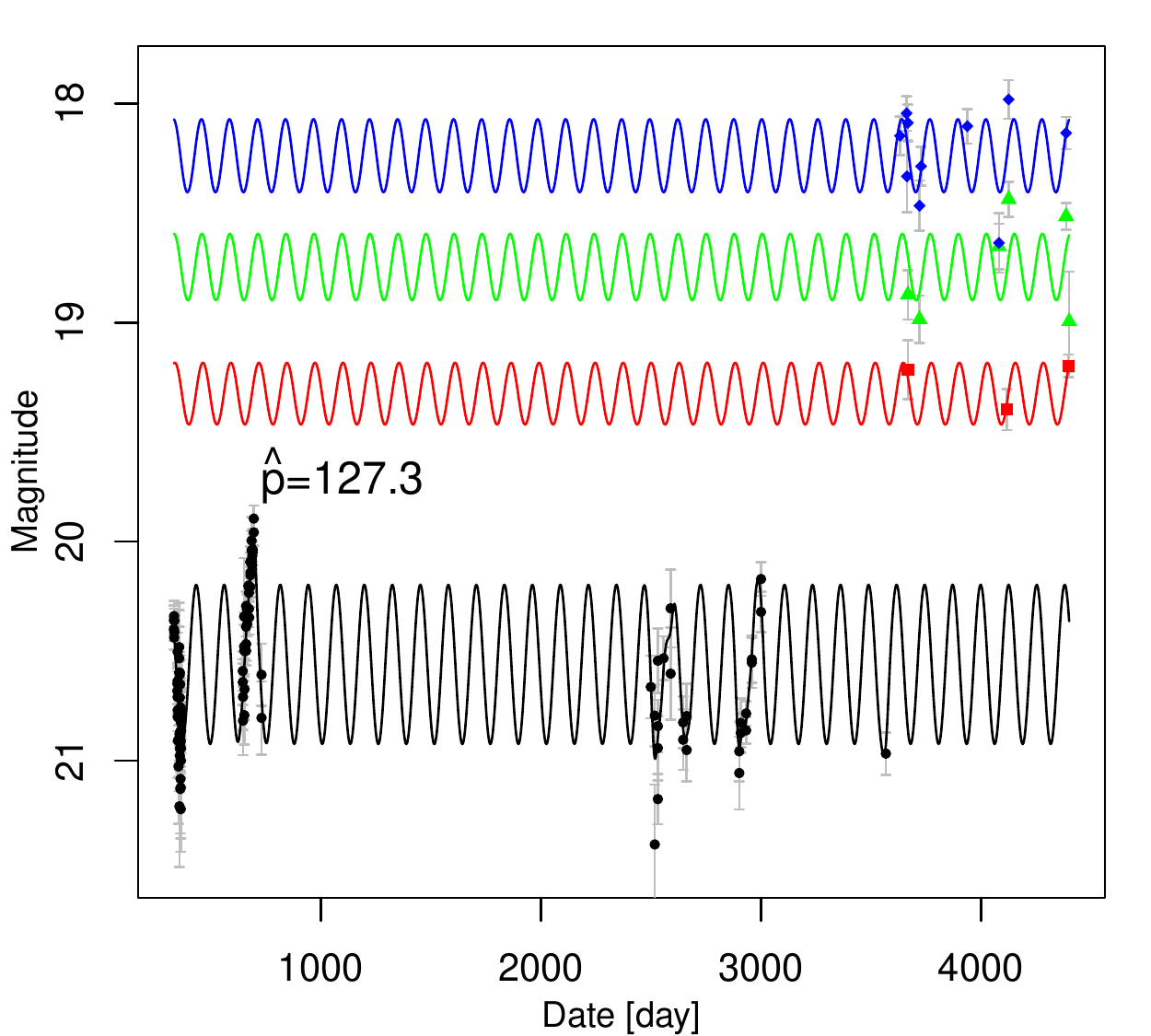}
	
	\caption{Light curve fitting for six randomly selected Miras in M33. The double exponential kernel is used for the Gaussian process.} \label{fig:realdata:fitlc2}
\end{figure}

The quality of light curve fitting using Gaussian process can be visually assessed by plotting the MAP estimator. Figure~\ref{fig:realdata:fitlc2} plots the data and fitted curves for six randomly selected Miras in M33. The light curves are fitted by our model with the double exponential kernel.  The model provides reasonable fits to the data. In particular, for the sparse M33 light curves, the $J,H,K_s$ bands show smaller variability beyond the sinusoidal periodical component. The $I$-band light curves have larger variability beyond the sinusoidal periodical component, and the corresponding kernel parameter $\tau_{b1}$ is larger than that of the other three bands. The kernel we used is 
$$k_{b}(u, u') = \tau_{b1} \cdot \exp\big\{-  (u-u')^2 /\tau_{b2} \big\} +  \tau_{b3}\cdot \mathrm{I}(u=u').$$
The table below summaries the kernel parameter fitted to each band of the real data. The values of $\tau_{b2}$ (or $\tau_{b3}$)  among different bands are close, the minor difference may not be easily explained. 

\begin{table}[h!]
	\centering
	\begin{tabular}{c|c|c|c}
		band $b$ & $\tau_{b1}$ & $\tau_{b2}$ & $\tau_{b3}$  \\
		\hline 
		$I$ & $e^{-3.33}$ &$e^{-4.20}$ & $e^{-4.21}$ \\
		$J$ & $e^{-10.48}$ &$e^{-3.00}$ & $e^{-5.28}$ \\
		$H$ & $e^{-10.00}$ &$e^{-2.57}$ & $e^{-3.90}$ \\
		$K_s$ & $e^{-10.59}$ &$e^{-4.39}$ & $e^{-4.18}$ \\
	\end{tabular}
\end{table}

\end{document}